\documentclass[11pt]{article}

\RequirePackage{amsthm,amsmath,amsfonts,amssymb}
\RequirePackage[numbers]{natbib}
\RequirePackage[colorlinks,citecolor=blue,urlcolor=blue]{hyperref}
\RequirePackage{graphicx}
\usepackage{xcolor}
\usepackage{authblk}

\newcommand{\field}[1]{\mathbb{#1}}

\newcommand{\F}{\field{F}}
\newcommand{\WC}{\textcolor{black}}
\newcommand{\p}{\field{P}}

\newcommand{\E}{\field{E}}
\newcommand{\pp}{\mathcal{P}}
\newcommand{\FF}{\mathcal{F}}

\newcommand{\bs}{\boldsymbol}

\def\argmax{\mathop{\mbox{argmax}}}
\def\argmin{\mathop{\mbox{argmin}}}

\allowdisplaybreaks 
\newtheorem{theorem}{Theorem}[section]
\newtheorem{example}{Example}[section]
\newtheorem{remark}{Remark}[section]
\newtheorem{lemma}{Lemma}[section]

\newtheorem{corol}{Corollary}[section]

\newtheorem{proposition}{Proposition}[section]

\newtheorem{corollary}{Corollary}[section]

\def\lf{\lfloor}
\def\rf{\rfloor}

\def\argmax{\mathop{\mbox{argmax}}}
\def\argmin{\mathop{\mbox{argmin}}}
\def\mf{\mathbf }
\def\bs{\boldsymbol}

\renewcommand{\thelemma}{\arabic{section}.\arabic{lemma}}
\renewcommand{\theproposition}{\arabic{section}.\arabic{proposition}}

\renewcommand{\theequation}{\arabic{section}.\arabic{equation}}
\renewcommand{\thetheorem}{\arabic{section}.\arabic{theorem}}

\renewcommand{\thesection}{\arabic{section}}  
\renewcommand{\theremark}{\arabic{section}.\arabic{remark}}

\begin{document}

	\title{Adaptive Estimation for  locally stationary Factor Models And A Test for Static Factor Loadings}

	\author[1]{Weichi Wu}
	\author[2]{Zhou Zhou}
	\affil[1]{\small Center for Statistical Science, Deparment of Industrial Engineering, Tsinghua University}
	\affil[2]{\small Department of Statistical Sciences, University of Toronto}
	\maketitle
\begin{abstract}
This paper considers the estimation and testing of a class of locally stationary time series factor models with evolutionary temporal dynamics. In particular, the entries and the dimension of the factor loading matrix are allowed to vary with time while the factors and the idiosyncratic noise components are locally stationary.   We propose an adaptive sieve estimator for the span of the varying loading matrix and the locally stationary factor processes. A uniformly consistent estimator of the effective number of factors is investigated via eigenanalysis of a non-negative definite time-varying matrix. A possibly high-dimensional bootstrap-assisted test for the hypothesis of static factor loadings is proposed by comparing the kernels of the covariance matrices of the whole time series with their local counterparts. We examine our estimator and test via simulation studies and real data analysis. Finally, all our results hold  at the following popular but distinct assumptions: (a) the white noise idiosyncratic errors with either fixed or diverging dimension, and (b) the correlated idiosyncratic errors with diverging dimension.
\end{abstract}

\section{Introduction}
Technology advancement has made it easy to record simultaneously a large number of stochastic processes of interest over a relatively long period of time where the underlying data generating mechanisms of the processes are likely to evolve over the long observation time span. As a result both high dimensional time series analysis (\cite{wei2018multivariate}) and locally stationary time series analysis (\cite{dahlhaus2012locally}) have undergone unprecedented developments over the last two decades.  This paper focuses on the following evolutionary linear factor model for a multivariate locally stationary time series:
\begin{eqnarray}\label{eq:fm}
\mf x_{i,n}=\mf A(i/n)\mf z_{i,n}+\mf e_{i,n},
\end{eqnarray}
where $\{\mf x_{i,n}\}_{i=1}^n$ is a $p$-dimensional observed time series, $\mf A(t)$: $[0,1]\rightarrow \field{R}^{p\times d(t)}$ is a matrix-valued function of possibly time-varying factor loadings and the number of factors $d(t)$ is assumed to be a piecewise constant function of time, $\{\mf z_{i,n}\}_{i=1}^n$ is a $d(i/n)$-dimensional unobserved sequence of common factors and $\{\mf e_{i,n}\}_{i=1}^n$ are the idiosyncratic components. Here $p=p_n$ may diverge to infinity with the time series length $n$ and $d(t)$ is typically much smaller than $p$ uniformly over $t$. Note that $\mf x_{i,n}$, $\mf z_{i,n}$ and $\mf e_{i,n}$ are allowed to be locally-stationary processes for which the generating mechanism varies with time, see \eqref{eq-2} for detailed formulation. Throughout the article we assume that $\{\mf e_{i,n}\}$ and $\{\mf z_{i,n}\}$ are centered. 

The version of model \eqref{eq:fm} with constant loadings is among the most popular dimension reduction tools for the analysis of multivariate stationary time series (\cite{stock2011dynamic}, \cite{tsay2013multivariate}, \cite{wei2018multivariate}). According to the model assumptions adapted and estimation methods used, it seems that recent literature on linear time series factor models mainly falls into two types. The  cross-sectional averaging method (summarized in \cite{stock2011dynamic}) which is popular in the econometric literature of linear factor models, exploits the assumption of weak dependence among the vector components of $\mf e_{i,n}$ and hence achieves de-noising via cross-sectional averaging. See for instance  \cite{stock2002forecasting}, \cite{bai2002determining}, \cite{bai2003inferential} and \cite{forni2000generalized} among many others. One advantage of the cross-sectional averaging method is that it allows for a very high dimensionality. In general the method requires that $p$ diverges to achieve consistency and the estimation accuracy improves as $p$ gets larger under the corresponding model assumptions. On the other hand,  the linear factor model can also be fitted by exploring the relationship between the factor loading space and the auto-covariance or the spectral density matrices of the time series under appropriate assumptions. This method dates back at least to the works of \cite{anderson1963use}, \cite{brillinger2001time} and \cite{pena1987identifying} among others for fixed dimensional multivariate time series and is extended to the high dimensional setting by the recent works of \cite{lam2011estimation}, \cite{lam2012factor}, \cite{wang2019factor} and others. The latter method allows for stronger contemporary dependence among the vector components and is consistent when $p$ is fixed under the requirement that the idiosyncratic components form a white noise.

To date, the literature on non-stationary linear factor models is relatively scarce and most existing results are focused on extensions of the cross-sectional averaging method. Among others, 
\WC{ \cite{pelger2022state-varying} 
	\cite{motta2011locally} and \cite{su2017time} considered evolutionary model \eqref{eq:fm} using the cross-sectional averaging method.} \cite{eichler2011fitting} and \cite{barigozzi2021time} studied non-stationary dynamic factor models. See also \cite{forni2000generalized} for the first use of Brillinger's spectral PCA approach to the analysis of dynamic factor models. 
\WC{Non-stationary factor models with time-varying loadings and diemionsality have also drawn some attention in Bayesian analysis. Prominent examples inculde  \cite{aguilar2000bayesian} and \cite{Nakajima2013}, among others.}
In this paper, we shall extend the second estimation method mentioned in the last paragraph to the case of evolutionary factor loadings with locally stationary  factor and  idiosyncratic component time series whose data generating mechanisms change smoothly over time while allowing for weakly correlated idiosyncratic components.
 Using this framework, our approach to the factor model estimation and the corresponding theory contribute to the literature mainly in the following three aspects. 
\begin{description}
\item (a) \WC{Our proposed adaptive estimator is proven to be consistent under two sets of assumptions. The first set allows for the dimension \( p \) to be either fixed or diverging, with the requirement that the noises \( \mathbf{e}_i \) form white noise. This assumption is commonly posited by most approaches that explore factor loading space through the structure of autocovariance or spectral density matrices. Meanwhile, the second set of assumptions permits weak correlation among the \( \mathbf{e}_i \)'s, accommodating many classic time series models, while requiring that the dimension \( p \) diverges. This set of conditions has been adopted by most literature that utilizes the cross-sectional averaging method. To the best of our knowledge, our method is the first to demonstrate consistency under both sets of popular conditions for locally stationary factor models with time-varying factor loadings, thereby offering a broad application scope.}

	\item (b) 
 To estimate the  time-varying loading matrix, the prevailing approach in the literature is the local-constant kernel estimator, see for example \cite{motta2011locally}, \cite{su2017time}. It seems that it is difficult to extend the local-constant method to general local polynomial methods for factor models under the cross-sectional averaging set-up and therefore the estimation accuracy of the existing methods is not adaptive to the smoothness (with respect to time) of the factor loading matrix function. In this paper, we propose an alternative adaptive estimation method based on the method of sieves ( \cite{chen2007large}). The sieve method is computationally simple to implement and has the advantage of being {\it adaptive to the unknown smoothness of the target function} if certain linear sieves such as the Fourier basis (for periodic functions), the Legendre polynomials or the orthogonal wavelets are used (\cite{chen2007large}, \cite{wang2012convergence}).
Specifically, we adapt the method of sieves to estimate the high-dimensional auto-covariance matrices of $\mf x_{i,n}$ at each time point and subsequently estimate \WC{the space spanned by the loadings $\mf A(t)$ at each $t$  exploiting the relationship between $\mf A(\cdot)$ and the kernel of the latter local auto-covariance matrices. We use "span" for "space spanned" in the remaining of the article for short. } We will show that the span of $\mf A(\cdot)$ can be estimated at a rate independent of $p$ uniformly over time provided that all factors are strong with order $p^{1/2}$ Euclidean norms, \WC{extending} the corresponding result for factor models with static loadings established in \cite{lam2011estimation}. 

\item (c) In most literature for time-varying factor models such as \cite{motta2011locally}, \cite{su2017time}, to estimate the time-varying loading matrix, it is assumed that the number of factors is constant over time.  Typically further assumptions on the factor process such as  independence or time-invariance of its covariance matrix were required. 
In this paper, we model the factor process as general locally stationary time series and allow the number of factors to be time-varying. 
Uniform consistency of the estimated span of the loading matrix as well as the number of factors will be established without assuming that the positive eigenvalues of the corresponding matrices are distinct which is commonly posited in the literature of factor models.

\end{description}
Testing whether $\mf A(\cdot)$ is constant over time is important in the application of \eqref{eq:fm}. In the literature, among others \cite{breitung2011testing} proposed LR, LM
and Wald statistics for testing static factor model against an alternative of piece-wise constant loadings, and \cite{yamamoto2015testing}  improved the power of \cite{breitung2011testing} by maximizing the test statistic over possible numbers of the original factors. Assuming piece-wise stationarity, \cite{barigozzi2018simultaneous} estimated the change points of a factor model via wavelet transformations. \cite{su2017time} considered an ${\cal L}^2$ test of static factor loadings under the cross-sectional averaging framework assuming that each component of $\mf e_{i,n}$ is a martingale difference sequence. 
\WC{To the best of our knowledge,  this paper is the first to propose a high-dimensional ${\cal  L}^\infty$ or maximum deviation test on the time-invariance of the span of $\mf A(\cdot)$ which utilizes the observation that the kernel of the full-sample auto-covariance matrices coincides with all of its local counterparts under the null hypothesis of static span of loadings while the latter observation is likely to fail when the span of $\mf A(\cdot)$ is time-varying. Using the uniform convergence rates of the estimated factor loadings established in this paper, the test statistic will be shown to be asymptotically equivalent to the maximum deviation of the sum of  a high-dimensional locally stationary time series under some mild conditions.  A multiplier bootstrap procedure with overlapping blocks is adapted to approximate the critical values of the test.}
The bootstrap will be shown to be asymptotically correct under the null and powerful under a large class of local alternatives.  The theory and methodology of our testing procedure contribute to the literature mainly in the following two aspects.
\begin{description}
	\item (i) Under the null hypothesis of constant $\mf A(\cdot)$, the common components of the time series, i.e. $\mf A(i/n)\mf z_{i,n}$, considered in the above-mentioned works are stationary or have time-invariant variance-covariance. Since $\mathbf z_{i,n}$ is assumed to be locally stationary in this paper, under the null hypothesis,  the common components are allowed to be locally stationary where their variance-covariance matrices can be smoothly time-varying. 
	\item (ii) 
	\WC{The validity of the tests of the above works was built on the divergence of both the length of time series $n$ and the dimension of the time series $p$. In contrast, our proposed tests is proved to be asymptotically correct under two sets of assumptions. The first set of assumption is that $p$ is fixed or diverging slowly with $n$ while the idiosyncratic errors are white nose, and the second  set of assumption is that both $p$ and $n$ diverge while idiosyncratic errors can be correlated. Our estimation is also shown to be consistent under both sets of assumptions. Hence our results have a wide application scope.}

\end{description}
 
 \WC{Other methodological and theoretical innovations include the following. 1) We introduce a penalized eigen-ratio estimator for the dimension of the locally stationary low-dimensional common factors. We combine techniques from \cite{bhatia1982analysis} and \cite{yu2015useful} to establish and justify a sieve estimator for the factor loadings under the two sets of assumptions mentioned in (a) above. 2) We introduce a state-of-the-art high-dimensional Gaussian approximation technique to construct a test for static factor loading. Notably, the test involves the estimated idiosyncratic errors. Although the convergence rate of these estimated idiosyncratic errors is not sufficiently fast for a direct plug-in, we demonstrate that when combined with the eigenvectors of the null space of the loading matrix, the approximation error can be controlled.}

The paper is organized as follows. Section \ref{sec::notation} introduces some  notation.  Sections \ref{Sec:estimate}  and \ref{sec:test_loading} discuss the estimation of the evolutionary factor loading matrices and the test of static factor loadings, respectively. Section \ref{tech-assumption} contains some technical assumptions while Section \ref{Sec:load} presents the theoretical results of estimation.  Section \ref{Sec:null-static} investigates the theoretical properties of test of the static factor loading. Section \ref{Sec::Pre} discusses the time varying dimension.  Section \ref{Sec::Tuning} gives out methods for tuning parameter selection.  Simulation studies are displayed in Section \ref{Simu-Results}, and a real data analysis is in Section  \ref{data-ana}. 
Section \ref{appendix} provides the proofs of
 Theorem \ref{Thm-approx}, Theorem \ref{Space-Distance} (i) as well as some important preliminary assumptions on locally stationary multivariate time series. 
  A class of high dimensional locally stationary time series examples, the proofs of the remaining theorems, propositions, lemmas and corollaries are relegated to the online supplemental material.

\section{Notation}\label{sec::notation}

 For two series $a_n$ and $b_n$, write $a_n \asymp b_n$ 
if the exists $0<m_1<m_2<\infty$ such that $m_1\leq \liminf \frac{|a_n|}{|b_n|}\leq \limsup \frac{|a_n|}{|b_n|}\leq m_2<\infty$. Write $a_n\lessapprox b_n$ ($a_n\gtrapprox b_n$) if there exists a uniform constant $M$ such that $a_n\leq Mb_n$ ($a_n\geq Mb_n$). Let $A:=B$ represent "A is define as B".
For any $p$ dimensional (random) vector $\mf v=(v_1,...,v_p)^\top$, write $\|\mf v\|_u=(\sum_{s=1}^p|v_s|^u)^{1/u}$, and the corresponding $\mathcal L^v$ norm $\|\mf v\|_{\mathcal L^v}=(\E(\|\mf v\|^v_2))^{1/v}$ for $v\ge 1$. For any real symmetric matrix $\mf F$ let $\lambda_{max}(\mf F)$ be its largest eigenvalue,  $\lambda_{min}(\mf F)$ be its smallest eigenvalue, and $\lambda_{k}(\mf F)$ be its $k_{th}$ largest eigenvalue.   For any matrix $\mf F$ let $\sigma_{k}(\mf F)$  be $\mf F$'s $k_{th}$ largest singular value. Let $\|\mf F\|_F=(\text{trace} (\mf F^\top\mf F))^{1/2}$ denote the Frobenius norm, and  $\|\mf F\|_{m}$ be the positive square root
of the minimum eigenvalue of $\mf F\mf F^\top$ or $\mf F^\top \mf F$, whichever is a smaller matrix.  Denote by $vec(\mf F)$ the vector obtained by stacking the columns of $\mf F$.  Let $\|\mf F\|_2=\sqrt{\lambda_{max}(\mathbf F\mathbf F^\top)}$. In particular, if $\mf F$ is a vector, then $\|\mf F\|_2=\|\mf F\|_F$. We also write $|\mf F|$ for $\|\mf F\|_2$ if $\mf F$ is a vector.  For any vector or matrix $\mathbf A=(a_{ij})$ let $|\mf A|_{\infty}=\max_{i,j}|a_{ij}|$. For any integer $v$ let $\mf I_v$ denote the $v\times v$ identity matrix. Write $|\mathcal I|$ for the length of the interval $\mathcal I$. Let  $\mathcal C^{K}(\tilde M)[0,1]$ be the collection of functions $f$ defined on $[0,1]$ such that the $K_{th}$ order derivative of $f$ is Lipschitz continuous with  Lipschitz constant $\tilde M$, $\tilde M>0$. 

\section{Model Estimation}\label{Sec:estimate}\setcounter{equation}{0}
 Adapting the formulation in \cite{zhou2009local}, we model the $p,d,q$ dimensional locally stationary time series $\mf x_{i,n}$, ${\mf z}_{i,n}$ and $\mf e_{i,n}$, $1\leq i\leq n$ as follows: 
\begin{align}\label{eq-2}
	\mf x_{i,n}=\mf G(i/n,\FF_i),~ {\mf z}_{i,n}=\mf Q(i/n,\FF_i),~ \mf e_{i,n}=\mf H(i/n,\FF_i)
\end{align}
where the filtration $\FF_i=(...,\bs \epsilon_{i-1},\bs \epsilon_i)$ with $\{\bs \epsilon_i\}_{i\in \mathbb Z}$ 
i.i.d. random elements in some measurable space $\mathcal S$, and $\mf G:[0,1]\times \mathcal S^{\mathbb Z}\to \mathbb R^p$, $\mf Q: [0,1]\times \mathcal S^{\mathbb Z}\to \mathbb R^d$ and $\mf H: [0,1]\times \mathcal S^{\mathbb Z}\to \mathbb R^p$ are $p$, $d$ and $p$ dimensional measurable nonlinear filters. 
Write $j_{th}$ entry of the time series 
$\mf x_{i,n}$, ${\mf z}_{i,n}$ and $\mf e_{i,n}$ as $x_{i,j,n}=G_j(i/n,\FF_i)$, $z_{i,j,n}=Q_j(i/n,\FF_i)$ and $e_{i,j,n}=H_j(i/n,\FF_i)$. Let  $\{\bs \epsilon'_i\}_{i\in \mathbb Z}$ be an independent copy of $\{\bs \epsilon_i\}_{i\in \mathbb Z}$ and let $\FF^{(h)}_{i}=(\bs \epsilon_{-\infty},...\bs \epsilon_{h-1},\bs \epsilon_h',\bs \epsilon_{h+1},...,\bs \epsilon_i)$ for $h\leq i$, and $\FF^{(h)}_{i}=\FF_i$ otherwise. 
The dependence measures for $\mf z_{i,n}$  and $\mf e_{i,n}$ in $\mathcal L^l$ norm are defined  as 
\begin{align*}
	\delta^z_{l}(k):=\max_{1\leq j\leq d}\delta^z_{l,j}(k):=\max_{1\leq j\leq d}\sup_{t\in[0,1],i\in \mathbb Z}\E^{1/l}(|Q_j(t,\FF_i)-Q_j(t,\FF_i^{(i-k)})|^l),\\
	\delta^e_{l}(k):=\max_{1\leq j\leq p}\delta^e_{l,j}(k):=\max_{1\leq j\leq p}\sup_{t\in[0,1],i\in \mathbb Z}\E^{1/l}(|H_j(t,\FF_i)-H_j(t,\FF_i^{(i-k)})|^l),
\end{align*}
which quantify the magnitude of change of systems $ \mf Q, \mf H$ in $\mathcal L^l$ norm when the inputs of the systems $k$ steps ahead are replaced by their $i.i.d.$ copies. \WC{We also refer to \cite{zhang2021convergence} for the definition of local stationarity and functional dependence for high dimensional time series. Due to the page limit, we move  the regularity conditions for $\mf Q, \mf H$, and $\mf G$ as well as an example of a high dimensional moving average process  to the Section \ref{appendix} and the online supplemental material, respectively. } 
Observe from equation (\ref{eq:fm}) that for $k\geq 1$
\begin{align*}
    \mf x_{i+k,n}\mf x_{i,n}^\top=\mf A(\tfrac{i+k}{n})\mf z_{i+k,n}\mf z_{i,n}^\top \mf A^\top (\tfrac{i}{n})+
    \mf A(\tfrac{i+k}{n})\mf z_{i+k,n}\mf e_{i,n}^\top+\mf e_{i+k,n}\mf z_{i,n}^\top \mf A^\top(\tfrac{i}{n})+\mf e_{i+k,n}\mf e_{i,n}^\top.
\end{align*}
In this paper we consider two set of assumptions. The first is that the idiosyncratic component are uncorrelated with past common factors and past idiosyncratic component as assumed by \cite{lam2012factor}, i.e.,  $\E(\mf e_{i+k,n}\mf e_{i,n}^\top)=0$ and  $\E(\mf e_{i+k,n}\mf z_{i,n}^\top)=0$. The dimension $p$ can be either diverging or fixed.  The second set of assumptions allows   $\E(\mf e_{i+k,n}\mf e_{i,n}^\top)\neq 0$, instead assuming $p\rightarrow \infty$, and $\|\bs\Sigma_{e}(t,k)\|_{2}:=\|\E(\mf H(t,\FF_{i+k})\mf H^\top(t,\FF_{i}))\|_2$ is bounded.   Denote the $k_{th}$ order auto-(cross)covariance by  
$\bs \Sigma_{z}(t,k)=\E({\mf Q}(t,\FF_{i+k}){\mf Q}^\top(t,\FF_i))$ 
and $\bs \Sigma_{ze}(t,k)=\E({\mf Q}(t,\FF_{i+k})\mf H^\top(t,\FF_i))$, $\bs \Sigma_{x}(t,k)=\E(\mf G(t,\FF_{i+k})\mf G^\top(t,\FF_i))$.  Under the first assumption, taking expectation on both sides of the above equation will yield for $ k\ge 1$,
\WC{\begin{eqnarray}\label{Lambdanull}
\bs \Sigma_x(i/n,k) \approx \E(\mf x_{i+k,n}\mf x_{i,n}^\top) = \mf A(\tfrac{i+k}{n})\E(\mf z_{i+k,n}\mf z_{i,n}^\top) \mf A^\top (\tfrac{i}{n})+
    \mf A(\tfrac{i+k}{n})\E(\mf z_{i+k,n}\mf e_{i,n}^\top)\notag\\
\approx  \mf A(i/n)\bs \Sigma_z(i/n,k)\mf A^\top(i/n)+\mf A(i/n)\bs \Sigma_{ze}(i/n,k).
\end{eqnarray}}
Further define $\mf \Lambda(t)=\sum_{k=1}^{k_0}\bs \Sigma_x(t,k)\bs \Sigma^\top_x(t,k)$ for some pre-specified integer $k_0$ and we  have $\mf \Lambda(t)\approx\mf \Lambda_1(t)$ where the positive semidefinite matrix  $\mathbf \Lambda_1(t)$ is defined as 
\begin{align}\label{DefLambda}
\mf \Lambda_1(t)=\mf A(t)\Big[\sum_{k=1}^{k_0}(\bs \Sigma_z(t,k)\mf A^\top(t)+\bs \Sigma_{ze}(t,k))(\mf A(t)\bs \Sigma^\top_z(t,k)+\bs \Sigma^\top_{ze}(t,k))\Big]\mf A^\top(t).
\end{align} Therefore in principle the span of $\mf A(t)$ can be identified by the null space of $\mf \Lambda(t)$. Under the second set of assumptions, since $\E(\mf e_{i+k,n}\mf e_{i,n}^\top)\neq 0$ we can write $\mf \Lambda(t)$ by $\mf \Lambda(t)=\mf \Lambda_1(t)+\mf \Lambda(t)-\bs \Lambda_1(t)$, where $\mf \Lambda(t)-\bs \Lambda_1(t)$ is a symmetric matrix. The boundedness of $\|\Sigma_{e}(t,k)\|_{2}$ will lead to  $\|\mf \Lambda(t)-\bs \Lambda_1(t)\|_2=o(\inf_t\lambda_{d(t)}(\mf \Lambda_1(t)))$. Therefore, an application of Davis-Khan theorem shows that the eigenvectors of null space of {\color{black}$\mf A^\top(t)$} will be close to the eigenvectors of $\mf \Lambda(t)$ with respect to its $(d+1)_{th},...p_{th}$ eigenvalue (in descending order).  
 The use of $\mf \Lambda(t)$ was advocated in \cite{lam2011estimation} under the first set of assumptions, and not considered under the second set of assumptions to the best of the authors' knowledge. In this paper we aim at estimating a set of time-varying orthonormal basis of this time-varying null space, which is identifiable up to rotation, to characterize $\mf A(t)$. 
 \WC{The identification of factors and factor numbers has drawn considerable research attention in  the literature of factor models. \cite{lam2011estimation} proposes conditions that the loading matrix is normalized for identification. For sparse factor models, \cite{fruhwirth2018sparse} develops a counting rule on the number of nonzero factor loadings as well as generalised lower triangular representation to resolve rotational invariance. For locally stationary factor models, the direct extension of the existing methods as above are not straightforward since the identification conditions should be also time-varying. Therefore we leave the identification of the factor loadings as a rewarding future work, and focus on the span of the factor loadings that does not rely on identification conditions in this paper.}
 As we discussed in the introduction, fitting factor models using relationships between the factor space and the null space of the auto-covariance matrices has a long history.  
In the following we shall propose  a  nonparametric sieve-based method for time-varying loading matrix estimation which is  adaptive to the smoothness (with respect to $t$) of the covariance function $\bs \Sigma_x(t,k)$. For a pre-selected set of orthonormal basis functions $\{B_j(t)\}_{j=1}^\infty$ 
we shall approximate $\bs \Sigma_x(t,k)$ by a finite but diverging order basis expansion  
\begin{align}\bs \Sigma_x(t,k)\approx \sum_{j=1}^{J_n}\left(\int_0^1\bs \Sigma_x(u,k)B_j(u)du\right)B_j(t),\label{sieveapprox}\end{align} 
where the order $J_n$ diverges to infinity. The speed of divergence is determined by the smoothness of $\bs \Sigma_x(t,k)$ with respect to $t$.  Motivated by \eqref{sieveapprox} we propose to estimate $\mf \Lambda(t)$ by the following $\hat{\mf \Lambda}(t)$:\begin{align}
&\hat {\mf \Lambda}(t)=\sum_{k=1}^{k_0}\hat {\mf M}(J_n,t,k)\hat {\mf M}^\top(J_n,t,k),~~ \text{where}~~ \hat {\mf M}(J_n,t,k)=\sum_{j=1}^{J_n} \tilde {\mf \Sigma}_{x,j,k}B_j(t) \label{11-12-11},\\\label{11-12-10}
&\tilde{\bs \Sigma}_{x,j,k}=\frac{1}{n}\sum_{i=1}^{n-k}\mf x_{i+k}\mf x_{i}^\top B_j(\frac{i}{n}).
\end{align}
In order to help the readers understand our main ideas better, we shall first assume that the number of factors $d(t)$ is constant over time, that is, $d(t)=d$, $\forall t\in[0,1]$. The more complicated case where $d(t)$ is time-varying will be discussed in Section \ref{Sec::Pre}.
Let $\bar {\hat {\mf \Lambda}}=\sum_{i=1}^n\hat {\mf \Lambda}(i/n)/n.$ Then we estimate $d$ by $\hat d_n$ defined as   
\begin{align}\label{Feb4-hatd}
\hat d_n=\argmin_{1\leq i\leq p}\lambda_{i+1}(\bar {\hat {\mf \Lambda}}+q_n) /(\lambda_{i}(\bar {\hat {\mf \Lambda}})+q_n).
\end{align}
where $q_n$ is the penalty which will be discussed in detail in Section \ref{Sec:load}. \WC{The eigenvalue ratio has been considered in the literature of econometrics, see for example \cite{Ahn2013eigenratio}.} 
 In fact, $\hat d_n$ is a penalized version of the eigen-ratio statistics advocated by \cite{lam2012factor} which assumes that $\E(\mf e_{i,n}\mf e^\top_{j,n})=0$. The introducing of the penalization extend the applicability of \cite{lam2012factor} to the model satisfying our second set of assumptions.
Further define $\hat {\mf V}(t)=(\hat {\bf v}_1(t),...,\hat {\bf v}_{\hat d_n}(t))$ where $\hat {\bf v}_i(t)'s$ are  the eigenvectors of $\hat{\mf \Lambda}(t)$ corresponding to  $\lambda_1(\hat {\mf \Lambda}(t))$,...,$\lambda_{\hat d_n}(\hat {\mf \Lambda}(t))$.
Then we estimate the column space of $\mf A(t)$ by \begin{align}\label{span-estimate}Span(\hat{\mf v}_1(t),...,\hat {\mf v}_{\hat d_n}(t)).\end{align}

\section{Test for Static Factor Loadings}\label{sec:test_loading}\setcounter{equation}{0}
It is of practical interest to test  $H_0$ : $\mbox{span}(\mf A(t))=\mbox{span}(\mf A)$, where $\mf A$ is a $p\times d$ matrix. In other words, one can find a time-invariant matrix $\mf A$ to represent the factor loading matrices throughout time.  Without loss of generality, we shall assume that $\mf A(t)=\mf A$ under the null hypothesis throughout the rest of the paper if no confusions will arise. 

Observe that testing $H_0$ is more subtle than testing covariance stationarity of $\mf x_{i,n}$ as both $\mf z_{i,n}$ and $\mf e_{i,n}$ can be locally stationary under the null. By equation \eqref{eq:fm},  assuming $\bs  \Sigma_{ez}(t,k)=0$ as in Section \ref{Sec:estimate},
\begin{align*}
	\int_{0}^1\bs \Sigma_x(t,k)\,dt= \int_{0}^1\mf A(t)(\bs \Sigma_z(t,k)\mf A^\top(t)+\bs \Sigma_{ze}(t,k))\,dt+\int \bs \Sigma_{e}(t,k)dt,\quad k> 0
	\\:=\int \tilde{\bs\Sigma}_x(t,k)dt+\int \bs \Sigma_{e}(t,k)dt,\quad k>0
	\end{align*}
where  $\tilde{\bs\Sigma}_x(t,k)=\mf A(t)(\bs \Sigma_z(t,k)\mf A^\top(t)+\bs \Sigma_{ze}(t,k)).$
As  in Section \ref{Sec:estimate}, we consider (i) $\bs \Sigma_{e}(t,k)=0$ and (ii)$\|\bs \Sigma_{e}(t,k)\|_{2}$ is bounded as $p\rightarrow \infty.$ Under both cases, it can be shown that the null space of $(\int_{0}^1\bs \Sigma_x(t,k))(\int_{0}^1\bs \Sigma_x(t,k))^\top$ is close to the null space of $(\int  \tilde{\bs\Sigma}_x(t,k)dt)(\int \tilde{\bs\Sigma}_x(t,k)dt)^\top$. Furthermore, under null hypothesis 
it's obvious that the null space of {\color{black}$\mf A^\top$} is the same  as the  null space of $(\int  \tilde{\bs\Sigma}_x(t,k)dt)(\int \tilde{\bs\Sigma}_x(t,k)dt)^\top$.

Consider the following quantity $\bs\Gamma_k$ and its estimate $\hat {\bs\Gamma}_k$: \begin{align*}
	\mf \Gamma_k=\int_0^1\bs \Sigma_x(t,k)\,dt \int_{0}^1\bs \Sigma^\top_x(t,k)\,dt, ~\hat {\mf \Gamma}_k=(\sum_{i=1}^{n-k}\mf x_{i+k,n}\mf x^\top_{i,n}/n)(\sum_{i=1}^{n-k}\mf x_{i+k}\mf x^\top_{i,n}/n)^\top.
\end{align*}

Let $\hat {\mf \Gamma}=\sum_{k=1}^{k_0} \hat {\mf \Gamma}_k$. Then the kernel space of {\color{black}$\mf A^\top$} can be estimated by the kernel of $\hat {\mf \Gamma}$ under $H_0$. Let $\tilde d_n$ be an estimate of $d$ which will be described at the end of this section. Let $\hat {\mf f}_{i}$ be the orthonormal eigenvectors of $\hat{\mf \Gamma}$  w.r.t. ($\lambda_{\tilde  d_n+1}(\hat {\mf \Gamma})$,...,$\lambda_p(\hat {\mf \Gamma})$). Write 
$\hat{\mathbf F}=(\hat{\mf f}_1,...,\hat {\mf f}_{p-\tilde d_n})$. The test is then constructed by segmenting the time series into non-overlapping equal-sized blocks of size $m_n$. Without loss of generality, consider $n=m_nN_n$ for integers $m_n$ and $N_n$. Define for $1\leq h\leq N_n$ the index set $b_h=((h-1)m_n+1,...,hm_n)$. The test statistics is 
\begin{align}\label{eq15hatTn}
	\hat T_n=\sqrt m_n\max_{1\leq h\leq N_n}\max_{1\leq i\leq p-\tilde d_n}|\hat{\mf f}_i^\top \mf S^X_h|
\end{align}
where $\mf S_h^X=\sum_{i\in b_h}\mf x_{i,n}/m_n$.   
Then under the null hypothesis, $\mf f_i^\top\mf S^X_h=\mf f_i^\top\mf S^e_h$ where $\mf S_h^e=\sum_{i\in b_h}\mf e_{i,n}/m_n$. Notice that under alternative,  $\mf f_i^\top\mf S^X_h=m_n^{-1}\sum_{s\in b_h}(\mf f_i^\top\mf A(s/n)\mf z_{s,n}+\mf f_i^\top\mf S^e_h)$ which will be large if the exists an eigenvector $\mf f_q$  such that $\mf f_q^\top\mf \sum_{s\in b_h}\mf A(s/n)\mf z_{s,n}/m_n$ is large for some $h$. Let $\hat {\mf e}_i=(\mf I_p-\hat {\mf V}(i/n) \hat {\mf V}^\top(i/n))\mf x_{i,n} $ where $\hat {\mf V}(i/n)$ is the eigenvector used to estimate the column space of $\mf A(t)$ in Section \ref{Sec:estimate}. To implement the test, we propose the following bootstrap procedure. 
Notice that Theorem 3 of \cite{bai2003inferential} specified the optimal rate of $\hat {\mf e}_i-\mf e_i$, which is $O_p(T^{-1/2}+p^{-1/2})$, is too slow for proving $\mf c^\top \hat{\mf e}_i$ well approximate $\mf c^\top \mf e_i$ with  any $|\mf c|$ via plugging in this rate. Nevertheless, using the property of null space we shall show in the online supplement that $\hat {\mf f}^\top_s \hat{\mf e}_i$ is a good proxy of $\mf f_s^\top {\mf e}_i$ for any $1\le s\leq p-d$, which motives us to develop the bootstrap procedure based on $\hat {\mf f}^\top_s \hat{\mf e}_i$ as follows.
Define for $1\leq s\leq N_n$ and $1\leq j\leq m_n$,
\begin{align}\label{hatz}
\hat{\bs l}_{j,s}=\Big(\hat {\mf f}_1^\top\hat{\mf e}_{j+(s-1)m_n},...,\hat {\mf f}_{p-\tilde d_n}^\top 
\hat {\mf e}_{j+(s-1)m_n}\Big)^\top.  
\end{align}
Further define 
\begin{align}\label{new.eq11}
\hat{\bs l}_i=(\hat{\bs l}^\top_{i,1},...,\hat{\bs l}^\top_{i,N_n})^\top
\end{align}
for $1\leq i\leq m_n$. 
Let $
\hat {\mf s}_{j,w_n}=\sum_{r=j}^{j+w_n-1}\hat {\bs l}_r$ and $
\hat {\mf s}_{m_n}=\sum_{r=1}^{m_n}\hat {\bs l}_r$ for $1\leq j\leq m_n$ where $w_n=o(m_n)$ and $w_n\rightarrow \infty$ is the window size. Define 
\begin{align}\label{kappan}
\bs \kappa_n=\frac{1}{\sqrt{w_n(m_n-w_n+1)}}\sum_{j=1}^{m_n-w_n+1}(\hat {\mf s}_{j,w_n}-\frac{w_n}{m_n}\hat {\mf s}_{m_n})R_j
\end{align}
where $\{R_i\}_{i\in\mathbb Z}$ are $i.i.d.$ $N(0,1)$ independent of $\{ {\mf x}_{i,n},1\leq i\leq n\}$. Then we have  the following algorithm for testing static factor loadings:\\\ \\
\noindent 
{\bf Algorithm for implementing the multiplier bootstrap:}
\begin{description}
	\item (1) Select $m_n$ and $w_n$ by the Minimal Volatility (MV) method that will be described in Section \ref{selectmn}.
	\item (2)  Generate $B$ (say 2000) conditionally $i.i.d.$ copies of  $K_r=|\bs \kappa_n^{(r)}|_\infty$, $r=1,...B$, where $\bs \kappa_n^{(r)}$ is obtained by \eqref{kappan} via the $r_{th}$ copy of $i.i.d.$ standard normal random variables $\{R_i^{(r)}\}_{i\in \mathbb Z}$.
	\item (3)  Let $K_{(r)}, 1\leq r\leq B$ be the order statistics for $K_r, 1\leq r\leq B$.  Then we reject $H_0$ at level $\alpha$ if $\hat T_n\geq K_{(\lf(1-\alpha)B \rf)}$. Let $B^*=\min \{r: K_{(r)}\geq \hat T_n\}$ and the corresponding $p$ value of the test can be approximated by $1-B^*/B$.
\end{description}
To implement our test, $d$ will be estimated by
\begin{align}\label{eigen-ratio-hatd}
\tilde d_n=mode(\tilde d^*_i)_{1\leq i\leq T},\notag\\ \text{where~~~~} \tilde d^*_i=\argmax_{1\leq j\leq p}(\lambda_{j+1}(\hat {\mf \Lambda}(i/n))+q_n) /(\lambda_{j}(\hat {\mf \Lambda}(i/n))+q_n)). 
\end{align}
where $q_n$ is the penalty  which will be discussed later.
\textcolor{black}{\begin{remark}\label{newrmk4.1}
	In this paper, we test whether the span of the factor loading matrix, denoted by {\bf span}, is time-varying. At each time \( t \), this span is uniquely determined by the matrix \( \mathbf{\Lambda}_1(t) \) defined in equation \eqref{DefLambda}, which can be uniquely determined without identification issues. A static matrix \( \mathbf{A} \) can represent this span when it is time-invariant. Once the loading matrix \( \mathbf{A} \) is determined to be static, the factors are then determined up to time-invariant rotation and dilation. Hence, in this scenario the factors should be modelled as locally stationary processes if the covariance structure of the time seires $\{\mf A\mf z_{i,n}\}$ is time-varying.  It is important to note that a model with time-varying factor loadings and stationary low-dimensional vectors is similar to, yet distinct from, the model with static factor loadings and locally stationary low-dimensional factors considered in the null hypothesis of our tests. The time-varying second-order structure of the latter model is determined by that of the low-dimensional common factors, whereas in the former model, the time-varying covariance structure is determined by the \( p \times d \) components of factor loading matrix, and the model complexity becomes large when $p$ is high. 
	As pointed out by a referee, besides testing static factor loading as considered in our paper, testing constancy of the auto-covariance strucuture of the factor process is also very important, especially for predicting the low-dimensional common factors. We leave this test as a promising direction for future work.
\end{remark}
}
\section{Technical Assumptions}\label{tech-assumption}
We  first discuss the conditions for the loading matrix of model \eqref{eq:fm}, which relates $\mf G$ to $\mf Q$ and $\mf H$.
\begin{description}
	\item (A1)  Let $a_{ij}(t)$, $1\leq i\leq p$, $1\leq j\leq d$ be the $(i,j)_{th}$ element of $\mf A(t)$. We assume there exists a sufficiently large constant $M$ such that
	\begin{align}
	\sup_{t\in [0,1]}|a^{}_{ij}(t)|\leq M.
	\end{align}

\end{description}

\begin{description}
	\item (A2)  $\mf A(t)$ is full rank. Write $\mf A(t)=(\mf a_1(t),....,\mf a_{d}(t))$ where $\mf a_s(t), 1\leq s\leq d$ are $p$ dimensional vectors. Then $\sup_{t\in [0,1]}\|\mf a_s(t)\|^2_2\asymp p$
		for $1\leq s\leq d$.
	Besides, 
	the matrix norm of $\mf A(t)$ satisfies
	\begin{align}
	\inf_{t\in [0,1]}\|\mf A(t)\|_F\asymp p^{\frac{1}{2}},\sup_{t\in [0,1]}\|\mf A(t)\|_F\asymp p^{\frac{1}{2}},
	\inf_{t\in [0,1]}\|\mf A(t)\|_m\geq \eta^{1/2}_n p^{\frac{1}{2}}
	\end{align}
	for a positive sequence $\eta_n=O(1)$. Note that $\eta_n$ is allowed to converge to 0.
\end{description}
Condition (A1) concerns the boundedness of the loading matrix, while condition (A2) assume strong  factor strength ( c.f. Section 2.3 of \cite{lam2011estimation}) for the ease of reading. We discuss weak factor strength in the proofs. Since we are only interested in identifying the linear span of $\mf A(t)$ in this paper, there is no need to assume that the matrix $\mf A(t)$ is standardized. 
 Notice that we identify the column space \eqref{span-estimate} via estimated dimensions and eigenvectors.  Larger   $\|\mf A(t)\|_m$ will yield better separation of null space and kernel which enables us to correctly identify the number of factors, as well as more accurate estimation of eigenvectors via the well-known Davis Kahan Theorem \cite{davis1970rotation}. Here we allow the sequence $\eta_n\rightarrow 0$, and the theoretical results  will rest on the magnitude of order of $\eta_n$. Such formulation is convenient for us 
 to further discuss the model \eqref{eq:fm} with time-varying $d=d(t)$ in detail, 
 in Section \ref{Sec::Pre}. 
    
We then postulate the following assumptions on the covariance matrices of the common factors $\mf z_{i,n}$ and the idiosyncratic components $\mf e_{i,n}$, which are needed for spectral decomposition.  Let $\bs \Sigma_{ez}(t,k)=\E(\mf H(t,\FF_{i+k}){\mf Q}^\top(t,\FF_i))$ and $\bs \Sigma_e(t,k)=\E(\mf H(t,\FF_{i+k})\mf H^\top(t,\FF_i))$. 

\begin{description}
		\item (S0) Let $\sigma_{x,u,v}(t,k)$ be the $(u,v)_{th}$ element of $\bs \Sigma_x(t,k)$. Assume $\sigma_{x,i,j}(t,k)$, $1\leq i\leq p$ and $1\leq j \leq p$, $1\leq k\leq k_0$ belongs to a common functional space $\Omega$ which is equipped with an orthonormal basis $B_j(t)$, i.e. $\int_{0}^1 B_m(t)B_n(t)dt=\mathbf 1(m=n)$, where $\mathbf 1(\cdot)$ is the  indicator function. Assume $\Omega \in \mathcal C^{K}(\tilde M)[0,1]$ for some $K\geq 2$. Moreover for $1\leq k\leq k_0$,
	\begin{align}\label{Basisapprox}
	\max_{1\le i\leq p,1\leq j\leq p}\sup_{t\in [0,1]}|\sigma_{x,i,j}(t,k)-\sum_{u=1}^{J_n} \tilde \sigma_{x,i,j,u}(k)B_u(t)|=O(g_{J_n,K, \tilde M}),
	\end{align}
	where $\tilde \sigma_{x,i,j,u}(k)=\int_0^1 \sigma_{x,i,j}(t,k)B_u(t)dt$, and $g_{J_n,K, \tilde M}\rightarrow 0$ as $J_n\rightarrow \infty$.  
	
	\item(S1) For $t\in [0,1]$ and $k=1,...,k_0$, all components of $\mf \Sigma_{e}(t,k)$ are $0$.
	\item (S1') $p\rightarrow \infty$,  $\|\bs \Sigma_e(t,k)\|_{2}\leq M$ for some constant $M$ for $k=1,...,k_0$.
	\item (S2) For $k=0,1,...,k_0$, $\mf \Sigma_z(t,k)$ is full ranked 
    such that $\inf_{0\leq k\leq k_0}\sigma_d(\mf \Sigma_z(t,k))> 0$.
	\item (S3) For $t\in [0,1]$ and $k=1,...,k_0$, all components of $\mf \Sigma_{ez}(t,k)\mf A^\top (t)$ are $0$. 
	\item (S4)  For $t\in [0,1]$ and $1\leq k\leq k_0$, $\|\mf \Sigma_{ze}(t,k)\|_{F}=o(\eta^{1/2}_n p^{\frac{1}{2}})$. 
\end{description}
 (S0) means $\bs \Sigma_x(t,k)$ can be approximated by the basis expansion. The approximation error rate $g_{J_n,K, \tilde M}$ diminishes as $J_n$ increases. Often higher differentiability yields more accurate approximation rate. 
 \WC{ We refer to \cite{wang2012convergence} and \cite{chen2007large} for evaluating $g_{J_n,K, \tilde M}$ when normalized Legendre polynomial, trigonometric polynomials or orthogonal wavelets are used for basis. } Condition (S1) indicates that $(\mf e_{i,n})$ does not have auto-covariance up to order $k_0$ which is slightly weaker than the requirement that $(\mf e_{i,n})$ is a white noise process used in the literature. In (S1) $p$ can either be fixed or divergent. Condition (S1') allows $(\mf e_{i,n})$ to have non-zero autocovariance but requires $p$ diverge while avoiding strong  cross-sectional dependence among $e_{i,n}$ and $e_{j,n}$ which is {\color{black}satisfied by many dynamic factor models, see for example \cite{stock2011dynamic}}. In this paper, we assume either (S1) or (S1'). Condition (S2) implies that for $1\leq i\leq n$, 
no linear combination of components of $\mf z_{i,n}$ is white noise that can be absorbed into $\mf e_{i,n}$. (S3) can be implied by $\bs\Sigma_{ez}(t,k)\equiv 0$, i.e., $\mf z_{i,n}$ and $\mf e_{i+k,n}$ are uncorrelated for any $k\ge 0$. Condition (S4) requires a  weak correlation between $\mf z_{i+k,n}$ and $\mf e_{i,n}$.  In fact, it is the locally stationary extension of Condition (i) in Theorem 1 of \cite{lam2011estimation} and condition (C6) of \cite{lam2012factor}. Though (C6) of \cite{lam2012factor} assumes a rate of $o(p^{1-\delta})$, it requires standardization of the factor loading matrix. 

\section{Asymptotic Results for Model Estimation}\label{Sec:load}\setcounter{equation}{0}
Theorem \ref{Thm-approx} provides the estimation accuracy of $\hat{\bs \Lambda}(t)$ by the sieve method.  Due to the page limit, we move conditions (M1)-(M3) to Section \ref{appendix}, which are standard assumptions for multivariate locally stationary time series.
\begin{theorem}\label{Thm-approx}
	Assume conditions (A1), (A2), (M1), (M2) ,(M3) and (S0), (S2)--(S4) hold. Define $\iota_n=\sup_{1\leq j\leq J_n}Lip_j+\sup_{t,1\leq j\leq J_n}|B_j(t)|$, where $Lip_j$ is the Lipschitz constant of basis function $B_j(t)$. Write 
	 $\nu_n=\frac{J_n\sup_{t,1\leq j\leq J_n}|B_j(t)|^2}{\sqrt n}+\frac{J_n\sup_{t,1\leq j\leq J_n}|B_j(t)|\iota_n}{n}+g_{J_n,K,\tilde M}$, where the quantity $g_{J_n,K,\tilde M}$ is defined in condition (A2). 
	 Then we have if (S1) hold
	\begin{align*}
	\Big\|\sup_{t\in [0,1]}\Big\|\hat {\mf \Lambda}(t)-\mf \Lambda_1(t)\Big\|_2\Big\|_{\mathcal L^1}=O(p^{2}\nu_n).
	\end{align*}
and  if (S1') hold, the rate will be $O(p^{2}\nu_n+p)$.
\end{theorem}
From the proof, we shall see that $\|\mf \Lambda_1(t)\|_F$ is of the order $p^{2}$ uniformly for $t\in[0,1]$. Hence under (S1) the approximation error of $\hat {\mf \Lambda}(t)$ is negligible compared with the magnitude of $\mf \Lambda_1(t)$. Under (S1') to achieve negligible approximation error we additionally require that $p\rightarrow \infty$.
For orthnormal Legendre polynomials and trigonometric polynomials it is easy to derive that $Lip_j=O(j^2)$. Similar calculations can be performed for a large class of frequently-used basis functions.  The first term of $\nu_n$ is due to the stochastic variation of $\hat {\mf M}(J_n,t,k)$ (see \eqref{11-12-11}), while the second and last terms are due to the basis approximation.

We now discuss the validity of estimator \eqref{span-estimate}.
 Write $\hat {\mf B}(t)=(\hat{\mf b}_{d+1}(t),...,\hat{\mf b}_{p}(t))$ where $\hat {\mf b}_{s}(t)$, $d+1\leq s\leq p$ are orthonormal eigenvectors of $\hat{\bs \Lambda}(t)$ corresponding to $\lambda_{d+1}(\hat {\mf \Lambda}(t))$,... ,$\lambda_p(\hat {\mf \Lambda}(t))$, and $\tilde {\mf V}(t)=(\hat{\mf v}_{1}(t),...,\hat{\mf v}_{d}(t))$ where $\hat {\mf v}_{s}(t)$, $1\leq s\leq d$ are orthonormal 
 eigenvectors of $\hat{\bs \Lambda}(t)$ corresponding to $\lambda_{s}(\hat {\mf \Lambda}(t))$, $1\leq s\leq d$. Hence
$ (\hat{\mf v}_1(t),...,  \hat{\mf v}_{d}(t),  \hat {\mf b}_{d+1}(t),...,  \hat{\mf b}_p(t))$ form a set of orthonormal basis of $\mathbb R^p$.  
Define $\mf V(t) =(\mf v_1(t),...,\mf v_{d}(t))$ where $\mf v_i(t)s$ are  the orthonormal  eigenvectors of $\bs \Lambda_1(t)$ corresponding to $\lambda_s(\mf \Lambda_1(t))$,  $1\leq s\leq d$,  and  $ {\mf B}(t)=( \mf b_{d+1}(t),...,\mf  b_{p}(t))$ with $ \mf b_{s}(t)$, $d+1\leq s\leq p$ being a set of orthonormal eigenvectors of ${\bs \Lambda}(t)$ corresponding to $\lambda_{d+1}( {\mf \Lambda}(t))$,... ,$\lambda_p( {\mf \Lambda}(t))$. Therefore
$(\mf v_1(t),..., \mf v_{d}(t),  \mf b_{d+1}(t),...,\mf  b_p(t))$ also form a set of orthonormal basis of $\mathbb R^p$. Notice that $\tilde {\mf V}(t)$ will coincide with $\hat{\mf V}(t)$ if $\hat d_n=d$. 

 \begin{theorem}\label{Space-Distance}
 	Under conditions of Theorem \ref{Thm-approx}, we have \begin{description}
 		\item (i) For each $t \in[0,1]$ there exist orthogonal matrices $\hat {\mf O}_1(t)\in \mathbb R^{d\times d}$ and  $\hat {\mf O}_2(t)\in \mathbb R^{(p-d)\times (p-d)}$ such that if (S1') holds,
 		\begin{align*}
 		\|\sup_{t\in [0,1]}\|\tilde {\mf V}(t) \hat{\mathbf O}_1(t)-\mf V(t)\|_F\|_{\mathcal L^1}=O(\eta_n^{-1} \nu_n+\eta_n^{-1}p^{-1}),\\
 		\|\sup_{t\in [0,1]}\|\hat {\mf B}(t) \hat{\mathbf O}_2(t)-\mf B(t)\|_F\|_{\mathcal L^1}=O(\eta_n^{-1}\nu_n+\eta_n^{-1}p^{-1}).
 		\end{align*}
 	If (S1) holds then the rate will be reduced to $O(\eta_n^{-1}\nu_n)$.
 		\item (ii) Furthermore, if (S1') holds. $\limsup_p\sup_{t\in[0,1]}\lambda_{max}(\E(\mf H(t,\FF_{i})\mf H^\top(t,\FF_i)))<\infty$ we have that for $i/n\in [0,1]$, $1\leq i\leq n$, \begin{align*}p
 			^{-1/2} \|\tilde{\mf V}(i/n)\tilde{\mf V}^\top(i/n)\mf x_{i,n}-\mf A(i/n)\mf z_{i,n}\|_2=O_p( \eta_n^{-1}\nu_n+p^{-1/2}+\eta_n^{-1}p^{-1}).\end{align*} 
 		If (S1) holds then the rate will be reduced to $O_p( \eta_n^{-1}\nu_n+p^{-1/2})$.
 	\end{description}
 \end{theorem}

 Assertion (i) follows from Theorem \ref{Thm-approx} and a variant of Davis Kahan Theorem (\cite{yu2015useful}) which does not require the separation of all non-zero eigenvalues.\  (i) involves orthogonal matrices $\hat  O_1(t)$ and $\hat O_2(t)$ since it allows multiple eigenvalues at certain time points, which yields the non-uniqueness of the eigen-decomposition. Moreover, under either (S1) or (S1'), if $\eta_n\gtrapprox 1$, the rate in (i) will not increase as $p$, and reduces to the uniform nonparametric sieve estimation rate for univariate smooth functions if (S1) holds, which coincides with the well-known "blessing of dimension" phenomenon for stationary factor models, see for example \cite{lam2011estimation}.
 
 \WC{The sieve approximation rates $\nu_n$ will be adaptive to the smoothness and will be slower when $\sigma_{x,i,j}(t,k)'s$ are less smooth in which case $g_{J_n,K, \tilde M}$ converges to zero at an adaptive but slower rate as $J_n$ increases.  	If we assume that  $\sigma_{x,i,j}(t,k)'s$ are real analytic and normalized Legendre polynomials or trigonometric polynomials (when all $\sigma_{x,i,j}(t,k)$ can be extended to periodic functions) are used as basis, we shall take $J_n=M\log n$ for some large constant $M$ to yield $\nu_n=\frac{\log n}{\sqrt n}.$}  

 The next proposition states that with high probability $\hat d_n=d$ if the penalization $q_n=c(p^{1-\delta}+\nu_np^2)\log p$ for some constant $c>0$. 

 \begin{proposition}\label{hatdrate}
 	Assume conditions (A1), (A2), (M1)-(M3), (S0)-(S4) (either (S1) or (S1') holds) hold, and that $\eta_n\gtrapprox 1$. Furthermore, under (S1), suppose that $c$ is a sufficiently small and positive constant  and that  $\nu_n\log n\rightarrow 0$.  Under (S1'),  assume that
 	$\frac{p^{2}}{(p^{2}\nu_n+p)\log n}\rightarrow \infty$, and that $n^a\lessapprox  p\lessapprox n^b$ for some $a<b$. Then

 	\begin{align}
 	\p(\hat d_n\neq d)=
 	O(
 	\eta_n^{-1}\theta(n,p)
 	)+O(\log^{-1/2} n)=o(1).
 	\end{align}
 where $\theta(n,p)=\nu_n$ under (S1) and $\nu_n+p^{-1}$ under (S1'). 
 \end{proposition}

\section{Theoretical Results for Testing Static Factor Loadings}\label{Sec:null-static}


We discuss the limiting behavior of $\hat T_n$ of \eqref{eq15hatTn} under $H_0$ in this section. Notice that under $H_0$ the dimension of the loading matrix $\mf A(t)\equiv \mf A$ is fixed.  Therefore for simplicity in this section we assume $\eta_n\equiv 1$ for $\eta_n$ in conditions (A2), (S3) and (S4).
 ~First, the following proposition indicates that with probability tending to one $\tilde d_n$ equals  $d$ under $H_0$.
	\begin{proposition}\label{prop1}
 		Assume conditions  of Proposition \ref{hatdrate} hold.
 		Then we have, under $H_0$, $$\p(\tilde d_n \neq d)=O(\theta(n,p))+O(\log^{-1/2} n)=o(1)$$ as $n\rightarrow \infty.$  
 	\end{proposition}
 By construction, Proposition \ref{prop1} is an  immediate consequence of Proposition \ref{hatdrate}. 
 To derive the asymptotic correctness of our bootstrap-assisted testing procedure, we further assume condition (M2') to replace (M2).
	\begin{description} 
	\item (M2') There exists constants $l\geq4$ and $M$, such that
	$\max_{1\leq j\leq d}\sum_{k=1}^\infty \delta^z_{l,j}(k)<\infty,$ $\max_{1\leq j\leq p}\sum_{k=1}^\infty \delta^e_{l,j}(k)<\infty$  and \begin{align*}
	\sup_{t\in[0,1]}\max_{1\leq u\leq d}\E | Q_u(t,\FF_0)|^{l}\leq M,~~ \sup_{t\in [0,1]}\max_{1\leq v\leq p}\E |H_v(t,\FF_0)|^{l}\le M.
	\end{align*}
\end{description}


Define   $\mf \Sigma_H(t)$ as  $\mf \Sigma_H(t)=\sum_{k\in \mathbb Z} Cov(\mf H(t,\FF_i), \mf H(t,\F_{i+k}))=\sum_{k\in \mathbb Z}\mf \Sigma_e(t,k),$ 
which is the long-run covariance matrix of $\mf H$.  We then have the following condition (M4)-(M8).
\begin{description}
\item (M4) There exists a constant $M_{l}$ depending on $l$ such that for $1\leq i\leq n$, and for all $p$ dimensional vector $\mf c$ such that $|\mf c|_2=1$, the inequality $\|\mf c^\top\mf e_{i,n}\|_{\mathcal L^{l}}\leq M_{l}\|\mf c^\top\mf e_{i,n}\|_{\mathcal L^2}$ holds. 
Also $
\max_{1\leq i\leq n}\lambda_{\max}(\E(\mf e_{i,n}\mf e^\top_{i,n}))$ is uniformly bounded as $n$ and $p$ diverges.
	\item (M5) There exist constants $c$ and $C$ such that
	\begin{align*}
		c\leq \lambda_{\min}(\mf \Sigma_H(t))\leq \lambda_{\max}(\mf \Sigma_H(t))\leq C.
	\end{align*}
\end{description}
\begin{description}
	\item(M6) 
	Write $\frac{\partial}{\partial t}\mf H(t,\FF_i)=(\frac{\partial}{\partial t}H_1(t,\FF_i),...,\frac{\partial}{\partial t}H_p(t,\FF_i))^\top$ where $H_s(t,\FF_i)$ is the $s_{th}$ entry of $\mf H(t,\FF_i)$, 	and \textcolor{black}{$\mf H'(s, \FF_i)=\frac{\partial}{\partial t}\mf H(t,\FF_i)|_{t=s}$}.
	 Assume that for all $t,s\in (0,1)$ and $u,v\in \mathbb Z$, 
	\begin{align}
		\|\E(\mf H(t,\FF_u)(\mf H'(s,\FF_v) )^\top)\|_2=O\Big(|(u-v)^{-2}\log^{-2}|u-v||\wedge 1 \Big)
	\end{align}

	\item (M7) For all $t\in[0,1]$, $\sum_{k\in \mathbb Z}k\|\Sigma_e(t,k)\|_2<\infty$. 
		\item (M8) There exists a $q\geq 3$, s.t. $\max_{|\mf c|=1}\|\mf c^\top  (\mf e_i-\mf e_i^*)\|_{\mathcal L^q}=O(\Delta_q(i))$ with  $\sum_{j\in \mathbb Z, j\geq 0}j\Delta_q(j)<\infty$. To save notation we assume $q=l$.
\end{description}

 Condition (M4) controls the magnitude of the $\mathcal L^{l}$ norm of projections of $\mf e_{i,n}$ by their ${\cal L}^{2}$ norm which essentially requires that the dependence among the components of $\mf e_{i,n}$ cannot be too strong. (M4) is mild in general and is satisfied, for instance, if a bounded number of components of $\mf e_{i,n}$ are dependent, or $\mf e_{i,n}$ has the form of $\mf M \bs \varepsilon_{i,n}$ for a $p\times p$ matrix $M$ and a random vector $\bs \varepsilon_{i,n}$ where $\|\mf M\|_2$ is bounded, and the component of $\bs \varepsilon_{i,n}$ are  independent sub-Gaussians with bounded variance proxy. 
 Suppose that $\delta^e_{2}(k)+\delta^{e'}_{2}(k)=o(k^{-2}\log^{-2} k)$ where $\delta_2^{e'}$ is the dependence measure of $\frac{\partial}{\partial t}\mf H(t,\FF_i)$,  then by Lemma 5 of \cite{zhou2010simultaneous} (M6) will hold for fixed $p$.  When $p$ diverges, (M6) will be satisfied if $\mf H$ is stationary.
 \textcolor{black}{For nonstationary $\mf H$, (M6) means weak cross-sectional dependence among components of $\mf H$, for example it holds if  there are at most a bounded number of components of $\mf H(\cdot,\FF_i)$ and $\mf H'(\cdot,\FF_j)$ that are correlated. Moreover, it can be verified for a general class of locally stationary high dimensional moving average models. See Section \ref{Modelassumptions} in the supplemental material for more detailed examples. }
(M7) posits a weak cross-sectional correlation for idiosyncratic error.  If the idiosyncratic error is white noise, then our theoretical results will hold without assuming (M6) and (M7).   (M8) will be fulfilled if $p$ is fixed or if the components of $\mf e_i$ are independent.
(M8) can be easily checked for high dimensional linear process. We refer to Proposition \ref{propexample} in the supplemental material for verifying (M8) for a large class of high dimensional moving average process.  

 Write $\mathbf F=(\mf f_1,...,\mf f_{p-d})$.
Define $ {\bs l}_i$ by replacing $\hat{\mf F}$ with $\mathbf F$ and $\mf e_i$ with $\hat {\mf e}_i$ in the definition of $\hat {\bs l}_i$ (c.f. \eqref{new.eq11}) with its $j_{th}$ element denoted by ${l}_{i,j}$. Then straightforward calculations indicate that  $\hat T\approx|\frac{\sum_{i=1}^{m_n}{\bs l}_i}{\sqrt{m_n}}|_\infty$ under null hypothesis. Therefore we can approximate $\hat T$ by the $\mathcal L^\infty$ norm of a certain mean zero Gaussian process via the recent development in high dimensional Gaussian approximation theory, see for instance \cite{chernozhukov2013gaussian} and \cite{zhang2018gaussian}. Let $\mf y_i=(y_{i1},...y_{i(N_n(p-d) )})$ be a centered $N_n(p-d) $ dimensional Gaussian random vectors that preserved the auto-covariance
structure of ${\bs l}_i$ for $1\leq i\leq m_n $ and write $\mf y=\sum_{i=1}^{m_n}\mf y_i/\sqrt{m_n}$. 
\begin{theorem}\label{Jan23-Thm4}
	Assume conditions of Proposition \ref{prop1} and 
	(M2'), (M4)-(M8) hold. Furthermore, suppose that $\Omega_n(M'):=\sqrt{\frac{M'}{m_n}}(N_np)^{1/l}+N_n^{1/l}\theta_0(n,p)p^{\frac{1}{2}}=o(1)$, where  $\theta_0(n,p)=
	1/\sqrt n $ under (S1), and 	$1/\sqrt n+p^{-1} $ under (S1'). Assume that there exists $k', 1\leq k'\leq k_0$ such that $\sigma_d(\int \bs \Sigma_x(t,k')dt)\geq \eta>0$.  
Then under null hypothesis
	 
	  \begin{align}\label{Jan23-84}
	 	\sup_{t\in \mathbb R}|\p(\hat T_n\leq t)&-\p(|\mf y|_{\infty}\leq t)|= O\Big(\log^{-1/2} n+\theta(n,p)\log n \notag\\&
	 	+(\Omega_n(M')^{\frac{l}{2l+1}}\sqrt{\log(n/\Omega_n(M'))}+\upsilon(m_n-2M', N_n,p,d,l))\Big),
	 \end{align}  
	 for any sequence $M'=o(m_n)$.
\end{theorem}
  Since the detailed form of $\upsilon(m_n, N_n,p,d,l)$ is complicated and long, we relegate its formula to Proposition \ref{definitioniota} in the online supplement.	
  
  \begin{remark}\label{rmk7.1}
  	The term $\upsilon(m_n-2M', N_n,p,d,l)=o(1)$ if $\Omega_n(M')/m_n^{-\epsilon}\rightarrow \infty$ for some $\epsilon>0$, $\Delta_l(j)=j^{-(1+\beta)}$ for some $\beta>2$, 
  	and $l$ is sufficiently large such that $(N_np)^{1/l}=O(m_n^{5/16-\iota_1})$ and $p^{1/l}=O(m_n^{\frac{0.5+\beta}{8}-\iota_2})$ for some $\iota_1, \iota_2>0$. 
  	 Furthermore, if 
  	$(N_np)^{1/4}\lessapprox m_n^{\frac{3-25\zeta}{32}} $ and  $N_np\lessapprox \exp(m_n^{\zeta})$ for some $0\leq \zeta< 1/11$, and $\Delta_l(k)=O(\chi_0^k)$ for some constant $\chi_0\in (0,1)$, then by setting $M'=\log m_n$ it follows that
  	$\upsilon(m_n-2M', N_n, k_0,p,d,l)=O(m_n^{-(1-11\zeta)/8})$ and $\Omega_n(M')=\sqrt{\frac{\log m_n}{m_n}}(N_np)^{1/l}+N_n^{1/l}\theta_0(n,p)p^{\frac{1}{2}}$.
  \end{remark}


\begin{remark}
	When $l$ is sufficiently large, the second term of $\Omega_n(M')$ in last line of Remark \ref{rmk7.1} is close to $p^{1/2}/\sqrt n$ under (S1) and $p^{1/2}/\sqrt n+p^{-1/2}$ under (S1'). Hence, in order for this term to vanish, $p$ can be as large as $O(n^{a})$ for any $a<1$.
\end{remark}


 \subsection{Block Multiplier Bootstrap}\label{boots}

The validity of the bootstrap procedure is supported by the following theorem.  Let $
\bar{\theta}(n,p,l,N_n,w_n)=\sqrt{w_n} N_n^{1/l}\theta(n,p)p^{1/2}$. 
 \begin{theorem}\label{Boots-thm5}
Let $W_{n,p}=(N_n(p-d))^2$. Assume that the conditions of Theorem \ref{Jan23-Thm4} hold,  $w_n\rightarrow \infty$, $w^2_n/m_n=o(1)$, $	\bar{\theta}(n,p,l,N_n,w_n)\log^{1/2} n=o(1)$ and that there exist $q^*\geq l$ and $\epsilon>0$ such that  $ \Theta_n:=w_n^{-1}+\sqrt{w_n/m_n}W_{n,p}^{2/{q^*}}\lessapprox W_{n,p}^{-\epsilon}$, $\upsilon(m_n-2M', N_n,p,d,l)=o(1)$, $n^{-\epsilon_1}\lessapprox\Omega_n(M')\lessapprox n^{-\epsilon_2}$ for some $M'=o(m_n)$ and $\epsilon_1,\epsilon_2>0$, 
and \begin{description} 
\item	(i)  $\|\mf c^\top\mf e_{i,n}\|_{\mathcal L^{q^*}}\leq M_{q^*}\|\mf c^\top\mf e_{i,n}\|_{\mathcal L^2}$ holds for all $|\mf c|=1$. 
\item (ii) $\Delta_{q^*}(j)=O(((j+1)\log (j+1))^{-2})$.
\item	(iii) $\lambda_{\max}(Var(\mf H(t,\FF_0)-\mf H(s,\FF_0) ))\leq C|t-s|$ for some constant $C$.
	
%
	
\end{description} Then  we have that conditional on $\mf x_{i,n}$ and under $H_0$,
\begin{align}\label{Oct16_26}
\sup_{t\in \mathbb R}|&\p(\hat T_n\leq t)-\p(|\bs \kappa_n|_{\infty}\leq t|{\mf x}_{i,n},1\leq i\leq n)|=o_p(1)
\end{align}

 \end{theorem}
The condition $w_n^{-1}+\sqrt{w_n/m_n}W_{n,p}^{2/{q^*}}\lessapprox W_{n,p}^{-\epsilon}$ holds if $w_n\gtrapprox (N_np)^{2\epsilon}$ and $\sqrt{\frac{w_n}{m_n}}W_{n,p}^{\frac{2}{q^*}+\epsilon}=o(1).$ 
 For the condition $\upsilon(m_n-2M', N_n,p,d,l)=o(1)$ we refer to Remark \ref{rmk7.1}. We also provide the detailed rate of \eqref{Oct16_26} in  the proof presented in the online supplement.

\subsection{Power}\label{Sec4}
In this section we discuss the  power of our bootstrap-assisted testing algorithm in Section \ref{boots} for testing static factor loadings.

\begin{theorem}\label{Power}
	Recall the bootstrap critical value $K_{(\lf (1-\alpha)B \rf)}$   defined in Section \ref{sec:test_loading}.	Suppose that the conditions of Theorem \ref{Boots-thm5} holds.
 Consider the following class of alternatives: \begin{align}\label{alternative_A}H_A: \mf A(t)=\mf A_n(t):=\mf A+\rho_n\mf D(t),\end{align}
where $\mf D(t)=(d_{ij}(t))$ is a $p\times d$ matrix \textcolor{black}{satisfying (A1) and (A2) (with $\eta_n\equiv 1$),  $\|\mf D(t)\|_2=1$ for identification, and 
$\rho_n=O(1)$ } controls the magnitude of deviation from the null. 	Let $ \tilde{\mf F}=(\tilde{\mf f}_{1},...,\tilde{\mf f}_{p-d'})$ be the eigenvectors of $\mf \Gamma$ where $d'$ is the rank of $\mf \Gamma$.  
\begin{description}

	\item (i)  
Assume that there exists some $q$ such that $\tilde {\mf f}_q$ satisfying $|\tilde {\mf f}_q^\top \mf A(t)|/\sqrt{\log (N_n p)}\rightarrow \infty$ as  $(n,p)$ diverges at some $t\in (0,1)$, and that $|\frac{\partial}{\partial t}|\tilde{ \mf f}_q^\top \mf A(t)||\leq M |\tilde{ \mf f}_q^\top \mf A(t)|$ for all $t\in [0,1]$ and some universal constant $M$.
	 Suppose that the long run covariance of $\mf z_{i,n}$ is not degenerated, i.e., 
	\begin{align}\label{D.56}
		\underline \lambda_z:=\inf_{t\in [0,1]}\lambda_{\min}(\sum_{k=-\infty}^{\infty}\bs \Sigma_z(t,k))>0
	\end{align}
	with  $m_n^{3/2}=o(n)$. 
	Then we have as $n\rightarrow \infty$,
	\begin{align}\label{new.7.8}
		\lim_{B\rightarrow \infty }\p(\hat T\geq K_{(\lf (1-\alpha)B \rf)}|\mf x_{i,n},1\leq i\leq n)\rightarrow_p 1.
	\end{align}
\item (ii) Suppose that $p$ is fixed, and there exists $q$, $1\leq q\leq p-d'$ such that
\begin{description}
   \item (a)  \textcolor{black}{The \eqref{D.56} holds, and that $\|\bs \Sigma_e(t,k)\|_F=0$ for all $t\in [0,1], k\geq 1$. Moreover, for each $n$ there  exists a union $\mathcal I$  of sub-intervals of $[0,1]$, such that
   $\min_{t\in \mathcal I}(|\tilde{\mf f}^\top_q\mf A(t)|)> (18+\gamma_0)^{1/2}	\underline \lambda^{-1/2}_z\sup_t \|Var(\mf H(t,\FF_0))\|^{1/2}_2 $ for some $\gamma_0>0$, and $|\mathcal I|\geq \gamma_1>0.$ }
    \item (b) Let $\tilde x_i=\tilde{\mf f}^\top_q\mf x_i:=\tilde { G}(i/n,\FF_i)$ where $\FF_i$ is defined in Section \ref{Sec:estimate}. Define $\delta^{\tilde G}_l(k)=\|\tilde{G}(t,\FF_i)-\tilde{G}(t,\FF_i^{(i-k)})\|_{\mathcal L^l}$. 
     Assume that $\delta^{\tilde G}_l(j)=O(j^{-1-\beta})$ for some $\beta>2$, and that $\E(|\tilde {G}(t,\FF_0)|^l)<\infty$ for some $l\geq 8$, $m_n\asymp n^\alpha$ for some $\alpha>\frac{16}{5l}$.  
\end{description}
Then 
we have
\eqref{new.7.8} still holds.
\end{description}
\end{theorem}
In fact, in Proposition \ref{PropG1} of the online supplement we show $d'=d$ under \eqref{alternative_A}.
 The condition that $|\frac{\partial}{\partial t}|\tilde{ \mf f}_q^\top \mf A(t)||\leq M |\tilde{ \mf f}_q^\top \mf A(t)|$ in (i) is mild. A sufficient condition is that for $1\leq i\leq p$ and $1\leq j\leq d$, $|\frac{\partial}{\partial t}a_{ij}(t)|\leq M |a_{ij}(t)|$ for some uniform constant $M$. Let  $\mf D$ be a subspace of null space of $\mf A^\top$, i.e., $\mf A^\top\mf D=\mf 0$, and consider $\mf  D(t)=a(t)\mf D$ with some non-constant function $a(t)\in \mathcal C^1(M_0)[0,1]$ for some constant $M_0>0$. Then there exists a union $\mathcal I$ of sub-interval of $[0,1]$ such that $|\tilde {\mf f}_q^\top\mf A(t)|$ will be the order of $\rho_np^{1/2}$ on $\mathcal I$ with $|\mathcal I|>0$. 

\section{Factor Loadings with Varying Dimensions}
\label{Sec::Pre}
We now discuss model \eqref{eq:fm} when the number of factors and the dimension of the loading matrix are time-varying. Since the number and the dimension are integers, it is sophisticated to define the "smoothly changing" factor number or "smoothly changing" dimensions directly, where the concept of "smoothly changing" is the key assumption of locally stationary models and is the key to the nonparametric smoothing approaches. Moreover, in current literature, many assumptions including stationarity and dependence strength, are not directly applicable to time series with possibly changing dimensions $d(t)$. 
To circumvent  this difficulty we consider such \eqref{eq:fm} that are generated from a  possibly \textit{unidentifiable} locally stationary factor model with fixed dimension defined as follows.  Let $d=\max_{t\in[0,1]}d(t)$, and in this paper we focus on the case that $d$ is fixed and independent of $p,n$. Let $i_0=\min\{i: d(i/n)=d\}$ and consider 
\begin{align}\label{eq:fix}
\mf x_{i,n}=\mf A^*(i/n)\mf z^*_{i,n}+\mf e_{i,n},
\end{align} 
where  $\mf z^*_{i,n}$ and  $\mf e_{i,n}$ are $d$ and $p$ dimensional locally stationary time series, and the $p\times d$ loading matrix $\mf A^*(t)$  
is not necessarily full rank over the interval $[0,1]$. We now posit assumptions for \eqref{eq:fm} with varying dimensions through connections to \eqref{eq:fix}.  Consider the case that $\inf_{t\in [0,1]}\|\mf A^*(t)\|_F\asymp p^{\frac{1}{2}},\sup_{t\in [0,1]}\|\mf A^*(t)\|_F\asymp p^{\frac{1}{2}}$. 
 Using singular value decomposition (SVD),  model \eqref{eq:fix} can be written as
\begin{align}\label{SVD}
\mf x_{i,n}=p^{\frac{1}{2}}\mf U(i/n)\bs \Sigma(i/n)\mf{ W^{\top}}(i/n)\mf z^*_{i,n}+\mf e_{i,n},
\end{align}
where $\bs\Sigma(i/n)$ is a $p\times d$  rectangular diagonal matrix  with diagonal $\bs \Sigma_{uu}(i/n)=\sigma_u(i/n)$ for $1\leq u\leq d$,  $(\sigma_u(i/n))_{1\leq u\leq d}$ are singular values of $\mf A^*(i/n)/p^{\frac{1}{2}}$, 
$\mf U(i/n)$ and $\mf W(i/n)$ are corresponding left and right singular vectors, respectively. The $(\sigma_u(i/n))_{1\leq u\leq d}$ are ordered such that $\sigma_1(i_0/n)\geq...\geq \sigma_d(i_0/n)> 0$. It is easy to see that $\max_{1\leq u\leq d}\sup_{t\in (0,1]}\sigma_u(t)$ is bounded, and $\mf x_{i,n}$ will be locally stationary if $\{\sigma_l(t)\}_{1\leq l\leq d}$, $\mf U(t)$ and $\mf W(t)$ are smoothly time varying. 
 The equation \eqref{SVD} can be further written as
\begin{align}\label{eq-equive}
\mf x_{i,n}=p^{\frac{1}{2}}\tilde{\mf U}(i/n)\tilde{\bs \Sigma}(i/n)\tilde{\mf{ W}}^{\top}(i/n)\mf z^*_{i,n}+\mf e_{i,n},
\end{align}
where $\tilde{\bs \Sigma}(i/n)$ is the matrix by deleting all $k_{th}$, $1\leq k\leq d$ rows and columns of $\bs \Sigma(i/n)$ if 
 $\sigma_{k}(i/n)=0$
and $\tilde{\mf U}(i/n)$ and $\tilde{\mf{ W}}^{\top}(i/n)$ are the matrices resulted from the deletion of $k_{th}$ columns of ${\mf U}(i/n)$ and $k_{th}$ rows of   and ${\mf{ W}}^{\top}(i/n)$, respectively. Then \eqref{eq:fix} has a form of  \eqref{eq:fm} by  setting 
\begin{align*}
\mf A(i/n)= p^{\frac{1}{2}}\tilde{\mf U}(i/n)\tilde{\bs \Sigma}(i/n),~
\mf z_{i,n}=\tilde{\mf{ W}}^{\top}(i/n)\mf z^*_{i,n}.
\end{align*}  
We then replace condition (A2) by (A2'),  which will be displayed in Section \ref{appendix} in detail.
	 Then the 
analogy of the theoretical results in Section \ref{Sec:load} where $d$ is replaced by $d(t)$ will hold if (i) we assume (A1), (A2') (S0)-(S4), (M1)-(M8) with $d$ therein replaced by $d(t)$, and (ii) the following estimator $\hat d_n(t)$ is used to estimate $d(t)$ instead of \eqref{Feb4-hatd}: 
	\begin{align}\label{tilded}
	\hat d_n(t)=\argmin_{1\leq i\leq p}(\lambda_{i+1}(\hat {\mf \Lambda}(t))+q_n )/(\lambda_{i}(\hat {\mf \Lambda}(t))+q_n).
	\end{align}

\section{Selection of Tuning Parameters}\label{Sec::Tuning}
\setcounter{equation}{0}
\subsection{Selection of $J_n$ for the estimation of time-varying factor loading matrices}
We discuss the selection of $J_n$ for the estimation of time-varying factors. Since in practice $g_{J_n,K,\tilde M}$ is unknown, a data-driven method to select $J_n$ is desired. Recall that the residuals are $\hat {\mf e}_{i,n}=\mf x_{i,n}-\hat {\mf
	V}(\frac{i}{n})\hat {\mf
	V}^\top(\frac{i}{n})\mf x_{i,n}$, and $\hat {\mf e}_{i,n}=(\hat e_{i,1,n},...,\hat e_{i,p,n})^\top$ is a $p$ dimensional vector. We select $J_n$ as the minimizer of the following cross validation standard $CV(J)$, 
\begin{align}\label{criterionCV}
CV(J)=\sum_{i=1}^n\sum_{s=1}^p\frac{\hat e^2_{i,s,n}(J)}{(1-v_{i,s}(J))^2}
\end{align} 
where $v_{i,s}(J)$ is the $s_{th}$ diagonal element of $\hat {\mf V}(\frac{i}{n})\hat {\mf V}^\top(\frac{i}{n})$ obtained by setting $J_n=J$, and $\hat e_{i,s,n}(J),$ $1\leq i\leq n$, $1\leq s \leq p$ are also the components of residuals calculated when $J_n=J$. The cross-validation has been widely used in the literature of sieve nonparametric estimation and has been advocated by for example \cite{hansen2014nonparametric}. 

\begin{remark}\label{newrmk9.1}
\WC{Although \eqref{criterionCV} works reasonably well in our numerical studies, as pointed out by one referee, the validity of this criterion has only been  theoretically justified for independent observations.  The theoretically justified cross-validation for locally stationary time series has attracted considerable research interest recently, see for example \cite{stefancrossvalidation2019}.  However, their results focus on local M-estimators for uni-variate time series.  We leave the development of theoretically justification for \eqref{criterionCV} or the development of such criterion for the estimation of time-varying parameters for high dimensional locally stationary time series as a rewarding future work.}
\end{remark}
\subsection{Selection of tuning parameters $m_n$ and $w_n$ for testing static factor loadings}\label{selectmn}
We select $m_n$ by first choosing $N_n$ and letting $m_n=\lf (n-k_0)/N_n \rf$. The $N_n$ is chosen by the minimal volatility method as follows. 
For a given data set,  let \begin{align}
	\hat T_n=\sqrt m_n\max_{1\leq h\leq N_n}\max_{1\leq i\leq p-\tilde d_n}|\hat{\mf f}_i^\top \mf S^X_h|
\end{align}
be the test statistic obtained by using $N_n$. 
Consider a set of possible values for $N_n$, which is denoted by $\{J_1,...,J_s\}$ where $J_s$ are positive integers. For each $J_v$, $1\leq v\leq s$ we calculate $\hat T(J_v)$ and hence the local standard error 
\begin{align}
SE(\hat T(J_l),b)=\left(\frac{1}{2b}\sum_{u=-b}^b \left(\hat T(J_{u+l})-\frac{1}{2b+1}\sum_{u=-b}^b\hat T(J_{u+l}) )\right)^2\right)^{1/2}\label{S105}
\end{align}
where  $1+b\leq l\leq s-b  $ and $b$ is a positive integer, say $1$.
We then select $N_n$ by
\begin{align}\label{S116}
\argmin_{1+b\leq l\leq s-b} SE(\hat T(J_l),b)
\end{align}
which stabilizes the test statistics. The idea behind the minimum volatility method is that the test statistic should behave stably as a function of $N_n$ when the latter parameter is in an appropriate range. In our empirical studies we find that the proposed method performs reasonably well, and the results are not sensitive to the choice of $N_n$  as long as $N_n$ used is not very different from that chosen by \eqref{S116}.

After choosing $N_n$ and hence $m_n$, we then further choose $w_n$ again by the minimal volatility method. In this case, we first obtain the $
N_n(p-\tilde d)$ dimensional vectors \{$\hat {\bs l}_i$, $1\leq i\leq m_n$\} defined in Section \ref{sec:test_loading}. Then we select $w_n$ by a multivariate extension of the minimal volatility method in \cite{zhou2013heteroscedasticity} as follows. We consider choosing $w_n$ from a grid $w_1\leq ...\leq w_r$. For each $w_n=w_i$, $1\leq i\leq r$ we calculate a $
N_n(p-\tilde d_n)$ dimensional vector $\mf b^o_{i,u}=\frac{1}{w_i(m_n-w_i+1)}\sum_{j=1}^{u}(\hat {\mf s}_{j,w_i}-\frac{w_i}{m_n}\hat {\mf s}_{m_n})^{\circ 2}$ where $\circ$ represents the Hadamard product and $1\leq u\leq m_n-w_r+1$. Let $\mf B^o_i=(\mf b^{o\top}_{i,1},...,\mf b^{o\top}_{i,m_n-w_r+1})^\top$ be a $
N_n(p-\tilde d) (m_n-w_r+1)$ dimensional vector, and $\mf B$ be a $
N_n(p-\tilde d) (m_n-w_r+1)\times r$ matrix with its $i_{th}$ column $\mf B^o_i$. Then for each row, say $i_{th}$ row $\mf B_{i,\cdot}$ of $\mf B$, we calculate the local standard error $SE(\mf B_{i,\cdot},h)$ for a given window size $h$, see \eqref{S105} for definition of $SE$ and therefore obtain a $r-2h$ length row vector $(SE(\mf B_{i,h+1},h),...SE(\mf B_{i,r-h},h))$. Stacking these row vectors we get a new  $
N_n(p-\tilde d)  (m_n-w_r+1)\times (r-2h)$ matrix $\mf B^\dag$. Let $colmax(\mf B^\dag)$ be a $(r-2h)$ length vector with its $i_{th}$ element being the maximum entry of the $i_{th}$ column of $\mf B^\dag$. Then we choose $w_n=w_{k+h}$ if the smallest entry  of $colmax(\mf B^\dag)$ is its $k_{th}$ element.  
Finally,  as a rule of thumb, we recommend to use $0.02(p+p^2/\sqrt n)\log p$ for penalty $q_n$. This choice works reasonably well in our simulation and data analysis.
\section{Simulation Studies}\label{Simu-Results}
\subsection{Estimating the time-varying factor models}\label{Simupart1}

In this subsection we shall examine the performance of our proposed estimator \eqref{span-estimate} for time-varying factor models, and compare it with that in \cite{lam2011estimation}. The latter is equivalent to fixing $J_n=0$ in  \eqref{sieveapprox}. We use normalized shifted Legendre polynomials as our basis throughout our empirical studies. The method studied in \cite{lam2011estimation} is developed under the assumption of stationarity  with static factor loadings and hence the purpose of our simulation is to 
illustrate that the methodology developed under stationarity does not directly carry over to the locally stationary setting. 
To demonstrate the advantage of the adaptive sieve method, our method is also compared with a  simple local estimator of $\mf {\Lambda}(t)$, which was considered in the data analysis section in \cite{lam2011estimation} and we shall call it the local PCA method in our paper. Specifically, for each $i$, $\mf{\Lambda}(\frac{i}{n})$  will be consistently estimated by
 	\begin{align}
 	\hat{\mf{\Lambda}}(\frac{i}{n})=\sum_{k=1}^{k_0}\hat{\mf M}(\frac{i}{n},k)\hat{\mf M}^\top(\frac{i}{n},k), ~~\hat{\mf M}(\frac{i}{n},k)=\frac{1}{2m+1}\sum_{j=i-m}^{j=i+m}\mf x_{j+k}\mf x_j^\top
 	\end{align}  
 	  where $m$ is the window size such that $m\rightarrow \infty$ and $m=o(n)$.  \WC{The $J_n$ of our method is selected by cross validation, while $m$ of the local PCA method is selected by the one which minimizes MSE. We find this $m$ by using the underlying model. In practice, it is unclear how to determine the optimal value for $m$. According to Definition 1 in Section 2.7.3 of \cite{Buja1989linear}, the "degrees of freedom"  of local PCA and our sieve method are the same, indicating the two methods have similar model complexity and therefore the comparison is meaningful.}

  Define the following smooth functions:
\begin{align*}
g_0(t)=0.4(0.4-0.2t), \alpha_1(t)=1.3\exp(t)-1,\alpha_2(t)=0.6\cos(\frac{\pi t}{6})+2t,\\
\alpha_3(t)=-(0.5+2t^2), \alpha_4(t)=2\cos(\frac{\pi t}{6})+0.6t.
\end{align*}
Let $\mf A=(a_1,..,a_p)^\top$ be a $p\times 1$ matrix with $a_i=1+0.2(i/p)^{0.5}$. Define the locally stationary process 
$ z_i=G_1(i/n,\FF_i)$ where $G_1(t,\FF_i)=\sum_{j=0}^\infty g^j_0(t)\epsilon_{i-j}$ where filtration $\FF_i=(\epsilon_{-\infty},...,\epsilon_i)$ and $(\epsilon_i)_{i\in \mathbb Z}$ is a sequence of $i.i.d.$ $N(0,1)$ random variables. We then define the time varying matrix
  \begin{align}\label{new}
\mf A(t)=
\Big(\mf A^\top_1\alpha_1(t)~~\mf A_2^\top\alpha_2(t)~~\mf A^\top_3\alpha_3(t)~~\mf A^\top_4\alpha_4(t)\Big)^\top
\end{align}
where $\mf A_{1}$, $\mf A_2$, $\mf A_3$ and $\mf A_4$ are  the sub-matrices of $\mf A$ which consist of the first $round(p/5)_{th}$ rows, the  $(round(p/5)+1)_{th}$ to  $round(2p/5)_{th}$, $(round(2p/5)+1)_{th}$ to $(round(3p/5))_{th}$ and  the $(round(3p/5)+1)_{th}$ to $p_{th}$ rows of $\mf A$, respectively. Let $\mf e_{i,n}=(e_{i,1},...,e_{i,p})^\top$ be a $p\times 1$ vector with independent components and are independent of $(\epsilon_{i})_{i\in \mathbb Z}$. Moreover,  for each $j$, $1\leq j\leq p$, $e_{i,j}=(\exp(0.5i/n)+1)Z_{ij}/4$, where $Z_{i,j_1}$ and $Z_{i,j_2}$ are independent if $j_1\neq j_2$, and each $(Z_{i,j})_{1\leq i\leq n}$  is generated from an AR(1) process with AR coefficient  $0.3$ with $i.i.d.$ $N(0,1)$s innovation.   

We consider the cases that $p=50,100,200,500$ and $n=1000, 1500$.  The performances of the methods are measured in terms of the Root-Mean-Square Error (RMSE)  and the average principal angle. The RMSE of the estimation is defined as \begin{align*}
	RMSE=\frac{1}{np}\sum_{i=1}^n\|\hat{\mf V}(i/n)\hat{\mf V}^\top(i/n)\mf x_{i,n}-\mf A(i/n)\mf z_{i,n}\|_2^2.
	\end{align*} The principle angle between $\mf A(i/n)$ and its estimate $\hat {\mf A}(i/n)$ is defined as follows. 
 Let $\sigma_{1,i}\geq ,...,\sigma_{d,i}$ be the singular values of $\hat {\mf A}^\top(i/n) \mf A(i/n)$, and  the principle angle  is defined as $\bs \Upsilon_{i,n}:=(cos^{-1}\sigma_{1,i},...,cos^{-1}\sigma_{d,i}) $, which is also a well-defined distance between spaces $span(\mf {A}(i/n))$ and  $span(\hat{\mf A}(i/n))$. Finally, the average magnitude of  the principle angle is defined as $\|\bar {\bs \Upsilon}\|=\frac{1}{n}\sum_{i=1}^n\|\bs \Upsilon_i\|$. We present the RMSE and the average  magnitude of  the principle angle of the three estimators using 800 simulation samples in Table \ref{Table-RMSE} and Table \ref{Table-angle}, respectively. Our method achieves the minimal RMSE and  average principle angle in all simulation scenarios among the three estimators. We choose $k_0=3$ in our simulation. Other choices  $k_0=1,2,4$ yield similar results and are not reported here. As predicted by Theorem \ref{Space-Distance}, RMSE in Table \ref{Table-RMSE} decreases as $n$, $p$ increases and the average principle angle decreases with $n$ increases, and is independent of $p$. 

\begin{table}[ht]
	\centering
	 	\caption{  Mean and standard errors (in brackets)  of simulated RMSE for our sieve method,  the static loading method ($J_n=0$) and Local PCA for model \eqref{new}. The results are multiplied by $1000$.}
	\begin{tabular}{|l|ccc||ccc|}
		\hline
		& \multicolumn{3}{c||}{$n=1000$} & \multicolumn{3}{c|}{$n=1500$} \\
		\hline
		& Sieve & $J_n=0$ & Local PCA & Sieve & $J_n=0$ & Local PCA \\ 
		\hline
		$p=50$ & $541.78_{(1.03)}$ & $645.57_{(1.58)}$ & $552.68_{(1.22)}$ & $524.06_{(0.79)}$ & $634.92_{(1.26)}$ & $535.20_{(0.92)}$ \\ 
		$p=100$ & $532.45_{(1.03)}$ & $635.85_{(1.51)}$ & $543.15_{(1.12)}$ & $516.26_{(0.81)}$ & $629.38_{(1.18)}$ & $528.78_{(0.96)}$ \\ 
		$p=200$ & $526.61_{(1.04)}$ & $634.30_{(1.47)}$ & $540.22_{(1.10)}$ & $512.53_{(0.78)}$ & $624.99_{(1.24)}$ & $524.46_{(0.91)}$ \\ 
		$p=500$ & $525.94_{(0.99)}$ & $631.27_{(1.49)}$ & $538.46_{(1.12)}$ & $510.88_{(0.77)}$ & $621.98_{(1.28)}$ & $521.20_{(0.87)}$ \\ 
		\hline
	\end{tabular}
\label{Table-RMSE}
\end{table}
\begin{table}[ht]
	\centering
		\caption{ Mean and standard errors (in brackets)  of simulated principle angles for our sieve method,  the static loading method ($J_n=0$) and Local PCA for model \eqref{new}. The results are multiplied by $1000$.}
	\begin{tabular}{|l|ccc||ccc|}
	\hline
	& \multicolumn{3}{c||}{$n=1000$} & \multicolumn{3}{c|}{$n=1500$} \\
	\hline
	& Sieve & $J_n=0$ & Local PCA & Sieve & $J_n=0$ & Local PCA \\ 
	\hline
		$p=50$ & $17.07_{(0.25)}$ & $46.54_{(0.28)}$ & $18.66_{(0.25)}$ & $12.51_{(0.18)}$ & $43.28_{(0.20)}$ & $14.50_{(0.19)}$ \\ 
		$p=100$ & $16.79_{(0.24)}$ & $47.17_{(0.30)}$ & $18.52_{(0.24)}$ & $12.68_{(0.18)}$ & $43.54_{(0.19)}$ & $14.90_{(0.19)}$ \\ 
		$p=200$ & $16.49_{(0.25)}$ & $46.90_{(0.26)}$ & $18.87_{(0.23)}$ & $12.85_{(0.18)}$ & $43.35_{(0.22)}$ & $14.82_{(0.18)}$ \\ 
		$p=500$ & $16.99_{(0.24)}$ & $46.59_{(0.28)}$ & $18.98_{(0.23)}$ & $12.94_{(0.18)}$ & $43.65_{(0.27)}$ & $14.70_{(0.18)}$ \\ 
		\hline
	\end{tabular}
	\label{Table-angle}
\end{table}
\subsection{Testing static loading matrix: type I error}\label{type1check}
We  now examine our testing procedure in Section \ref{sec:test_loading} to test the hypothesis of static factor loadings via $B=2000$ bootstrap samples. 
Define
\begin{align*}
g_1(t)=0.1+0.06t^2, g_2(t)=0.12+0.04t,g_3(t)\equiv 0.15,\\
	\alpha_1(t,D)=0.8+2cos(\pi t/2)D ,
\alpha_2(t,D)=0.9-6(t-0.5)^2D,
\alpha_3(t,D)=(1+1.6tD)
\end{align*}
Let $\mf A$ be a $p\times 3$ matrix with each element generated from $2U(-1,1)$,  and  \begin{align}
\mf A(t,D)=\begin{pmatrix}
\mf A_1 \alpha_1(t,D)\\\mf A_2\alpha_2(t,D)\\\mf A_3 \alpha_3(t,D)
\end{pmatrix}
\end{align}
where $\mf A_{1}$, $\mf A_2$ and $\mf A_3$ are  the sub-matrices of $\mf A$  which consist of its first $round(p/3)_{th}$ rows, the  $(round(p/3)+1)_{th}$ to  $round(2p/3)_{th}$ rows, and $(round(2p/3)+1)_{th}$ to  $p_{th}$ rows, respectively. By construction, $\tilde {\mf A}=\mf A(t,0)$ is time-invariant and  to examine type I error, we consider the null hypothesis that the loading matrix is $\tilde{\mf A}$. The factors $\mf z_{i,n}=(z_{i,1,n},z_{i,2,n},z_{i,3,n})^\top$ where $z_{i,k,n}=3\sum_{j=0}^\infty g_k^{j}(i/n)\epsilon_{i-j,k}$ for $k=1,2,3$, and $\{\epsilon_{i,k}\}$ are $i.i.d.$ standard normal. We consider the following two models for errors. Let $\bs e_{i,n}=(e_{i,1,n},...,e_{i,p,n})^\top$, where each component series $e_{i,s_1,n}$ and $e_{i,s_2,n}$ are independent if $s_1\neq s_2$. The first is the locally stationary high dimensional autoregressive model. For $1\leq s\leq p$, 
\begin{align}
	e_{i,s,n}=(0.5+0.2(i/n)^2)\tilde e_{i,s,n}
\end{align}
and $\tilde e_{i,s,n}$ is generated (independently w.r.t. $s$) from a stationary AR(1) process with AR coefficient $0.3$ and $i.i.d.$ $0.9\tilde t_8$ innovations. Here $\tilde t_8$ refers to standardized student $t$ distribution with degrees of freedom $8$, i.e., $\sqrt{0.8}t_8$. The second is the locally stationary white noise, i.e., for $1\leq s\leq p$,  $e_{i,s,n}=(0.5+0.2(i/n)^2)\tilde e_{i,s,n}\tilde e_{i-1,s,n}$ where $\tilde e_{i,s,n}$ are $i.i.d.$ standard normal. We  examine the type 1 error of our methods in the following Table \ref{Tablenew1} via 2000 simulated samples, and find that the simulated  type 1 error is reasonably close to their nominal level.
\begin{table}[ht]
	\centering
	\begin{tabular}{|r|rr|rr|rr|rr|}
		\hline
	&\multicolumn{4}{|c|}{High dimensional AR}&\multicolumn{4}{|c|}{High dimensional white noise}\\
		\hline
	&	\multicolumn{2}{|c|}{$T=1000$}&\multicolumn{2}{c}{$T=1500$}&
	\multicolumn{2}{|c|}{$T=1000$}&\multicolumn{2}{c|}{$T=1500$}\\
		\hline
		& 5\% & 10\% & 5\% & \%10 &\%5 &\% 10\% & 5\% & 10\% \\ 
		\hline
		$p=20$ & 4.9 & 9.5 & 4.9 & 9.9 & 5.35 & 9.6 & 4.6 & 9.6  \\ 
		$p=50$ &4.85 & 10.2 & 5.25 & 10.6 & 5.3 & 10.6 & 4.9 & 9.55 \\ 
		$p=100$ &5.65 & 11.4 & 5.75 & 11.15 & 5.75 & 10.15 & 4.95 & 9.25 \\ 
		\hline
	\end{tabular}
\caption{Simulated type 1 errors for high dimensional AR and White noise model, respectively.}\label{Tablenew1}
\end{table}

\subsection{Testing static factor loadings: power}
\begin{figure}[h]
	
	\begin{minipage}[c]{0.5\linewidth}
		\centering   
		\includegraphics[width=0.9\linewidth, height=0.3\textheight]{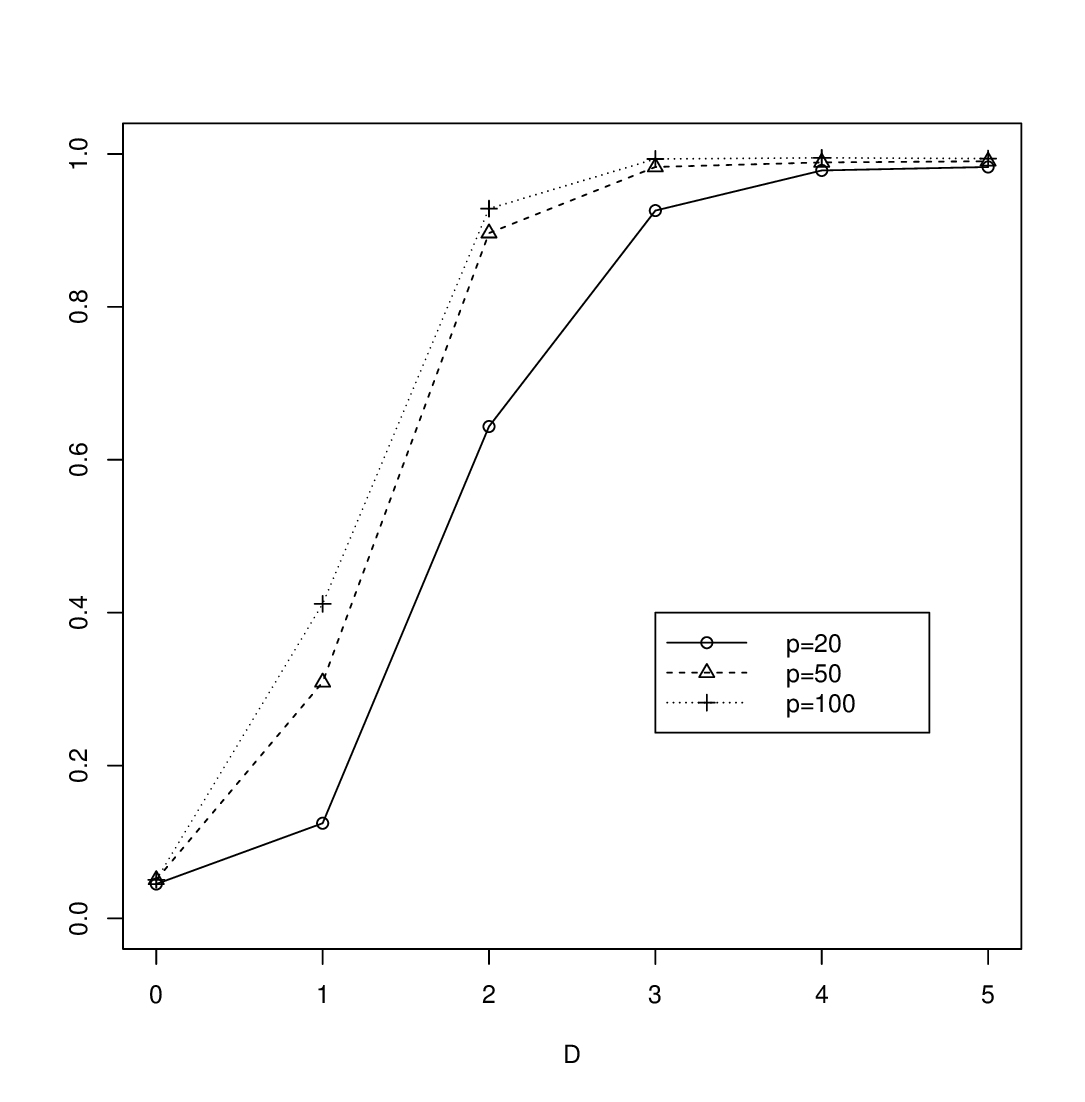}   
		\caption{\small Simulated power, $T=1000$}
		\label{powerT1000}   
	\end{minipage}%
	\begin{minipage}{0.5\linewidth}   
		\centering   
		\includegraphics[width=0.9\linewidth, height=0.3\textheight]{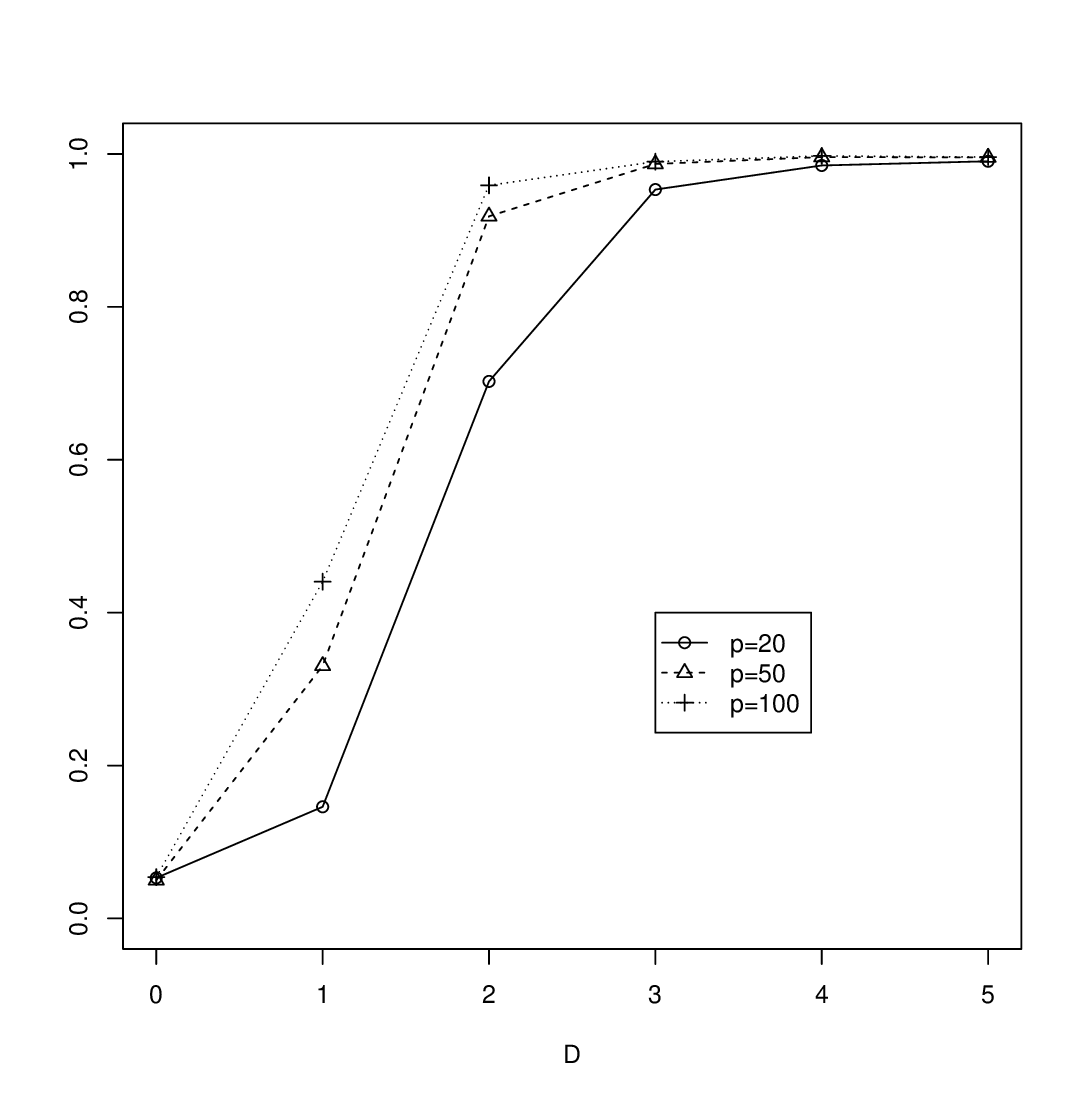}     
		\caption{ \small Simulated power, $T=1500$}
		\label{PowerT1500}   
	\end{minipage}   
\end{figure}

In this subsection, we examine the power performance of our testing procedure in Section \ref{sec:test_loading} via $B=2000$ bootstrap samples. We consider examining the empirical rejection rates of the model considered in Section \ref{type1check} with $\mf A(t,D)$ for different $D's$ and  with the high dimensional locally stationary AR error.  We consider $p=20, 50, 100$, $T=1000, 1500$ and $D$ varies from $0$ to $0.5$. The results are based on 2000 simulation samples, while the critical value in each run is generated from $2000$ bootstrap samples. The results are summarized in Figure \ref{powerT1000} for $T=1000$ and Figure \ref{PowerT1500} for $T=1500$. The empirical outcome evidences that our method has good power performance. The power of our methods increases as dimension expanse or sample size enlarges.

\section{Analysis of UK temperature data}\label{data-ana}
\setcounter{table}{0}
\setcounter{figure}{0}
To illustrate the usefulness of our method we investigate the UK historical station monthly  temperature data, which can be downloaded from https://www.metoffice.gov.uk/research/climate/maps-and-data/historic-station-data.  We consider stations with monthly temperate recorded in every year during Jan. 1979- May. 2023. and we have $33$ stations in total. We consider the monthly highest temperature and lowest temperature series, both forming a 33-dimensional time series with length $533$. There
are also 77 missing data in the  two series. The missing data are imputed by interpolating the trend component of the seasonal decomposition of the corresponding time series via implementing R package "imputeTS". 

For each series, we study the error processes after removing the seasonal trends which are obtained by the R command `stl'.
We first examine whether the 33-dimensional errors  have a  static loading matrix by performing our test procedure in Section \ref{sec:test_loading}.  In our data analysis we choose $k_0=3$. Recall  $N_n$ is the number of non-overlapping equal-sized blocks, and $w_n$ is the window size. Using the minimal volatility method stated in Section \ref{Sec::Tuning} we select $N_s
=16$ and $w_n=5$  for the monthly highest temperature which yields a $p$ value $=2.78\%$, $N_s
=15$ and $w_n=6$ 
for the monthly minimal temperature which yields a $p$ value  $61.25\%$. Both the $p$ values are derived from $B=10000$ bootstrap samples. The small $p$ value for the  monthly maximum temperature provides a strong evidence against the null hypothesis of static factor loadings, while for the monthly minimal temperature, the test is insignificant. \WC{Our findings reveal that the co-movements of the highest monthly temperatures among different weather stations in UK are significantly different from those of the lowest temperatures, with the former being time-varying and the latter being static over the considered time-span.} 



\begin{figure}[h]
	
	\begin{minipage}[c]{0.5\linewidth}
		\centering   
		\includegraphics[width=0.9\linewidth, height=0.3\textheight]{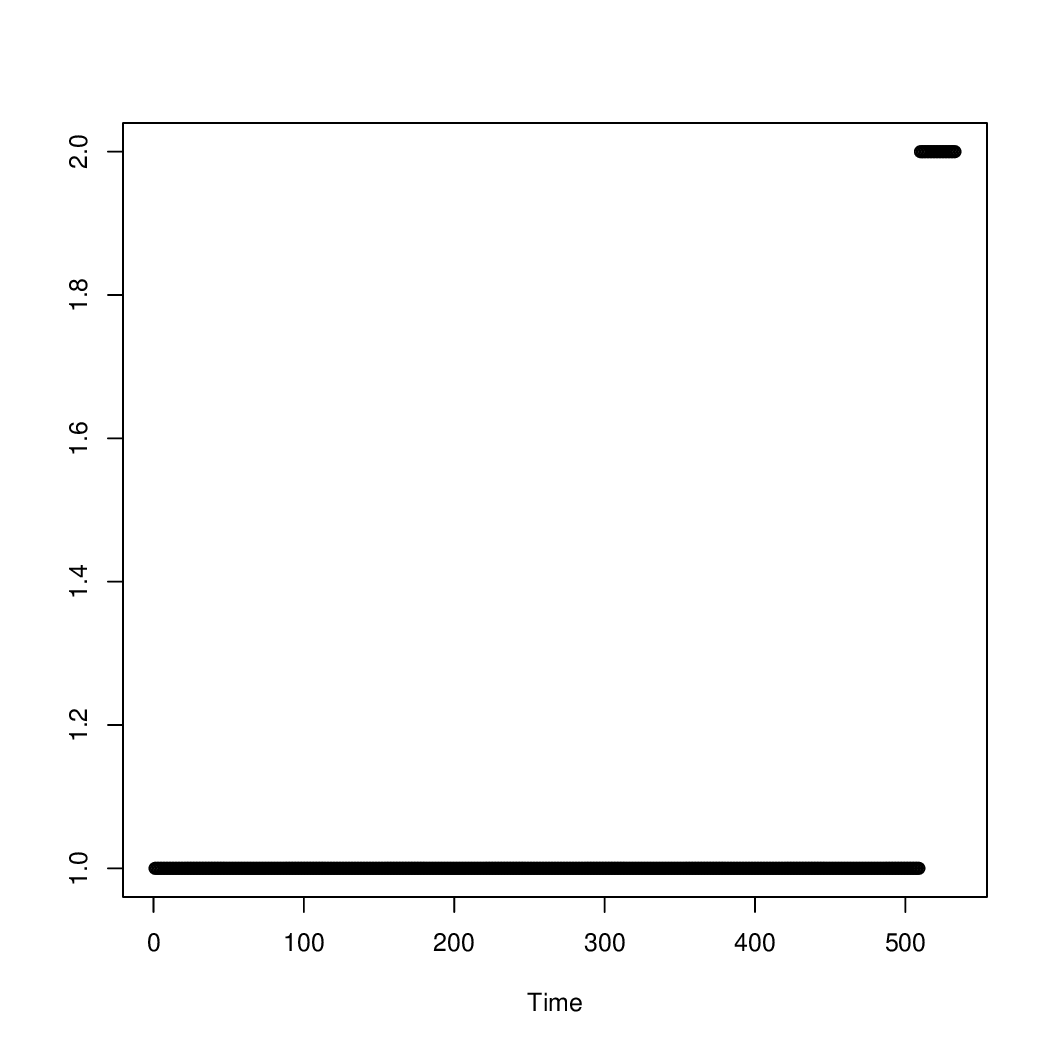}   
		\caption{\small Number of  factors of the monthly highest temperature}
		\label{eigen-plot}   
	\end{minipage}%
	\begin{minipage}{0.5\linewidth}   
		\centering   
		\includegraphics[width=0.9\linewidth, height=0.3\textheight]{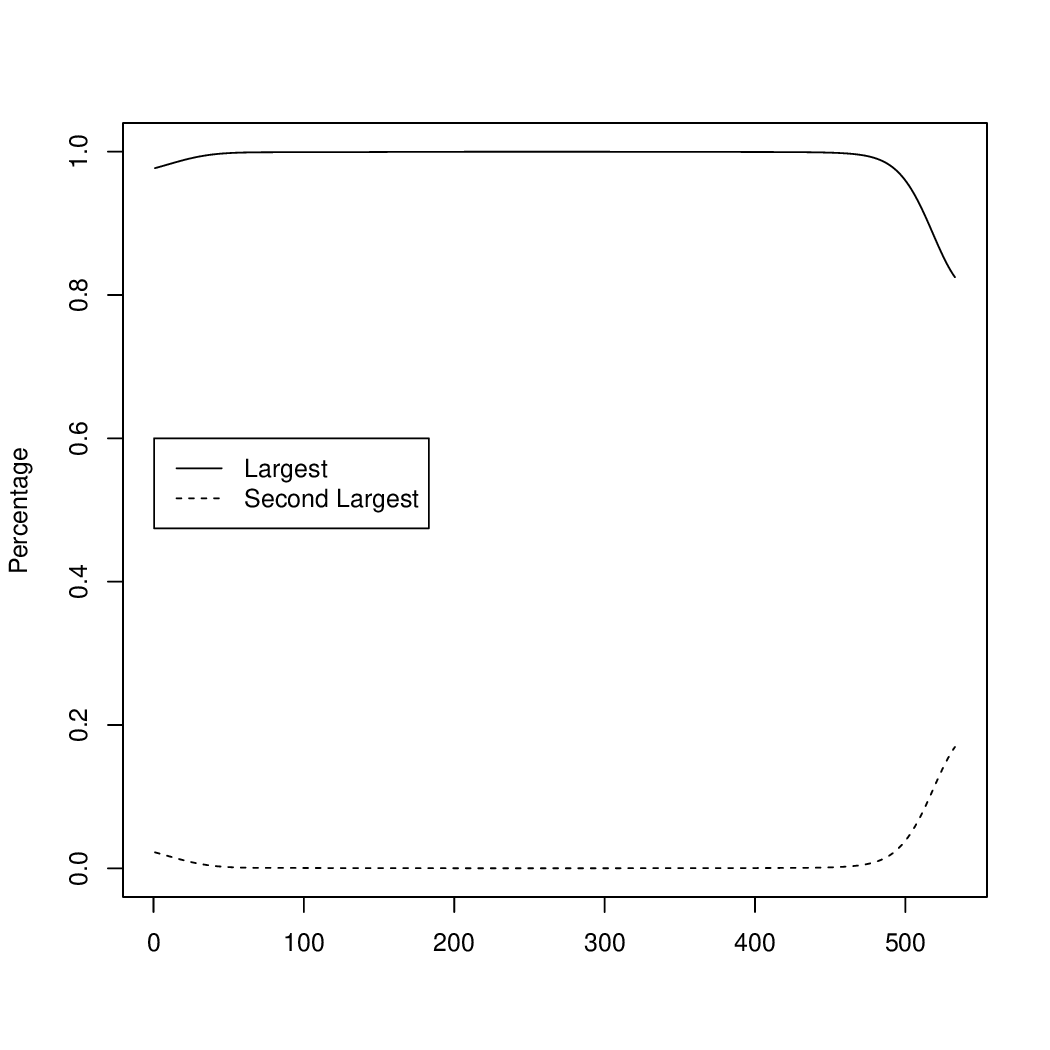}   
		\caption{ \small Proportion of trace of $\mf \Lambda_1(t)$ that can be explained by  the leading two eigenvectors}
		\label{explainvar}   
	\end{minipage}   
\end{figure}

We then apply our sieve estimator in Section \ref{Sec:estimate}  to estimating the time-varying loading matrix for the monthly highest temperature. The cross validation method suggests the use of the normalized shifted Legendre polynomial basis up to $3_{rd}$ order. We find that during the considered period the
number of factors is varying between $1$ and $2$. In Figure \ref{eigen-plot}  we display the estimated
number of factors at each time, and in Figure \ref{explainvar} we show the percentage of trace of $\mf \Lambda_1(t)$
that is explained by the eigenvectors corresponding to the first and second largest eigenvalues,
which reflects the time-varying structure of the loading matrix. The results underpin that the
loading matrix is time-varying. \WC{As pointed  out by a referee, it is important to test whether $\mf A(i/n)\mf z_{i,n}$ and $\mf e_{i+k,n}$ for $k=1,2,3$ are uncorrelated due to \eqref{Lambdanull} such that (S3) holds. This is equivalent to testing whether $\E((\mf G(t,\FF_i)-\mf H(t,\FF_i))\mf H^\top (t,\FF_{i+k}))=\mf 0$. This can be examined by our proposed test in Section \ref{sec:test_loading} with $\hat {\bs l}_i$ redefined as $Vec((\mf x_{i+(s-1)m_n}-\hat{\mf e}_{i+(s-1)m_n})\hat{\mf e}_{i+(s-1)m_n+k}^\top)$, $k=1,2,3$, and the $Vec$ stands for vectorization. The resulting test yields a $p$ value of $0.99$, indicating no evidence against uncorrelatedness.}

We examine the performance of our method on one-step prediction of the UK monthly highest temperature and compare it with the (local) PCA method (See \cite{lam2011estimation}).  Given $\mf Y_1$,...,$\mf Y_i$, the predictor $\hat{\mf Y}_{i+1}$ is obtained as follows.
\begin{description}
    \item  (1) Apply component-wise seasonal decomposition of $(\mf Y_s)_{1\leq s\leq i}$ to obtain the seasonality, trend and residuals
    $(\bs s_v)_{1\leq v\leq i}$, $(\bs t_v)_{1\leq v\leq i}$ and $(\mf X_v)_{1\leq v\leq i}$, respectively.
    \item (2) Apply our method to  $(\mf X_v)_{1\leq v\leq i}$ and obtain 
   $ \hat d(s/i), 1\leq s\leq i$ following \eqref{tilded}. 
    Let
    $\hat d_{\max}=\max_{1\leq s\leq i} \hat d(s/i)$.
    
    \item (3) Let $\check{\mf A}_v$ be a $p\times \hat d_{\max}$ matrix, where its $j_{th}$ column is the eigenvector of $\bs \Lambda(v/i)$ with respect to its $j_{th}$ largest eigenvalue, $1\leq j\leq \hat d_{max}$.  Then calculate $\hat {\mf z}_v
        =(\check{\mf A}_s)^\top \mf X_s.$
    \item (4) We then forecast the vector $\hat {\mf z}_{i+1}$ based on $\hat {\mf z}_v$, $v=1,..,i$. We consider two methods.
    \begin{description}
    \item (4a) Predict $\hat {\mf z}_v$ via a stationary vector AR model, using R package {\it vars}. 
    \item (4b)  Predict $\hat {\mf z}_v$ via a locally stationary time-varying vector AR model, using R package {\it tvReg} which implements \cite{casas2021time}.  
        \end{description}
    \item (5) Predict each component of $\mf e_i$ as in 4(b).

    \item (6) If $\hat d(1)<\hat d_{max}$, set the $l_{th}$ component of $\hat {\mf z}_{i+1}$ as zeros for $\hat d(1)+1\leq l\leq \hat d_{max}$. Then we predict $\mf X_{i+1}$ by 
    $\hat {\mf X}_{i+1}=\check{\mf A}_i \hat{\mf z}_{i+1}.$
    \item (7) Finally we predict $\mf Y_{i+1}$ by 
    $\hat{\mf Y}_{i+1}=\hat{\mf X}_{i+1}+\bs t_i+\bs s_{i-11}+\hat {\mf e}_{i+1}$.
\end{description}
The reason we consider time-varying vector AR model for the common factor is due to \cite{krampe2022inverse} which extends a recent work of \cite{dingzhouapprox} from univariate time series to the multivariate setting and proves that under certain conditions, a locally stationary vector time series has locally stationary vector $AR(\infty)$ representation with approximately smooth coefficients and can be further 
 approximated by locally stationary vector AR models with finite orders.  

 The performance is evaluated by the following squared mean prediction error (MSPE). For a given period $j,...,T$, the MSPE
 \begin{align}\label{MSPE}
 MSPE=\frac{\sum_{s=j}^T\|\hat{\mf Y}_{s+1}-\mf Y_{s+1}\|_2^2}{p(T-j+1)}
 \end{align}
 where $p=33$ is the dimension of the temperature vector.
  For comparison, we also consider $\hat {\mf Y}_{i+1}$ resulting from predicting $\mf X_{s+1}$ by the local PCA method advocated in \cite{lam2011estimation}.
 The period we consider for MSPE \eqref{MSPE} starts from the Jul. 2008, which corresponds to $j=355$ at the $2/3$ length of data, and ends at May. 2023 corresponding to $T=533$. 
 To apply the local PCA  method, we use data at $s-L\leq t\leq s$ to forecast $Y_{s+1}$ with $L=100$ and $L=200$.
 We summarize the results in Table \eqref{tabPred}. In general our method which predicts both $\mf z_i$ and $\mf e_i$ achieves the smallest MSPE. A benchmark procedure for forecast is to use the highest temperature of the same month in the previous year as a one-step prediction of the highest temperature of the corresponding month this year. The MSPE for this benchmark is $4.114$. 
  
  \begin{table}[htbp]
  	\centering
  	\caption{MSPE for Sieve PCA-based method and Local PCA based method. The first row corresponds to predicting $\mf z_{i+1}$ using stationary VAR fit, and the second row  corresponds to the use of time-varying locally stationary VAR fit. The third additionally predicts $\mf e_i$  } 
  	\begin{tabular}{lrrr}\hline
  		& \multicolumn{1}{l}{Our Method} & \multicolumn{1}{l}{Local PCA (L=100)} & \multicolumn{1}{l}{Local PCA (L=200)} \\
  		\hline
  		AR    & 2.286  &  2.191& 2.143\\
  		LS & 2.106  &   2.127 &  2.210\\
  		LS+predict&2.102* & 2.125 &2.137\\
  		\hline
  	\end{tabular}%
  	\label{tabPred}%
  \end{table}%

\section*{Acknowledgment}
	The work of the first author was supported by  NSFC  12271287 and 11901337. The work of the second author was supported by NSERC of Canada.
\section*{Online supplement}
The online supplement contains examples of locally stationary time series satisfying conditions in this paper, and the proofs (all under $\delta\geq 0$) of Theorem \ref{Space-Distance} (ii), auxiliary lemmas for Theorems \ref{Thm-approx}, \ref{Space-Distance} and Theorem \ref{Jan23-Thm4}, as well as the proofs of  Propositions \ref{hatdrate}, 
Theorems  \ref{Boots-thm5} and \ref{Power}. 

\setcounter{equation}{0}


\renewcommand{\theequation}{S.\arabic{equation}}

\section{Proof of Theorems \ref{Thm-approx}, \ref{Space-Distance}} \label{appendix}
\setcounter{equation}{0}
To prove the results in Section \ref{Sec:load}, we consider showing the results when the number of factors or the dimension of the loading matrix, which is denoted by $d(t)$, is allowed to be {\it time-varying}. The situation when the dimension is fixed can be shown in a similar but easier argument.  Recall $d=\max_{t\in [0,1]}d(t)$. In this section, we provide proofs for factor strength $\delta\geq 0$ for varying-dimension, i.e., assume the following (A2') instead of (A2), and also the following modified (S4) which  relies on $\delta$.

\begin{description}
    \item (A2')Assume  for some constant $\delta\in [0,1]$,  $\sup_{t\in [0,1]}\|\mf a_s(t)\|^2_2\asymp p^{1-\delta}$
for $1\leq s\leq d(t)\leq d$, and that
the matrix norm of $\mf A(t)=(\mf a_1(t),...,\mf a_{d(t)}(t))$ satisfies
\begin{align*}
	\inf_{t\in [0,1]}\|\mf A(t)\|_F\asymp p^{\frac{1-\delta}{2}},\sup_{t\in [0,1]}\|\mf A(t)\|_F\asymp p^{\frac{1-\delta}{2}},
	\inf_{t\in \mathcal T_{\eta_n}}\|\mf A(t)\|_m\gtrapprox \eta^{1/2}_n p^{\frac{1-\delta}{2}} 
\end{align*}
for a positive sequence $\eta_n=O(1)$ on a collection of intervals $\mathcal T_{\eta_n}\subset [0,1]$. Besides,  $\mf A(t)$ is full rank on $\mathcal T_{\eta_n}$.
\item(S4) for $t\in [0,1]$ and $1\leq k\leq k_0$, $\|\mf \Sigma_{ze}(t,k)\|_{F}=o(\eta^{1/2}_n p^{\frac{1-\delta}{2}})$.
\end{description}
If $\delta=0$ and $d(t)\equiv d$, the above (A2') and (S4) will be the same as (A2) and (S4) in Section \ref{tech-assumption}.
 As in \cite{lam2011estimation} and \cite{lam2012factor}, $\delta=0$ and $\delta>0$  correspond to strong and weak factor strengths, respectively. If $d(t)$ is piecewise constant with a bounded number of change points, then $|\mathcal T_{\eta_n}|\rightarrow 1$ as $n\rightarrow \infty$ and $\eta_n\rightarrow 0$ according to the connection to model\eqref{eq:fix} in Section \ref{Sec::Pre}. If $d(t)\equiv d$ we can assume that $\mathcal T_{\eta_n}=[0,1]$ for some sufficiently small positive $\eta_n:=\eta>0$.

For completeness we summarize the short memory and stochastic Lipschitz continuous, and moment conditions for $\mf z_{i,n}$  and $\mf e_{i,n}$:
\begin{description}
	\item (M1) The short-range dependence conditions hold for  both $\mf z_{i,n}$ and $\mf e_{i,n}$ in $\mathcal L^l$ norm , i.e.  \begin{align}
		\max_{1\leq j\leq d} \delta^z_{l,j}(k)=O((k\log k)^{-2}),\ \  \max_{1\leq j\leq p}\delta^e_{l,j}(k)=O((k\log k)^{-2})
	\end{align}
	for some constant $l\geq 4$. 
	\item (M2) There exists a constant $M$ such that 
	\begin{align*}
		\sup_{t\in[0,1]}\max_{1\leq u\leq d}\E | Q_u(t,\FF_0)|^4\leq M, ~\sup_{t\in [0,1]}\max_{1\leq v\leq p}\E |H_v(t,\FF_0)|^4\le M.
	\end{align*}
	\item (M3) For $t,s\in [0,1]$, there exists a constant $M$ such that 
	\begin{align}&\left(\E |Q_u(t,\FF_0)-Q_u(s,\FF_0)|^2\right)^{1/2}\leq M|t-s|, 1\leq u\leq d,\label{8-11-10}
		\\&\left(\E |H_v(t,\FF_0)-H_v(s,\FF_0)|^2\right)^{1/2}\leq M|t-s|, 1\leq v\leq p,\label{8-11-11}\\
		&\left(\E |G_v(t,\FF_0)-G_v(s,\FF_0)|^2\right)^{1/2}\leq M|t-s|, 1\leq v\leq p.\label{8-11-12}
	\end{align}
\end{description}
Conditions (M1)-(M3) mean that each coordinate process of ${\mf z}_{i,n}$ and $\mf e_{i,n}$, as well as that of $\mf x_{i,n}$ (see Lemma \ref{LS-G} ) is a standard short memory locally stationary time series defined in the literature. 

\medskip
\noindent {\bf Proof of Theorem \ref{Thm-approx}.} 
We shall prove that, if (S1) hold
\begin{align*}
	\Big\|\sup_{t\in [0,1]}\Big\|\hat {\mf \Lambda}(t)-\mf \Lambda_1(t)\Big\|_2\Big\|_{\mathcal L^1}=O(p^{2-\delta}\nu_n).
\end{align*}
and  if (S1') hold, the rate will be $O(p^{2-\delta}\nu_n+p^{1-\delta})$ under the condition that $p^\delta \nu_n=o(1)$ which is always satisfied when $\delta=0$. Notice that when $\delta=0$ and $d(t)\equiv d$ the results of Theorem \ref{Thm-approx} hold.
Notice that under condition (S) (either (S1) or (S1') holds) we have \begin{align}\label{S1}
	\bs \Sigma_x(t,x)=\mf A(t)\bs \Sigma_z(t,k)\mf A^\top(t)+\mf A(t)\bs \Sigma_{ze}(t,k)+\bs \Sigma_e(t,k).
\end{align}
and that for each $k\in 1,...,k_0$, we have that
\begin{align}\label{11-29-eq6}
\sup_{t\in[0,1]}&\|\hat {\mf M}(J_n,t,k)\hat {\mf M}^\top(J_n,t,k)-\bs \Sigma_x(t,k)\bs \Sigma_x^\top(t,k)\|_{F}\leq \\&2\|\bs \Sigma_x^\top(t,k)\|_{F}\|\hat {\mf M}(J_n,t,k)-\bs \Sigma_x^\top(t,k)\|_{F}+\|\hat {\mf M}(J_n,t,k)-\bs \Sigma_x^\top(t,k)\|^2_{F}.\notag
\end{align}

By \eqref{S1}, condition (A2), (S1) or (S1'), (S4) and the submultiplicity of Frobenious norm, we have for $t\in [0,1]$
\begin{align}\label{11-29-eq7}
&	\notag\|\bs \Sigma_x(t,k)\|_F\\&\leq \|\mf A(t)\|_F^2\|\bs \Sigma_z(t,k)\|_F+
	\|\mf A(t)\|_F\|\bs \Sigma_{ze}(t,k)\|_F+\|\bs \Sigma_e(t,k)\|_F
	\leq C d p^{1-\delta}
\end{align}
for some sufficiently large constant $C$ 
 which depends on the constant $M$ in condition (M2). On the other hand, by Lemmas  \ref{11-29-lemma3}, \ref{TildeSigma} and  \ref{11-29-lemma5} in the online supplement we have that
\begin{align}
\|\sup_{t\in [0,1]}\|\hat {\mf M}(J_n,t,k)-\bs \Sigma_x(t,k)\|_F\|_{\mathcal L^2}=O(p\nu_n).\label{11-29-eq8}
\end{align}
Then  it follows from equations \eqref{11-29-eq6}, \eqref{11-29-eq7} and \eqref{11-29-eq8} that
	\begin{align}\label{Final1}
	\Big\|\sup_{t\in [0,1]}\Big\|\hat {\mf \Lambda}(t)-\mf \Lambda(t)\Big\|_2\Big\|_{\mathcal L^1}\leq \Big\|\sup_{t\in [0,1]}\Big\|\hat {\mf \Lambda}(t)-\mf \Lambda(t)\Big\|_F\Big\|_{\mathcal L^1}=O(p^{2-\delta}\nu_n).
	\end{align}
Therefore, under (S1) the Theorem holds. We now show the theorem under (S1'). Elementary calculations show that $\mf \Lambda(t)-\mf \Lambda_1(t)=\mf \Lambda_2(t)+\mf \Lambda_3(t)$, where 
\begin{align*}
	\mf \Lambda_2(t)&=\sum_{k=1}^{k_0}\bs \Sigma_e(t,k)[\mf A(t)\bs  \Sigma_z^\top(t,k)\mf A^\top (t)+\bs\Sigma_{ze}^\top(t,k)\mf A^\top (t)]\notag\\&+\sum_{k=1}^{k_0}[\mf A(t)\bs \Sigma_z(t,k)\mf A^\top(t)+\mf A(t)\bs \Sigma_{ze}(t,k)]\bs \Sigma_{e}^\top(t,k),\\
	\mf \Lambda_3(t)&=\sum_{k=1}^{k_0}\bs \Sigma_e(t,k)\bs \Sigma_e^\top(t,k).
\end{align*}
Notice that $\sup_{t\in [0,1]}\|\mf \Lambda_3(t)\|_2=O(1)$, and for all $t\in[0,1]$
\begin{align}
		&\|\mf \Lambda_2(t)\|_2\leq 2\|\sum_{k=1}^{k_0}\bs \Sigma_e(t,k)[\mf A(t)\bs  \Sigma_z^\top(t,k)\mf A^\top (t)+\bs \Sigma_{ze}^\top(t,k)\mf A^\top (t)]\|_2\notag\\
		&\leq 2\sum_{k=1}^{k_0}\|\bs \Sigma_e(t,k)\|_2\|\mf A(t)\bs  \Sigma_z^\top(t,k)\mf A^\top (t)+\bs \Sigma_{ze}^\top(t,k)\mf A^\top (t)\|_F=O(p^{1-\delta}).
\end{align}
Together with \eqref{Final1} the theorem follows.
 \hfill $\Box$

\medskip
\noindent {\bf Proof of Theorem \ref{Space-Distance}.} 

 We shall prove the following results for $\delta\geq 0$. For (i) if (S1) holds then the rate will be reduced to $O(\eta_n^{-1}p^{\delta}\nu_n)$ and  if (S1') holds the estimation rate is $O(\eta_n^{-1}p^{\delta}\nu_n+\eta_n^{-1}p^{\delta-1})$.
 For (ii),  
if (S1) holds the rate is $O_p( \eta_n^{-1}p^{\delta}\nu_n+p^{-1/2})$ and if (S1') holds the rate is $O_p( \eta_n^{-1}p^{\delta}\nu_n+p^{-1/2}+\eta_n^{-1}p^{\delta-1})$ under the condition that $p^\delta \nu_n=o(1)$ which is always satisfied when $\delta=0$. The results of Theorem \ref{Space-Distance} correspond to $\delta=0$.

For simplicity, we only show under (S1'). The proof under (S1) is similar. 

We first prove (i).
It suffices to show that
the $d(t)_{th}$ largest eigenvalue of $\mf \Lambda_1(t)$ satisfies \begin{align} \inf_{t\in\mathcal T_{\eta_n}}\lambda_{d(t)}(\mf \Lambda_1(t)) \gtrapprox \eta_n p^{2-2\delta}.\label{lambdad}\end{align} Then the theorem follows from Theorem \ref{Thm-approx} (more precisely the results with $\delta\geq 0$, i.e., result \eqref{S1}), \eqref{lambdad} and Theorem 2 of \cite{yu2015useful}. We now show \eqref{lambdad}.
Consider the QR decomposition of $\mf A(t)$ such that $\mf A(t)=\mf Q(t)\mf R(t)$ where $\mf Q(t)^\top\mf Q(t)=\mf I_{d(t)}$ and $\mf I_{d(t)}$ is a $d(t)\times d(t)$ identity matrix. Here $\mf Q(t)$ is a $p\times d(t)$ matrix and $\mf R(t)$ is a $d(t)\times d(t)$ matrix.
Then \eqref{DefLambda}  can be written as
\begin{align}\label{Lambda1}
\mf \Lambda_1(t)=\mf Q(t) \tilde{\mf \Lambda}_1(t)\mf Q^\top(t),
\end{align}
where the $d(t)\times d(t)$ matrix \begin{align}
\tilde{ \mf  \Lambda}_1(t)=\mf R(t)\Big[\sum_{k=1}^{k_0}(\mf \Sigma_z(t,k)\mf A^\top(t)+\mf \Sigma_{ze}(t,k))(\mf A(t)\mf \Sigma^\top_z(t,k)+\mf \Sigma^\top_{ze}(t,k))\Big]\mf R^\top(t).
\end{align}


Since if $\mf v$ is an eigenvector of $\tilde{\mf \Lambda}_1(t)$ then $\mf Q\mf v$ is an eigenvector of $\mf \Lambda_1(t)$ with the same eigenvalue, we shall see that 
$\lambda_{\min}(\tilde {\mf \Lambda}_1(t))=\lambda_{d}({\mf \Lambda}_1(t))$.
 By (A2') we have
\begin{align}\label{R(t)}\inf_{t\in \mathcal T_{\eta_n}}\|\mf R(t)\|_m=\inf_{t\in \mathcal T_{\eta_n}}\|\mf Q^\top (t)\mf A(t)\|_m\gtrapprox \eta^{1/2}_n p^{\frac{1-\delta}{2}}
\end{align} 
where we have used the fact that $\|\mf A\mf B\|_m\geq \|\mf A\|_m\|\mf B\|_m$. The proof of this fact can be found in proof of Lemma  1 \cite{lam2011estimation}.
 Notice that via Weyl's inequality, and positive-definiteness of summands of $\tilde{ \mf \Lambda}_1(t)$, and the definition of $\|\cdot\|_m$
\begin{align}\label{tildeLambda1}
	&\lambda_{\min}(\tilde{ \mf \Lambda}_1(t))\notag\\ &\geq \inf_t\lambda_{\min}\Big(\mf R(t)\Big[(\mf \Sigma_z(t,k)\mf A^\top(t)+\mf \Sigma_{ze}(t,k))(\mf A(t)\mf \Sigma^\top_z(t,k)+\mf \Sigma^\top_{ze}(t,k))\Big]\mf R^\top(t)\Big)\notag\\&=\inf_t \Big\|\mf R(t)(\mf \Sigma_z(t,k)\mf A^\top(t)+\mf \Sigma_{ze}(t,k))\Big\|^2_m
	\notag\\&\geq \inf_t \|\mf R(t)\|^2_m\|\mf \Sigma_z(t,k)\mf A^\top(t)+\mf \Sigma_{ze}(t,k))\|^2_m\notag
\end{align}
On the other hand, via Weyl inequality and the fact that $\|\mf \Sigma_z(t,k)) \mf A^\top(t)\|_m=\sigma_d(\mf \Sigma_z(t,k) \mf A^\top(t))$, $\|\mf \Sigma_z(t,k)) \|_m=\sigma_d(\mf \Sigma_z(t,k)) $, and $\|\mf A(t))\|_m=\sigma_d(\mf A(t))$, we have
\begin{align}
\sup_t|\sigma_d(\mf \Sigma_z(t,k)\mf A^\top(t)+\mf \Sigma_{ze}(t,k)))-\sigma_d(\mf \Sigma_z(t,k)\mf A^\top(t))|\leq \sup_t\|\mf \Sigma_{ze}(t,k))\|_2,\notag\\
\sigma_d(\mf \Sigma_z(t,k)) \|\mf A(t)\|_m\leq \sigma_d(\mf \Sigma_z(t,k)\mf A^\top(t))\leq \sigma_d(\mf A(t))\|\mf \Sigma_z(t,k)\|_2
\end{align}
Combining conditions (S2), (S4) and (A2') we have that
\begin{align}
	\sigma_d(\mf \Sigma_z(t,k)\mf A^\top(t)+\mf \Sigma_{ze}(t,k)))\gtrapprox \eta_n^{1/2}p^{\frac{1-\delta}{2}}
\end{align}
Together with \eqref{R(t)} and \eqref{tildeLambda1} we have
\begin{align}\label{lambdaminD}\inf_{t\in \mathcal T_{\eta_n}}\lambda_{\min}(\tilde{\mf \Lambda}_1(t))\gtrapprox \eta_n p^{2-2\delta}.\end{align} This shows \eqref{lambdad} 
and the assertion (i) of the Theorem follows. 
\ \\

Due to the page limit, we move the proof of assertion (ii) to the supplemental material. \hfil $\Box$

\medskip



\def\theequation{\Alph{section}.\arabic{equation}}

\def\thetheorem{\Alph{section}.\arabic{theorem}}
\def\thelemma{\Alph{section}.\arabic{lemma}}
\def\thecorollary{\Alph{section}.\arabic{corollary}}
\def\theremark{\Alph{section}.\arabic{remark}}
\def\theproposition{\Alph{section}.\arabic{proposition}}
\def\thesection{\Alph{section}}

\setcounter{equation}{0}

\begin{center}{ Supplemental Material for ``Adaptive Estimation for  Locally Stationary Factor Models And A Test for Static Factor Loadings"}\\\ \\\ \\
	Weichi Wu and Zhou Zhou
\end{center}
\begin{abstract}
	Section \ref{Modelassumptions} provides an example of a high dimensional time series which satisfy the conditions of this paper. Section \ref{Sec::convergence} includes the theoretical results of eigenanalysis. Section \ref{Sec::Est} contains the proofs Theorem  \ref{Space-Distance} (ii) and of auxiliary lemmas for Theorem \ref{Thm-approx}, \ref{Space-Distance} and results in Section \ref{Sec::convergence}. 
	Section \ref{Sec3proof} includes the proof  of  Theorem \ref{Jan23-Thm4} and Theorem \ref{Boots-thm5} for testing the static factor loading, as well as auxiliary results. Finally, Section \ref{Proof-Power} proves 
	Theorem \ref{Power} for power analysis.
\end{abstract}
\setcounterpageref{section}{0}
\renewcommand{\theequation}{\Alph{section}.\arabic{equation}}

Let $\pp_j=\E(\cdot|\FF_j)-\E(\cdot|\FF_{j-1})$ be the projection operator. In the proof, we consider (S4) depending on $\delta$ defined in Section \ref{appendix} in the main article. In the proof, we focus on the general case allowing non-zero $\delta$. For this purpose define $\theta(n,p)=p^{\delta}\nu_n$ under (S1) and $p^{\delta}\nu_n+p^{\delta-1}$ under (S1'), 
and $\theta_0(n,p)=
p^\delta/\sqrt n $ under (S1) and $p^\delta/\sqrt n+p^{\delta-1} $ under (S1'). Observe that when $\delta=0$ these quantities reduce to their counterparts in the main article. To save notation, we omit the subscript $p$ of $\mathbf I_p$ for the $p\times p$ dimensional diagonal matrix, when the dimension $p$ is clear in the context.

\section{Preliminary: locally stationary multivariate time series}\label{Modelassumptions}
We discuss a prominent example for $\mf e_{i,n}$  as follows. 
\begin{example}[High dimensional moving average processes]\label{HDMA}
	Let $\FF_{i}=(\bs \epsilon_{-\infty},...\bs \epsilon_i)$ where $\bs \epsilon_i=(\epsilon_{i,1},...,\epsilon_{i,p'})^\top$ for some $p'>0$ ($p'$ can possibly diverge as $p$), and $(\epsilon_{i,s})_{i\in \mathbb Z, 1\leq s\leq p'}$ are $i.i.d.$ random variables with finite $\max(l,4)_{th}$ moment.
	Consider for $t\in [0,1]$,
	\begin{align}\label{eqA5}
		\mf H(t,\FF_i)=\sum_{j=0}^{\infty} \mf M_{j}(t)\bs \epsilon_{i-j}=(H_1(t,\FF_i),...,H_p(t,\FF_i))^\top
	\end{align}
	where $\mf M_j(t)$, $j\in \mathbb Z$, are smoothly varying $p\times p'$ matrices, and for  $1\leq v\leq p$,
	\begin{align}
		\label{A2new}	H_v(t,\FF_i)=\sum_{j=0}^\infty \sum_{s=1}^{p'} m_{j,v,s}(t)\epsilon_{i-j,s}=\sum_{j=-\infty}^{i}\sum_{s=1}^{p'} m_{i-j,v,s}(t) \epsilon_{j,s}
	\end{align}
	where $\mf m_{j,v}(t)=(m_{j,v,1}(t),....,m_{j,v,p'}(t))$ is the $v_{th}$ row of the matrix $\mf M_j(t)$. Observe that $(\sum_{s=1}^{p'} m_{i-j,v,s}(t) \epsilon_{j,s})_{j\leq i}$ are mean $0$ random variables and are independent of each other. Therefore by Burkholder inequality (see equation (15) in \cite{wu2005nonlinear}, we have for some large constant $M$,
	\begin{align}\label{eq3.9}
		\|H_v(t,\FF_i)\|_{\mathcal L^4}\leq M \sum_{j=-\infty }^i\sum_{s=1}^{p'}\|m_{i-j,v,s}(t)\epsilon_{j,s}\|_{\mathcal L^4}^2
	\end{align}
	As a consequence, (M2) will be satisfied if for $1\leq v\leq p$, 
	$\sup_{t\in [0,1]}\sum_{j=-\infty }^i\sum_{s=1}^{p'}|m_{i-j,v,s}(t)|^2\leq M$ for some constant $M$. For (M1), by definition, for $1\leq v\leq p$, 
	\begin{align}
		\delta^e_{l,v}(k)=\sup_{t\in[0,1]}\|\sum_{s=1}^{p'} m_{k,v,s}(t)(\epsilon_{0,s}-\epsilon_{0,s}')\|_{\mathcal L^l}=O(\sup_{t\in [0,1]}(\sum_{s=1}^{p'} m^2_{k,v,s}(t))^{1/2}).
	\end{align}
	Therefore, (M1) will hold if
	$\sup_{t\in [0,1]}(\sum_{s=1}^{p'} m^2_{k,v,s}(t))^{1/2}$ is $O((k\log k)^{-2})$ for $1\leq v\leq p$. Via using \eqref{A2new} and similar argument yielding \eqref{eq3.9}, \eqref{8-11-11} will be full-filled if 
	for $1\leq v\leq p$, 
	$$\sum_{j=-\infty }^i\sum_{s=1}^{p'}\sup_{t\in [0,1]}|\frac{\partial}{\partial t}m_{i-j,v,s}(t)|^2\leq M$$ for some constant $M$.
\end{example}

We now verify (M6). Rewrite \eqref{eqA5} as $\mf H(t,\FF_u)=\sum_{j=-\infty}^u \mf M_{u-j}(t)\bs \epsilon_{j}$, and hence 
$$\frac{\partial}{\partial t}\mf H(t,\FF_v)=\sum_{j=-\infty}^v \mf M_{v-j}'(t)\bs \epsilon_{j}.$$
where $\mf M_{v-j}'(t)=(\frac{\partial}{\partial t}m_{v-j,s_1,s_2}(t))_{1\leq s_1\leq p,1\leq s_2\leq p'}$ is a $p\times p'$ matrix. Then 
\begin{align}
	\|\E(\mf H(t,\FF_u)(\mf H'(s,\FF_v))^\top )\|_2=\|\sum_{j=-\infty}^{u\wedge v}\mf M_{u-j}(t)(\mf M'_{v-j}(s))^\top\|_2Var(\epsilon_{1,1}).
\end{align}
Then a sufficient condition for (M6) to hold is that $\sup_t\|\mf M_u(t)\|_2=O(u^{-2}\log^{-2} u)$ and $\sup_t\|\mf M'_u(t)\|_2=O(u^{-2}\log^{-2} u)$.

We now verify (M8).
We say $X~subG(\sigma^2)$ if for any $s\in\mathbb R$, $\E \exp(sX)\leq \exp(\frac{\sigma^2s^2}{2})$ where $\sigma^2$ is the variance proxy of $X$.

\begin{proposition}\label{propexample}
	Consider example \ref{HDMA}. If  $\sup_t\|\mf M_j(t)\|_2=O(\Delta(j))$ with $\sum_{j\in \mathbb Z, j\geq 0}j\Delta_q(j)<\infty$, 
	and $\epsilon_{ij}$, $1\leq i\leq p, 1\leq j\leq p'$ are $i.i.d.$ $subG(\sigma^2)$, then  (M8) will hold.
\end{proposition}
{\it Proof.} 
It is easy to verify that (a) if $X_i\sim subG(\sigma_i^2)$ and $X_i's$ are independent of each other, then  $\sum_i X_i\sim subG(\sum_i\sigma_i^2)$. (b)  $cX_i\sim subG(c^2\sigma^2_i)$. We now show (c)  if $X\sim subG(\sigma^2)$, then $\|X\|_{\mathcal L^q}\leq C \sqrt q \sigma$ for some uniform constant $C$. To see (c) using $(x/q)^q\leq e^x$ for $q\geq 1$ and $x\geq 0$, we have for $s\neq 0$
\begin{align} 
	\E(|X|^q)\leq \frac{q^q}{|s|^q}\E e^{|s||X|}\leq  \frac{q^q}{|s|^q}(\E \exp(sX)+\E \exp(-sX))\leq \frac{2q^q}{|s|^q}\exp(\frac{\sigma^2s^2}{2}).
\end{align}
Hence for $q\geq 1$ and $s\neq 0$,
\begin{align}\label{A10}
	\|X\|_{\mathcal L^q}\leq \frac{2^{1/q}q}{|s|}\exp(\frac{\sigma^2s^2}{2q})
\end{align} 
Take $s=\sqrt q\sigma^{-1}$ we have prove (c).

On the other hand, notice that $\mf e_j=\mf H(j/n,\FF_j)$ where $\mf H(t,\FF_j)$ is defined in \eqref{eqA5}. Recall that $\bs \epsilon'_0$ is an $i.i.d.$ copy of $\bs \epsilon_0$.  Then 
\begin{align}
	\|\mf c^\top(	\mf e_i-\mf e_i^*)\|_{\mathcal L^q}=\|\mf c^\top\mf M_i(i/n)(\bs \epsilon_0-\bs \epsilon'_0)\|_{\mathcal L^q}\leq
	\|\mf c^\top\mf M_i(i/n)\bs \epsilon_0\|_{\mathcal L^q}+\|\mf c^\top\mf M_i(i/n)\bs \epsilon'_0\|_{\mathcal L^q}.
\end{align}
By (a) and (b), since the component of $\bs \epsilon_0$ are $i.i.d.$ $subG(\sigma^2)$, which shows that $\mf c^\top\mf M_i(i/n)\bs \epsilon_0\sim subG(\|\mf c^\top \mf M_i(i/n)\|_2^2)$. By (c) and the fact that $|\mathbf c|=1$,  it follows that 
\begin{align}
	\|\mf c^\top\mf M_i(i/n)\bs \epsilon_0\|_{\mathcal L^q}\leq C\sqrt q \|\mf c^\top\mf M_i(i/n)\|_2=O(\Delta_q(i)).
\end{align}
By a similar argument applied to $\mf c^\top\mf M_i(i/n)\bs \epsilon'_0$ the proposition follows. \hfill $\Box$

\section{Results for eigenvalues and proof of  Proposition \ref{hatdrate}}\label{Sec::convergence}

\setcounter{equation}{0}

\subsection{Theorem \ref{eigentheorem} and its Proof and discussion}
In this section we allow the dimension of factor loading matrix and the number of factors of model \eqref{eq:fm} to vary with time. Thus we assume (A2') instead of (A2).
To show proposition \ref{tildedrate}, we first prove the following theorem, which investigates the eigenvalue of $\hat {\mf \Lambda}(t)$ and is of separate interest.
\begin{theorem}\label{eigentheorem}
	Assume that (A1), (A2'), (M1), (M2), (M3) and (S0)--(S4) (either (S1) or (S1') holds),  
	and that under (S1) $\eta_n(p^\delta\nu_n)^{-1}\rightarrow \infty$, or under (S1')	$\frac{\eta_np^{2-2\delta}}{(p^{2-\delta}\nu_n+p^{1-\delta})}\rightarrow \infty$,
	then  
	we have that
	\begin{description}
		\item (i) $\|\sup_{t\in (0,1)}\max_{1\leq j\leq p}| \lambda_j(\hat{\mf \Lambda}(t))-\lambda_j(\mf \Lambda_1 (t))|\|_{\mathcal L^1}=O(p^{2-\delta}\nu_n+p^{1-\delta})$ under (S1'), and the term $p^{1-\delta}$ vanishes under (S1).	
		\item (ii)  
		There exist constants $m<M$ such that 
		$$\p(m\eta_np^{2-2\delta}\leq \lambda_j(\hat {\mf \Lambda}(t))\leq Mp^{2-2\delta}, 1\leq j\leq d(t), \forall t\in \mathcal T_{\eta_n} )=1-O(\frac{(p^{2-\delta}\nu_n+p^{1-\delta})}{\eta_np^{2-2\delta}})=1-o(1)$$ under (S1'), and the term $p^{1-\delta}$ varnishes under (S1).
		\item (iii) 
		There exists a constant  $C$ such that 
		$$\p(0\leq \lambda_j(\hat {\mf \Lambda}(t))\leq (p^{2-\delta}\nu_n+p^{1-\delta})\log^{1/2} n, d(t)+1\leq j\leq p, \forall t\in \mathcal T_{\eta_n} )=1-O(\log^{-1/2} n)$$ under (S1'), and the term $p^{1-\delta}$ varnishes under (S1). 
	\end{description}
\end{theorem}

{\it Proofs.} (i) follows immediately from Lemma \ref{Bhatia} and Theorem \ref{Thm-approx} (more precisely the results with $\delta$ in Section \ref{appendix} of the main article).

		We now prove (ii).
		From \eqref{11-29-eq7} and \eqref{Final1} in the proof of Theorem \ref{Thm-approx}, we have
		\begin{align}\label{new.1}
			\|\|\lambda_1(\mf \Lambda(t))\|_F\|_{\mathcal L_1}\leq \|\|\hat{\mf \Lambda}(t)\|_F\|_{\mathcal L_1}=O(p^{2-2\delta})
		\end{align}
		By definition of $\mf \Lambda_1(t)$ in \eqref{DefLambda} of the main article we shall have
		\begin{align}\label{new.2}
			\lambda_1(\mf \Lambda_1(t))\leq \|\mf \Lambda_1(t)\|_2^2=O(p^{2-2\delta})
		\end{align}
		Observe that by \eqref{new.1} and \eqref{new.2}, the event $\{m\eta_np^{2-2\delta}\leq \lambda_j(\hat {\mf \Lambda}(t))\leq Mp^{2-2\delta}, 1\leq j\leq d(t), \forall t\in \mathcal T_{\eta_n}\}$ for some $m$ and $M$ will hold if $$\sup_{t\in (0,1)}\max_{1\leq j\leq d(t)}| \lambda_j(\hat{\mf \Lambda}(t))-\lambda_j(\mf \Lambda_1 (t))|\leq c_0m\eta_np^{2-2\delta}$$ for some small positive $c_0$ such that $c_0\leq \frac{1}{2}\inf_{t\in \mathcal T_{\eta_n}}\lambda_{d(t)}(\mf \Lambda_1(t))$.  Then (ii) follows in view of (i) and Markov inequality. 
		
		Now we show (iii). Without loss of generality we assume under (S1'). Since $\mf A(t)$ is a matrix of $p\times d(t)$, it follows that $\lambda_j(\mf \Lambda_1(t))=0$ for $d(t)+1\leq j\leq p$. Notice that $\hat{\mf \Lambda}(t)$ is a positive semidefinite matrix by construction. Hence the event $\{0\leq \lambda_j(\hat {\mf \Lambda}(t))\leq (p^{2-\delta}\nu_n+p^{1-\delta})\log^{1/2} n, d(t)+1\leq j\leq p, \forall t\in \mathcal T_{\eta_n}\}$ will hold if 
		$$\sup_{t\in (0,1)}\max_{d(t)+1\leq j\leq p}| \lambda_j(\hat{\mf \Lambda}(t))-\lambda_j(\mf \Lambda_1 (t))|\leq (p^{2-\delta}\nu_n+p^{1-\delta})\log^{1/2} n/2 $$ and if $n$ is sufficiently large.
		By assertion (i) and Markov inequality, the result (iii) follows.
			\hfill $\Box$\\

				The next proposition states that with high probability $\hat d_n=d_n$ if $q_n=c(p^{1-\delta}+\nu_np^{2-\delta})\log p$ for some constant $c>0$. Together with  Theorem \ref{Space-Distance},  it follows that the estimator \eqref{span-estimate} is consistent if 	$\frac{\eta^2_np^{2-2\delta}}{(p^{2-\delta}\nu_n+p^{1-\delta})\log n}\rightarrow \infty$.

				\subsection{Proof of Proposition \ref{hatdrate}}\label{proofhatdrate}
				In the remaining of proof, we consider the following version of (A2) which accommodates $\delta$.
				\begin{description}
					\item (A2) $\mf A(t)$ is full rank. Write $\mf A(t)=(\mf a_1(t),....,\mf a_{d}(t))$ where $\mf a_s(t), 1\leq s\leq d$ are $p$ dimensional vectors. Then $\sup_{t\in [0,1]}\|\mf a_s(t)\|^2_2\asymp p$
					for $1\leq s\leq d$.
					Besides, 
					the matrix norm of $\mf A(t)$ satisfies
					\begin{align}
						\inf_{t\in [0,1]}\|\mf A(t)\|_F\asymp p^{\frac{1-\delta}{2}},\sup_{t\in [0,1]}\|\mf A(t)\|_F\asymp p^{\frac{1-\delta}{2}},
						\inf_{t\in [0,1]}\|\mf A(t)\|_m\geq \eta^{1/2}_n p^{\frac{1-\delta}{2}}
					\end{align}
					for a positive sequence $\eta_n=O(1)$.
				\end{description}
				We now state the complete version of Proposition \ref{hatdrate} with possibly positive $\delta$ as follows. 
				
				Assume conditions (A1), (A2), (M1)-(M3), (S0)-(S4) (either (S1) or (S1') holds) hold, and that $\eta_n\gtrapprox 1$. Furthermore, Under (S1), suppose that $c$ is a sufficiently small and positive constant,  $\frac{1}{p^\delta\nu_n\log n}\rightarrow \infty$.  Under (S1'),  assume that
				$\frac{p^{2-2\delta}}{(p^{2-\delta}\nu_n+p^{1-\delta})\log n}\rightarrow \infty$, and that $n^a\lessapprox  p\lessapprox n^b$ for some $a<b$. 
				\begin{align}
					\p(\hat d_n\neq d)=
					O\Big(
					\eta_n^{-1}\theta(n,p)
					\Big)+O(\log^{-1/2} n)=o(1).
				\end{align}
				
				We shall prove a more general version of Proposition \ref{hatdrate}, which is Proposition 
				\ref{tildedrate} in the next subsection that allows time-varying $d(t)$ and allows $\eta_n=o(1)$. Then Proposition \ref{hatdrate} will follow from the same argument as in that of proof of  Proposition \ref{tildedrate}. \hfill $\Box$

				
				\subsection{ Proposition \ref{tildedrate}}

				In the proposition, we consider the following estimator $\hat d_n(t)$ for $d(t)$ when dimension $d(t)$ is allowed to vary with time and the factor strength is $\delta$. Setting $\delta=0$ will yield the estimator \eqref{tilded} in the main article.
				\begin{align}
					\hat d_n(t)=\argmin_{1\leq i\leq p}(\lambda_{i+1}(\hat {\mf \Lambda}(t))+q_n )/(\lambda_{i}(\hat {\mf \Lambda}(t))+q_n).
				\end{align}
				where $q_n=c_n(p^{1-\delta}+\nu_np^{2-\delta})\log p$. 
				\begin{proposition}\label{tildedrate}
					Assume conditions (A1), (A2'), (M1)-(M3), (S0)-(S4) (either (S1) or (S1') holds) hold. Furthermore, under (S1), we assume that $c_n\leq c\eta^2_n$ for some sufficiently small but positive constant $c$, and that  
					$c_n^{-1}\nu_n\log^{1/2}n=o\Big(\frac{c_n^{\frac{1}{1-\delta}}\log p}{(\eta_n\log^{1/2}n)^{\frac{1}{1-\delta}}}\Big)$, $\frac{\eta^2_n}{p^\delta\nu_n\log n}\rightarrow \infty$ and $\frac{\eta_nc_n\log p}{p^\delta \nu_n \log n}\rightarrow \infty $.  Under (S1'),  assume that
					$\frac{\eta_n\min(\eta_n,\eta_n')p^{2-2\delta}}{(p^{2-\delta}\nu_n+p^{1-\delta})\log n}\rightarrow \infty$ where $\eta_n'=\frac{ac_n\log^{1/2}n}{ac_n\log^{1/2}n+1}$, and that $n^a\lessapprox  p\lessapprox n^b$ for some $a<b$.  Then we have 
					\begin{align}
						\p(\exists t\in \mathcal{T}_{\eta_n}, \hat d_n(t)\neq d(t))=O\Big(\frac{(p^{2-\delta}\nu_n+p^{1-\delta})}{\eta_np^{2-2\delta}}\Big)+O(\log^{-1/2} n)=o(1)
					\end{align}
					where the term $p^{1-\delta}$ varnishes under (S1).
				\end{proposition}
				Proposition \ref{tildedrate} demonstrates that  $\hat d_n(t)$ is uniformly consistent on ${\cal T}_{\eta_n}$, and the results in Theorems  \ref{hatdrate}  and \ref{Space-Distance} are still valid if $d$ therein is replaced by $d(t)$ and $[0,1]$ is replaced by $\mathcal T_{\eta_n}$. In particular, multiple eigenvalues in ${\mf \Lambda}(t)$ are allowed.
				
				{\it Proof.}	
				On the event $A_1:=\{m\eta_np^{2-2\delta}\leq \lambda_j(\hat {\mf \Lambda}(t))\leq Mp^{2-2\delta}, 1\leq j\leq d(t), \forall t\in \mathcal T_{\eta_n}\}$, it is easy to verify that under either (S1) or (S1')  (noticing that under (S1) we consider $c$ is small such that if $c_n\leq c\eta^2_n$ s.t. $c_n(p^{2-\delta}\nu_n+p^{1-\delta})\log p< \inf_{t\in \mathcal T_n}\lambda_{d(t)}(\mf \Lambda_1(t))$),
				\begin{align}\label{piece1}
					\inf_{1\leq j\leq d(t)-1}\frac{\lambda_{j+1}(\hat{\mf \Lambda}(t))+q_n}{\lambda_{j}(\hat{\mf \Lambda}(t))+q_n}\gtrapprox \eta_n, \quad \forall t\in \mathcal T_{\eta_n}.
				\end{align}
				Define the events  $A_2(S1):=\{0\leq \lambda_j(\hat {\mf \Lambda}(t))\leq p^{2-\delta}\nu_n\log^{1/2} n, d(t)+1\leq j\leq p, \forall t\in \mathcal T_{\eta_n}\}$ and $A_2(S1'):=\{0\leq \lambda_j(\hat {\mf \Lambda}(t))\leq (p^{2-\delta}\nu_n+p^{1-\delta})\log^{1/2} n, d(t)+1\leq j\leq p, \forall t\in \mathcal T_{\eta_n}\}$. Then on $A_2(S_1')$ and under condition (S1'), we  have since $n^a\lessapprox p\lessapprox n^b$,
				\begin{align}\label{piece2}
					\inf_{d(t)+1\leq j\leq p}\frac{\lambda_{j+1}(\hat{\mf \Lambda}(t))+q_n}{\lambda_{j}(\hat{\mf \Lambda}(t))+q_n}\gtrapprox \frac{q_n}{\lambda_{j}(\hat{\mf \Lambda}(t))+q_n}\gtrapprox \frac{c_n\log p}{c_n\log p+\log^{1/2}n}\gtrapprox \frac{ac_n\log^{1/2}n}{ac_n\log^{1/2}n+1}, \quad \forall t\in \mathcal T_{\eta_n}.
				\end{align}
				If  on $A_2(S1)$ under condition (S1), 
				\begin{align}\label{piece2S1}
					\inf_{d(t)+1\leq j\leq p}\frac{\lambda_{j+1}(\hat{\mf \Lambda}(t))+q_n}{\lambda_{j}(\hat{\mf \Lambda}(t))+q_n}\gtrapprox \frac{c_n(p^{2-\delta}\nu_n+p^{1-\delta})\log p}{c_n(p^{2-\delta}\nu_n+p^{1-\delta})\log p+p^{2-\delta  }\nu_n\log^{1/2} n} 
					\quad \forall t\in \mathcal T_{\eta_n},
				\end{align}

				We first prove under (S1'). Notice that under (S1') and on the interception event $A_1\cap A_2(S1')$,  for $t\in \mathcal T_{\eta_n}$ we have $\lambda_{d(t)}(\hat{\mf \Lambda}(t))\geq m\eta_n p^{2-2\delta}$ and    $\lambda_{d(t)+1}(\hat{\mf \Lambda}(t))\leq (p^{2-\delta}\nu_n+p^{1-\delta})\log^{1/2} n$. Hence for $t\in\mathcal T_{\eta_n}$
				\begin{align}\label{piece3}
					\frac{\lambda_{d(t)+1}(\hat{\mf \Lambda}(t))+q_n}{\lambda_{d(t)}(\hat{\mf \Lambda}(t))+q_n}\lessapprox 
					\frac{(p^{2-\delta}\nu_n+p^{1-\delta})(\log^{1/2} n+c_n\log p)}{\eta_n p^{2-2\delta}}.
				\end{align}
				Recall that $\eta_n'=\frac{ac_n\log^{1/2}n}{ac_n\log^{1/2}n+1}$.
				Notice that if	$\frac{\eta_n\min(\eta_n,\eta_n')p^{2-2\delta}}{(p^{2-\delta}\nu_n+p^{1-\delta})\log n}\rightarrow \infty$, then \eqref{piece1},  \eqref{piece2},  \eqref{piece3} indicates that for sufficiently large $n$, $p$, $d(t)$ will be correctly identified on $\mathcal T_{\eta_n}$. Following the proof of Theorem \ref{eigentheorem} (ii) (iii) via Markov inequality, we shall see that
				\begin{align}\label{finalpieceS1'}
					\p(A_1\cap A_2(S1'))=1-O\Big(\frac{(p^{2-\delta}\nu_n+p^{1-\delta})}{\eta_np^{2-2\delta}}\Big)-O(\log^{-1/2} n).
				\end{align}
				We now prove the proposition under (S1). Note that under (S1), on the interception event $A_1\cap A_2(S1)$,  for $t\in \mathcal T_{\eta_n}$ we have $\lambda_{d(t)}(\hat{\mf \Lambda}(t))\geq m\eta_n p^{2-2\delta}$ and    $\lambda_{d(t)+1}(\hat{\mf \Lambda}(t))\leq p^{2-\delta}\nu_n\log^{1/2} n$. Hence for $t\in\mathcal T_{\eta_n}$
				\begin{align}\label{piece3S1}
					\frac{\lambda_{d(t)+1}(\hat{\mf \Lambda}(t))+q_n}{\lambda_{d(t)}(\hat{\mf \Lambda}(t))+q_n}\lessapprox 
					\frac{p^{2-\delta}\nu_n\log^{1/2} n+c_n(p^{2-\delta}\nu_n+p^{1-\delta})\log p}{m\eta_n p^{2-2\delta}+c_n(p^{2-\delta}\nu_n+p^{1-\delta})\log p}\notag\\\lessapprox\frac{p^\delta \nu_n (c_n\log p+\log^{1/2}n)}
					{\eta_n}+\frac{c_n\log p}{\eta_n p^{1-\delta}}.
				\end{align}
				Observing the right hand side of \eqref{piece2S1}, if (a) $c_n(p^{2-\delta}\nu_n+p^{1-\delta})\log p\gtrapprox p^{2-\delta}\nu_n\log^{1/2}n$, then 
				\begin{align}\label{piece3S2}
					\frac{c_n(p^{2-\delta}\nu_n+p^{1-\delta})\log p}{c_n(p^{2-\delta}\nu_n+p^{1-\delta})\log p+p^{2-\delta  }\nu_n\log^{1/2} n} 
					\gtrapprox 1
				\end{align} 
				and if (b)
				$c_n(p^{2-\delta}\nu_n+p^{1-\delta})\log p\lessapprox p^{2-\delta}\nu_n\log^{1/2}n$, then 
				\begin{align}\label{piece3S3}
					\frac{c_n(p^{2-\delta}\nu_n+p^{1-\delta})\log p}{c_n(p^{2-\delta}\nu_n+p^{1-\delta})\log p+p^{2-\delta  }\nu_n\log^{1/2} n} 
					\gtrapprox  \frac{c_n(p^{2-\delta}\nu_n+p^{1-\delta})\log p}{p^{2-\delta}\nu_n\log^{1/2}n}\gtrapprox  \frac{c_n\log p}{\log^{1/2} n}.
				\end{align}
				Recall the assumption that  $c_n^{-1}\nu_n\log^{1/2}n=o(\frac{c_n^{\frac{1}{1-\delta}}\log p}{(\eta_n\log^{1/2}n)^{\frac{1}{1-\delta}}})$, $\frac{\eta^2_n}{p^\delta\nu_n\log n}\rightarrow \infty$ and $\frac{\eta_nc_n\log p}{p^\delta \nu_n \log n}\rightarrow \infty $.
				Therefore, if $c_n\log p\gtrapprox p\nu_n \log^{1/2} n$ or $c_n\log p\gtrapprox \log^{1/2} n$ such that (a) holds,  then straightforward calculations show that (in the following using $c_n\leq c\eta_n^2$ for the first inequality and $\frac{\eta^2_n}{p^\delta\nu_n\log n}\rightarrow \infty$ for the second line)
				\begin{align}\label{A11}
					\frac{p^\delta \nu_n (c_n\log p+\log^{1/2}n)}
					{\eta_n}+\frac{c_n\log p}{\eta_n p^{1-\delta}}
					\leq c\eta_np^\delta \nu_n\log p+p^{\delta}\nu_n \log^{1/2}n/\eta_n+c\eta_n \log p/p^{1-\delta}
					\notag\\=o(c\eta^3_n\log p/\log n)+o(\eta_n/\log^{1/2}n)+c\eta_n \log p/p^{1-\delta}.
				\end{align}
				Hence in this case the eigen-ratio in \eqref{piece3S1} can be smaller than $c_1\eta_n$ for any small positive $c_1$ if $c$ is sufficiently small and $n$ is sufficiently large. On the other hand,  if $c_n\log p\lessapprox p\nu_n \log^{1/2} n$ and $c_n\log p\lessapprox \log^{1/2} n$ such that (b) holds, then
				\begin{align}
					\frac{p^\delta \nu_n (c_n\log p+\log^{1/2}n)}
					{\eta_n}+\frac{c_n\log p}{\eta_n p^{1-\delta}}
					\lessapprox \frac{p^\delta \nu_n \log^{1/2}n}
					{\eta_n}+c\eta_n \log p/p^{1-\delta}\notag\\=o(c_n\log p/\log^{1/2}n)+c\eta_n \log p/p^{1-\delta}
					=o(c_n\log p/\log^{1/2}n)+c\eta_n \log^\delta p (\frac{\log p}{p})^{1-\delta}\notag\\\lessapprox o(c_n\log p/\log^{1/2}n)+
					c\eta_n\log^\delta p(c_n^{-1}\nu_n \log^{1/2}n)^{1-\delta}=o(c_n\log p/\log ^{1/2}n)
				\end{align}
				where we have used $\frac{\eta_nc_n\log p}{p^\delta \nu_n \log n}\rightarrow \infty $ for the second line, and  $c_n\log p\lessapprox p\nu_n \log^{1/2} n$ for the third line, and the fact that $c_n^{-1}\nu_n\log^{1/2}n=o(\frac{c_n^{\frac{1}{1-\delta}}\log p}{(\eta_n\log^{1/2}n)^{\frac{1}{1-\delta}}})$ and $c_n\leq c\eta_n^2$ for the final conclusion.
				
				As a result, by 
				\eqref{piece3S1}, \eqref{piece3S2} and \eqref{piece3S3} it follows that for sufficiently large $n$, $d(t)$ will be correctly identified on $\mathcal T_{\eta_n}$. Again by the proof of Theorem \ref{eigentheorem}, 
				\begin{align}\label{finalpieceS1}
					\p(A_1\cap A_2(S1))=1-O\Big(\frac{p^\delta\nu_n}{\eta_n}\Big)-O(\log^{-1/2} n).
				\end{align}
				The proposition follows from \eqref{finalpieceS1} and \eqref{finalpieceS1'}.
				\hfill $\Box$

				\begin{remark} \label{remarketa}
					In practice to apply Theorems \ref{Thm-approx} and \ref{Space-Distance}, if the estimated  number of factors $\tilde d_n(t)$ does not change over time, then we can consider $\hat {\mathcal T}_{\eta_n}=[0,1]$. Otherwise one can consider
					\begin{align}\label{Tetan}
						\hat {\mathcal T}_{\eta_n}=(0,1)\cap_{s=1}^r\left(\hat  t_s-\frac{1}{\log^{2} n}, \hat t_s+\frac{1}{\log^{2} n}\right)
					\end{align}
					where $\hat t_s$, $s=1,...r$ are the time points when $\tilde d_n(t)$ changes. In fact, by condition (A2'),  \eqref{Tetan} corresponds to setting $\eta_n\asymp\frac{1}{\log^4 n}$ when the eigenvalues of $\mf A(t)/p^{(1-\delta)/2}$ are Lipschitz continuous.     	\end{remark}
				
				\section{Proof of Theorem  \ref{Space-Distance} (ii) and auxiliary technical results for theoretical analysis in Section \ref{Thm-approx}  and in Section \ref{Sec::convergence}}\label{Sec::Est}
				\setcounter{equation}{0}
				\textbf{Proof of Theorem  \ref{Space-Distance} (ii): }
				To show the assertion (ii),
				we apply eigen-decomposition to $\tilde  {\mf \Lambda}_1(t)$  to obtain that
				\begin{align}
					\mf \Lambda_1(t)
					=\mf Q(t)\mf U(t)\mf D(t)\mf U^\top(t)\mf Q^\top (t)
				\end{align}
				where $\mf U(t)$ are the orthnormal matrix consists of eigenvectors of 
				$\tilde {\mf \Lambda}_1(t)$ and $\mf D(t)$ is the diagnol matrix of which the diagnol elements are eigenvalue of $\tilde {\mf \Lambda}_1(t)$. By definition, $\mf V(t)=\mf Q(t)\mf U(t)$ and $\mf V(t) \mf V^\top(t)=\mf Q(t)\mf Q^\top(t)$. As a consequence 
				\begin{align*}
					{\mf V}(i/n)	 {\mf V}^\top(i/n)\mf x_{i,n}&=\mf Q(i/n)\mf Q(i/n)^\top(\mf Q(i/n)\mf R(i/n)\mf z_{i,n}+\mf e_{i,n})\notag\\
					&=\mf A(i/n)\mf z_{i,n}+\mf V(i/n)\mf V^\top(i/n) \mf e_{i,n}.
				\end{align*}
				Hence  together with $\mf x_{i,n}=\mf A(i/n)\mf z_{i,n}+\mf e_{i,n}$ it follows that
				\begin{align*}
					&\|	\tilde {\mf V}(i/n)	\tilde {\mf V}^\top(i/n)\mf x_{i,n}- \mf A(i/n)\mf z_{i,n}\|_2\notag\\	&=\notag\|\tilde {\mf V}(i/n)	\tilde {\mf V}^\top(i/n)\mf x_{i,n}
					-	 {\mf V}(i/n) {\mf V}^\top(i/n)\mf x_{i,n}+{\mf V}(i/n) {\mf V}^\top(i/n) \mf e_{i,n}\|_2
					\notag\\&\leq  \|\tilde {\mf V}(i/n)\hat {\mf O}_1(i/n)	\hat {\mf O}^\top_1(i/n)	\tilde {\mf V}^\top(i/n)\mf A(i/n)\mf z_{i,n}
					-	 {\mf V}(i/n) {\mf V}^\top(i/n)\mf A(i/n)\mf z_{i,n}\|_2\notag\\&+\|\tilde {\mf V}(i/n)\tilde {\mf V}^\top(i/n) \mf e_{i,n}\|_2:=\|I(i/n)\|_2+\|\tilde {\mf V}(i/n)\tilde {\mf V}^\top(i/n) \mf e_{i,n}\|_2.
				\end{align*}
				where $I(i/n)$  is defined in an obvious way. Furthermore, $\|I(i/n)\|_2$ is bounded by $\|I_1(i/n)\|_2+\|I_2(i/n)\|_2$ where 
				\begin{align*}
					I_1(i/n)=\tilde{\mf V}(i/n)\hat {\mf O}_1(i/n)[\hat {\mf O}_1^\top(i/n)\tilde {\mf V}^\top(i/n)-\mf V^\top(i/n)]\mf A(i/n)\mf z_{i,n},
					\\I_2(i/n)=[\tilde {\mf V}(i/n)\hat {\mf O}_1(i/n)-\mf V(i/n)]\mf V^\top(i/n)\mf A(i/n)\mf z_{i,n}.
				\end{align*}
				It is easy to verify that for $\tilde{\mf V}(i/n)\hat {\mf O}_1(i/n)$ and  $\mf V(i/n)$ their operator  norms are $1$. By condition (M2), $\|\mf z_{i,n}\|_2=O_p(1)$.   By assertion (i) we have $\|I_1(i/n)\|_2$ and $\|I_2(i/n)\|_2$ is  $O_p(\eta_n^{-1}p^{\frac{1-\delta}{2}}(p^{\delta}\nu_n+p^{\delta-1}))$ so
				\begin{align}
					\|I(i/n)\|_2=O_p(\eta_n^{-1}p^{\frac{1-\delta}{2}}(p^{\delta}\nu_n+p^{\delta-1})).
				\end{align}
				On the other hand, notice that $\|\tilde{\mf V}(t)\mf O(t)\|_2=1$ so that 
				$\|\tilde {\mf V}(i/n)\tilde {\mf V}^\top(i/n) \mf e_{i,n}\|_2\leq \|\hat {\mf O}^\top(i/n)\tilde {\mf V}^\top(i/n) \mf e_{i,n}\|_2$. Via claim (i) and the fact 
				\begin{align}\label{New.S16}
					\|\tilde {\mf V}(i/n)\tilde {\mf V}^\top(i/n) \mf e_{i,n}\|_2\leq \|(\hat {\mf O}^\top(i/n)\tilde {\mf V}^\top(i/n)-\mf V^\top(i/n)) \mf e_{i,n}\|_2+\| {\mf V}^\top(i/n) \mf e_{i,n}\|_2\notag\\=\|(\hat {\mf O}^\top(i/n)\tilde {\mf V}^\top(i/n)-\mf V^\top(i/n)) \mf e_{i,n}\|_2+(\sum_{s=1}^{d(i/n)} ({\mf v}_s^\top(i/n)\mf e_{i,n})^2)^{1/2}\notag\\
					=O_p((\eta_n^{-1} p^{\delta}\nu_n+\eta_n^{-1}p^{\delta-1})p^{1/2})+(\sum_{s=1}^{d(i/n)} ({\mf v}_s^\top(i/n)\mf e_{i,n})^2)^{1/2}
				\end{align}
				where we have used the fact that $\|\mf e_{i,n}\|_2=O_p(p^{1/2})$, ${\mf v}_s(t)$, $1\leq s\leq d(t)$ is the $s_{th}$ column of ${\mf V}(t)$, i.e., ${\mf V}(t)=({\mf v}_1(t),...,{\mf v}_d(i/n))$.
				By definition, for $1\leq s\leq p$ \begin{align}
					Var( {\mf v}_s^\top(i/n)\mf e_{i,n})\leq \limsup_p\sup_{t\in[0,1]}\lambda_{max}(\E(\mf H(t,\FF_{i})\mf H^\top(t,\FF_i)))\end{align}
				hence $	\|{\mf V}^\top(i/n) \mf e_{i,n}\|_2=O_p(1)$ via condition (M4). Therefore we show (ii).
				
				\hfill $\Box$
				
				\noindent The following lemma from \cite{bhatia1982analysis} is useful for proving Theorem \ref{eigentheorem}.%
				
				\begin{lemma}\label{Bhatia}
					Let $M(n)$ be the space of all $n \times n$ (complex) matrices.  A norm $\|\cdot\|$ on $M(n)$ is said to be unitary-invariant if $\|A\|=\|U A V\|$ for any two unitary matrices $U$ and $V .$  We denote by Eig $A$ the unordered $n$-tuple consisting of the eigenvalues of $A,$ each counted as many times as its multiplicity. Let $D(A)$ be a diagonal matrix whose diagonal entries are the elements of Eig $A .$ For any norm on $M(n)$ define
					$$
					\|(\operatorname{Eig} A, \operatorname{Eig} B)\|=\min _{W}\left\|D(A)-W D(B) W^{-1}\right\|
					$$
					where the minimum is taken over all permutation matrices $W .$ 
					If $A, B$ are Hermitian matrices, we have for all unitary-invariant norms (including the  Frobenius norm and the operator norm) the inequality
					$$
					\|(\operatorname{Eig} A, \operatorname{Eig} B)\| \leqslant\|A-B\|.
					$$
				\end{lemma}
				Recall $d=\max_t d(t)$ and $\sigma_u(t)$
				defined in  Section \ref{Sec::Pre} of the main article. In the remainder of this section, we consider the equivalent model in Section \ref{Sec::Pre}. With a little abuse of notation, consider and $\mf A(t)=(a_{uv}(t))_{1\leq u\leq p,1\leq v\leq d}$, where $a_{uv}(t)=0$ if $\sigma_v(t)=0$ for  $1\leq v\leq d$, and are the same as the corresponding elements in the varying dimension loading matrix. Then $\mf x_{i,n}=\mf A_{i,n}\mf z_{i,n}+\mf e_{i,n}$.
				\begin{lemma}\label{LS-G}
					Define the dependence measure for $\mf x_{i,n}$ in $\mathcal L^l$ norm as 
					\begin{align}
						\delta^G_{l}(k):=\max_{ 1\leq j\leq p}\delta^G_{l,j}(k):=\max_{ 1\leq j\leq p}\sup_{t\in[0,1],i\in \mathbb Z}\E^{1/l}(|G_j(t,\FF_i)-G_j(t,\FF_i^{(i-k)})|^l).
					\end{align}
				Under conditions (A1), and (M1)--(M3),  there exists a sufficiently large constant $M_0$, such that uniformly for $1\leq u\leq p$ and $t\in [0,1]$,\begin{align}
					\E | G_u(t,\FF_0)|^4\leq M_0, \label{state1}\\
					\delta^G_{l}(k)=O(d\delta^z_{l}(k)+\delta^e_{l}(k)).\label{state2}
				\end{align}
				
			\end{lemma}
			{\it Proof.}  By definition we have that for $1\leq u\leq p$,
			\begin{align*}
				G_u(t,\FF_i)=\sum_{v=1}^da_{uv}(t)Q_v(t,\FF_i)+H_u(t,\FF_i).
			\end{align*}
			Notice that here $d=\sup_{0\leq t\leq 1}d(t)$ is fixed.
			Therefore, assumptions (A1), (M2) and triangle inequality lead to the first statement of boundedness of fourth moment of \eqref{state1}. Finally (A1) and (M3) lead to the  assertion \eqref{state2}. \hfill $\Box$ 

			\begin{lemma}\label{Lemma-Bound1}
				Consider the process $z_{i,u,n}z_{i+k,v,n}$ for some $k>0$, and $1\leq u,v\leq d$. Then under conditions (A2'), (M1)--(M3) we have:\\ i) $\zeta_{i,u,v} =:\zeta_{u,v}(\frac{i}{n},\FF_i)=z_{i,u,n}z_{i+k,v,n}$ is a  locally stationary process with associated dependence measures \begin{align}
					\delta_{\zeta_{u,v},2}(h)\leq C(\delta^z_{4}(h)+\delta^z_{4}(h+k))
				\end{align}
				for some universal constant $C>0$ independent of $u,v$ and any integer $h$;\\ (ii) For any series of  numbers $a_i, 1\leq i\leq n$, we have  for some universal large positive constant $M$,
				\begin{align}
					\Big(	\E\Big|\frac{1}{n}\sum_{i=1}^na_i(\zeta_{i,u,v}-\E \zeta_{i,u,v})\Big|^2\Big)^{1/2}\leq \frac{MC}{n}\big(\sum_{i=1}^na_i^2\big)^\frac{1}{2}
				\end{align}
			\end{lemma}
			{\it Proof}. i) is a consequence of the Cauchy-Schwarz inequality, triangle inequality and condition $(M2)$. For ii), notice that 
			\begin{align}\label{Nov-24-1}
				\sum_{i=1}^na_i(\zeta_{i,u,v}-\E \zeta_{i,u,v})=\sum_{i=1}^na_i(\sum_{s=0}^\infty \pp_{i+k-s}\zeta_{i,u,v})=\sum_{s=0}^\infty\sum_{i=1}^na_i\pp_{i+k-s}\zeta_{i,u,v}.
			\end{align}
			By the property of martingale difference and i) of this lemma we have
			\begin{align}\label{Nov-24-2}
				\|\sum_{i=1}^na_i\pp_{i+k-s}\zeta_{i,u,v}\|^2_{\mathcal L^2}=\sum_{i=1}^na_i^2 \|\pp_{i+k-s}\zeta_{i,u,v}\|_{\mathcal L^2}^2\leq C^2\sum_{i=1}^na_i^2(\delta^z_{4}(s)+\delta^z_{4}(s-k))^2.
			\end{align}
			By triangle inequality, inequalities \eqref{Nov-24-1}, \eqref{Nov-24-2} and the fact that $\delta^z_{4}(k)=0$ if $k<0$, and condition (M1) the lemma follows. \hfill $\Box$

			\begin{corol}\label{Corol-Bound1}
				Under conditions (A2'), (M1) and (M2) we have for each fixed $k>0$, $1\leq u,v\leq p$ and $1\leq w\leq d$,  $\psi_i=:\psi(\frac{i}{n},\FF_i)=e_{i+k,u,n}e_{i,v,n}$,  
				$\phi_i=:\phi(\frac{i}{n},\FF_i)=z_{i+k,w,n}e_{i,v,n}$ and
				$\iota_i=:\iota(\frac{i}{n},\FF_i)=e_{i+k,u,n}z_{i,w,n}$  are  locally stationary processes with associated dependence measures 
				\begin{align}
					\max(\delta_{\psi,2}(h),\delta_{\phi,2}(h),\delta_{\iota,2}(h))\leq C(\delta^z_{4}(h)+\delta^e_{4}(h)+\delta^z_{4}(h+k)+\delta^e_{4}(h+k)). 
				\end{align}
				for some universal constant $C>0$ independent of $u$, $v$ and $w$.
			\end{corol}
			{\it Proof.} The corollary follows from the same proof of Lemma \ref{Lemma-Bound1}. \hfill $\Box$
			
			To save notation in the following proofs, for given $J_n,k$ write $\hat {\mf M}(J_n,t,k)$ as $\hat {\mf M}$ if no confusion arises. Recall the definition of $\tilde {\bs \Sigma}_{x,j,k}$ and $\hat {\mf M}(J_n,t,k)$ in \eqref{11-12-10} and \eqref{11-12-11} in the main article.
			Observe the following decompositions
			\begin{align}\label{Nov-23-1}
				\tilde {\bs \Sigma}_{x,j,k}=\mf V_{1,j,k}+\mf V_{2,j,k}+\mf V_{3,j,k}+\mf V_{4,j,k},
				\\\mf V_{1,j,k}=\frac{1}{n}\sum_{i=1}^{n-k}\mf A(\frac{i+k}{n})\mf {z}_{i+k,n}\mf {z}^\top_{i,n}\mf A^\top(\frac{i}{n})B_j(\frac{i}{n}),\\\label{Nov-23-2}
				\mf V_{2,j,k}=\frac{1}{n}\sum_{i=1}^{n-k}\mf A(\frac{i+k}{n})\mf z_{i+k,n}\mf e_{i,n}^\top B_j(\frac{i}{n}), 
				\\\label{Nov-23-3} \mf V_{3,j,k}=\frac{1}{n}\sum_{i=1}^{n-k}\mf e_{i+k}\mf z^\top_{i,n}
				\mf A^\top(\frac{i}{n})B_j(\frac{i}{n}),\mf V_{4,j,k}=\frac{1}{n}\sum_{i=1}^{n-k}\mf e_{i+k,n}\mf e^\top_{i,n}B_j(\frac{i}{n}).
			\end{align}

			\begin{lemma}\label{11-29-lemma3}
				Under conditions (A2'), (M1), (M2) and (M3) we have that $$\|\sup_{t\in [0,1]}\|\hat {\mf M}(J_n,t,k)-\E \hat {\mf M}(J_n,t,k)\|_F\|_{\mathcal L^2}=O(\frac{J_np\sup_{t,1\leq j\leq J_n}|B_j(t)|^2}{\sqrt n}).$$
			\end{lemma}
			{\it Proof.} Using equations \eqref{Nov-23-1}-\eqref{Nov-23-3} we have that
			\begin{align}
				\hat {\mf M}(J_n,t,k)-\E \hat {\mf M}(J_n,t,k)=\sum_{j=1}^{J_n}\sum_{s=1}^4(\mf V_{s,j,k}-\E(\mf V_{s,j,k}))B_j(t):=\sum_{s=1}^4\tilde {\mf V}_{s}(t),
			\end{align}
			where $\tilde{\mf V}_s(t)=\sum_{j=1}^{J_n} (\mf V_{s,j,k}-\E(\mf V_{s,j,k}))B_j(t)$, for $s=1,2,3,4$.
			Consider the $s=1$ case and then
			\begin{align}\notag
				&\tilde {\mf V}_1(t):=\sum_{j=1}^{J_n} (\mf V_{1,j,k}-\E(\mf V_{1,j,k}))B_j(t)\\=&\frac{1}{n}\sum_{j=1}^{J_n}B_j(t)\sum_{i=1}^{n-k}\mf A(\frac{i+k}{n})(\mf z_{i+k,n}\mf z^\top_{i,n}-\E(\mf z_{i+k,n}\mf z^\top_{i,n}))
				\mf A^\top(\frac{i}{n})B_j(\frac{i}{n}).
			\end{align}
			Further define $\tilde { \mf M}_j=\frac{1}{n}\sum_{i=1}^{n-k}\mf A(\frac{i+k}{n})(\mf z_{i+k,n}\mf z^\top_{i,n}-\E(\mf z_{i+k,n}\mf z^\top_{i,n}))
			\mf A^\top(\frac{i}{n})B_j(\frac{i}{n})$. Its $(u,v)_{th}$, $1\leq u\leq p$, $1\leq v\leq p$ element is 
			\begin{align}
				\tilde M_{j,u,v}=\frac{1}{n}\sum_{i=1}^{n-k}\sum_{u'=1}^d\sum_{v'=1}^d a_{uu'}(\frac{i+k}{n})( z_{i+k,u',n} z_{i,v',n}-\E(z_{i+k,u',n}z_{i,v'n}))
				a_{vv'}(\frac{i}{n})B_j(\frac{i}{n}).
			\end{align}
			Therefore it follows from the triangle inequality and Lemma \ref{Lemma-Bound1} that, 
			\begin{align}
				\Big\|\tilde M_{j,u,v}\Big\|_{\mathcal L^2}\leq \frac{C\sup_{j,t}|B_j(t)|}{n}\sum_{v'=1}^d\sum_{u'=1}^d\sqrt {\sum_{i=1}^{n-k}a^2_{uu'}(\frac{i+k}{n}) a^2_{vv'}(\frac{i}{n})}
			\end{align}
			for some sufficiently large constant $C$.
			Consequently by (A2') and Jansen's inequality, we get
			\begin{align}\label{tildeMF}
				\E\left(\|\tilde {\mf M}_j\|^2_F\right) &\leq \frac{C^2\sup_{j,t}|B_j(t)|^2}{n^2}\sum_{u=1}^p\sum_{v=1}^p\left(\sum_{v'=1}^d\sum_{u'=1}^d\sqrt {\sum_{i=1}^{n-k}a^2_{uu'}(\frac{i+k}{n}) a^2_{vv'}(\frac{i}{n})}\right)^2\notag
				\\&\leq\frac{C^2d^2\sup_{j,t}|B_j(t)|^2}{n^2}\sum_{i=1}^{n-k}\sum_{u=1}^p\sum_{v=1}^p\sum_{v'=1}^d\sum_{u'=1}^da^2_{uu'}(\frac{i+k}{n}) a^2_{vv'}(\frac{i}{n})\notag\\&\asymp \frac{d^2\sup_{j,t}|B_j(t)|^2p^{2-2\delta}}{n}
			\end{align}
			for $1\leq j\leq J_n$.
			On the other hand, since $(u,v)_{th}$ element of $ \tilde{\mf V}_1(t)$, which is denoted by $\tilde V_{1,u,v}(t)$, satisfies \begin{align}
				\tilde V_{1,u,v}(t)=\sum_{j=1}^{J_n}B_j(t)\tilde M_{j,u,v}.
			\end{align} 
			Therefore by Jansen's inequality it follows that
			\begin{align}
				\sup_{t\in [0,1]}\|\tilde{\mf V}_1(t)\|_F^2&=\sup_{t}\sum_{u=1}^p\sum_{v=1}^p\left(\sum_{j=1}^{J_n}B_j(t)\tilde M_{j,u,v}\right)^2\notag\\
				&\leq  \sup_{t, 1\leq j\le J_n }|B_j(t)|^2\sum_{u=1}^p\sum_{v=1}^p(\sum_{j=1}^{J_n}|\tilde M_{j,u,v}|)^2\notag\\
				&\leq \sup_{t, 1\leq j\le J_n}|B_j(t)|^2\sum_{u=1}^p\sum_{v=1}^pJ_n\sum_{j=1}^{J_n}|\tilde M_{j,u,v}|^2\notag\\
				&\leq \sup_{t, 1\leq j\le J_n}|B_j(t)|^2J_n\sum_{j=1}^{J_n}\|\tilde {\mf M}_j\|^2_F.
			\end{align}
			Therefore we have
			\begin{align}
				\E(\sup_{t\in [0,1]}\|\tilde{\mf V}_1(t)\|_F^2)\leq \sup_{t, 1\leq j\le J_n}|B_j(t)|^2J_n\sum_{j=1}^{J_n}\E(\|\tilde {\mf M}_j\|^2_F)
			\end{align} Combining \eqref{tildeMF} we have that
			\begin{align}\label{2019-Nov-25-1}
				\E(\sup_{t\in [0,1]}\|\tilde{\mf V}_1(t)\|_F^2)^{1/2}=O\left(\frac{J_n\sup_{t, 1\leq j\le J_n}|B_j(t)|^2p^{1-\delta}}{\sqrt n}\right).
			\end{align}
			Similarly using Corollary \ref{Corol-Bound1} we have that 
			\begin{align}\label{2019-Nov-25-2}
				\E\big(\sup_{t\in [0,1]}\|\tilde {\mf V}_s(t)\|_F^2\big)^{\frac{1}{2}}=O\left( \frac{J_n\sup_{t, 1\leq j\le J_n}|B_j(t)|^2p^{1-\delta/2}}{\sqrt n}\right), s=2,3,
			\end{align}
			and 
			\begin{align}\label{2019-Nov-25-3}
				\E\big(\sup_{t\in [0,1]}\|\tilde {\mf V}_4(t)\|_F^2\big)^{\frac{1}{2}}=O\left(\frac{J_np\sup_{t, 1\leq j\le J_n}|B_j(t)|^2}{\sqrt n}\right).
			\end{align}
			Then the lemma follows from \eqref{2019-Nov-25-1}, \eqref{2019-Nov-25-2}, \eqref{2019-Nov-25-3} and triangle inequality. \hfill $\Box$
			\begin{lemma}\label{TildeSigma}
				Under conditions (A1), (A2'), (S0), (M1), (M2) and (M3) we have that for $1\leq k\leq k_0$,
				\begin{align*}
					\|\sup_{t\in[0,1]}(\E \hat {\mf M}(J_n,t,k)-\mf \Sigma^*_k(t))\|_F=O(J_n\sup_{t,1\leq j\leq J_n}|B_j(t)|^2p/n)
				\end{align*}
				where $\mf \Sigma^*_k(t)=\frac{1}{n}\sum_{j=1}^{J_n}\sum_{i=1}^{n}\E(\mf G(\frac{i}{n},\FF_{i+k})\mf G(\frac{i}{n},\FF_{i})^\top)B_j(\frac{i}{n})B_j(t)$.
			\end{lemma}
			{\it Proof.} Consider the $(u,v)_{th}$ element of $(\E \hat {\mf M}(J_n,t,k)-\mf \Sigma^*_k(t))$, which ise denoted by $(\E \hat {\mf M}(J_n,t,k)-\mf \Sigma^*_k(t))_{u,v}$. Recall that $G_u(t,\FF_i)$ is the $u_{th}$ componentnt of $\mf G(t,\FF_i)$. By definition, we have for $1\leq u,v\leq p$,
			\begin{align}
				(\E \hat {\mf M}(J_n,t,k)-\mf \Sigma^*_k(t))_{uv}
				&=\frac{1}{n}\sum_{j=1}^{J_n}\sum_{i=1}^{n-k}\E\big(\big(G_u(\frac{i+k}{n},\FF_{i+k})-G_u(\frac{i}{n},\FF_{i+k})\big)G_v(\frac{i}{n},\FF_i)\big)B_j(\frac{i}{n})B_j(t)\notag\\
				&+\frac{1}{n}\sum_{j=1}^{J_n}\sum_{i=n-k+1}^{n}\E\big(G_u(\frac{i}{n},\FF_{i+k})G_v(\frac{i}{n},\FF_i)\big)B_j(\frac{i}{n})B_j(t).
			\end{align}
			
			By condition (M3), Lemma \ref{LS-G} 
			we have that uniformly for $1\leq u,v\leq p$,
			\begin{align}\sup_{t\in [0,1]}|(\E \hat {\mf M}(J_n,t,k)-\mf \Sigma^*_k(t))_{uv}|\leq M' J_n\sup_{t,1\leq j\leq J_n}|B_j(t)|^2k/n\end{align} for some sufficiently large constant $M'$ independent of $u$ and $v$. Therefore by the definition of Frobenius norm, and the fact that $k\leq k_0$ the lemma follows.
			\hfill $\Box$
			\begin{lemma}\label{11-29-lemma5}
				Let $\iota_n=\sup_{1\leq j\leq J_n}Lip_j+\sup_{t,1\leq j\leq J_n}|B_j(t)|$ where $Lip_j$ is the Lipschitz constant of the basis function $B_j(t)$. Then under conditions (A1), (A2'), (S0), (M1)--(M3) we have that
				\begin{align*}
					\sup_{t\in [0,1]}\|\bs \Sigma_k^*(t)-\bs \Sigma_x(t,k)\|_F=O\Big(\frac{J_n\sup_{t,1\leq j\leq J_n} |B_j(t)|p\iota_n}{n}+pg_{J_n,K, \tilde M}\Big),
				\end{align*}
				where $\bs \Sigma_k^*$ is defined in Lemma \ref{TildeSigma}.
			\end{lemma}
			{\it Proof.} Notice that by definition we have that
			\begin{align}
				\bs \Sigma_k^*(t)=\frac{1}{n}\sum_{j=1}^{J_n}\sum_{i=1}^n\bs \Sigma_x(\frac{i}{n},k)B_j(\frac{i}{n})B_j(t).
			\end{align}
			Define that $\tilde{\bs \Sigma}_k^*(t)=\sum_{j=1}^{J_n}\int_0^1 \bs \Sigma_x(s,k)B_j(s)dsB_j(t)$. Notice that the $(u,v)_{th}$ element of 
			$\bs \Sigma_k^*(t)-\tilde {\bs \Sigma}_k^*(t)$ is
			\begin{align}
				(\bs \Sigma_k^*(t)-\tilde {\bs\Sigma}_k^*(t))_{u,v} =\sum_{j=1}^{J_n}\Big(\frac{1}{n}\sum_{i=1}^n \E(G_u(\frac{i}{n},\FF_{i+k})G_v(\frac{i}{n},\FF_{i}))B_j(\frac{i}{n})\notag\\
				-\int_0^1 \E(G_u(s,\FF_{i+k})G_v(s,\FF_{i}))B_j(s)ds\Big)B_j(t).
			\end{align}
			Notice that Lemma \ref{LS-G} and Condition (M3) imply that there exists a sufficiently large constant $M'$ depending on $M_0$ of Lemma \ref{LS-G}, such that those Lipschitz constants of the functions
			\begin{align*}
				\E(G_u(s,\FF_{i+k})G_v(s,\FF_{i}))B_j(s)
			\end{align*}
			are bounded by 
			$M'\iota_n$ for all $1\leq k\leq k_0$, $1\leq u,v\leq p$. Then using similar argument to the proof of Lemma \ref{TildeSigma}, we obtain that
			\begin{align}
				\sup_{t\in [0,1]}\|\bs\Sigma_k^*(t)-\tilde {\bs\Sigma}_k^*(t)\|_F=O\Big(\frac{J_n\sup_{t,1\leq j\leq J_n} |B_j(t)|p\iota_n}{n}\Big).
			\end{align}
			Similarly by using basis expansion \eqref{Basisapprox} in condition (S0) of the main article we have that 
			\begin{align}
				\sup_{t\in [0,1]}\|\tilde{\bs  \Sigma}_k^*(t)-\bs \Sigma_x(t,k)\|_F=O(pg_{J_n,K,\tilde M})
			\end{align}
			which completes the proof. \hfill $\Box$
			
			
			\section{ 
				Proof of Theorem \ref{Jan23-Thm4}, and Proof of Theorem \ref{Boots-thm5}. }\label{Sec3proof}\setcounter{equation}{0}
			
			
			Recall $\tilde {\bs \Sigma}_x(t,k)$, $\hat{\mf \Gamma}$, $\hat {\mf \Gamma}_k$ and $\mf \Gamma_k$ defined in Section \ref{sec:test_loading}. Define $$\tilde {\bs \Gamma}=\sum_{k=1}^{k_0} (\int \tilde{\bs  \Sigma}_{x}(t,k)dt)(\int \tilde{\bs  \Sigma}_{x}(t,k)dt)^\top,  {\bs \Gamma}=\sum_{k=1}^{k_0} (\int {\bs  \Sigma}_{x}(t,k)dt)(\int {\bs  \Sigma}_{x}(t,k)dt)^\top$$  
			Let $\tilde{\mathbf W}=(\tilde{\mf w}_1,...,\tilde{\mf w}_d)$ be a set of orthonormal eigenvectors of 
			$\tilde{\bs \Gamma}$
			with respect to its $d$ positive eigenvalues: ($\lambda_1(\tilde{\bs \Gamma })$,...,$\lambda_d(\tilde{\bs \Gamma })$), and  $\mf G=(\mf g_1,...,\mf g_{p-d})$ be a set of orthnormal basis of null space of $\mf A$. Therefore, $((\tilde{\mf w}_i, 1\leq i\leq d), (\mf g_i, 1\leq i\leq p-d))$ is an  orthonormal bases for $\mathbb R^p$. Similarly define $\mf W_i=(\mf w_1,...,\mf w_d)$ where  $ {\mf w}_i$, $i=1,...,d$ are the orthonormal eigenvectors of 
			${\bs \Gamma}$
			with respect to its $d$ positive eigenvalues, and let $\mathbf F=(\mf f_1,...,\mf f_{p-d})$ be a set of orthnormal basis of null space of $\bs \Gamma$.
			Consequently $((\mf w_i, 1\leq i\leq d), (\mf f_i, 1\leq i\leq p-d))$ is a set of  orthonormal bases for $\mathbb R^p$. Let $\hat {\mathbf F}=(\hat {\mf f}_1,..., \hat {\mf f}_{p-d})$ be a basis of null space of $\hat {\bs \Gamma}=\sum_{k=1}^{k_0}\hat{\bs \Gamma}_k$. We consider condition (A2) in Section \ref{proofhatdrate} which allows non-zero $\delta$.
			\begin{corollary}\label{Jan-Corol1}
				Assume (A1), (A2), (S0), (M1)--(M3). 
				\begin{align}
					\|	\|\bs \Gamma-\hat {\bs\Gamma}\|_F\|_{\mathcal L^1}=O(\frac{p^{2-\delta}}{\sqrt n} )
				\end{align}
			\end{corollary}
			{\it Proof.} 
			It suffices to show   uniformly for $1\leq k\leq k_0$, \begin{align}\label{gamma_k}
				\|\|\bs \Gamma_k-\hat {\bs \Gamma}_k\|_F\|_{\mathcal L^1}=O(\frac{p^{2-\delta}}{\sqrt n} ).
			\end{align} 
			By the proof of Lemma \ref{11-29-lemma3}, it follows that for $1\leq k\leq k_0$,
			\begin{align}\label{newF3}
				\left\|\left\|\frac{1}{n}\sum_{i=1}^{n-k}\mf x_{i+k,n}\mf x_{i,n}^\top-\frac{1}{n}\E(\sum_{i=1}^{n-k}\mf x_{i+k,n}\mf x_{i,n}^\top)\right\|_F\right\|_{\mathcal L^2}=O(\frac{p}{\sqrt n}).
			\end{align}
			By the proof of Lemma \ref{TildeSigma} and Lemma \ref{11-29-lemma5}, it follows that for $1\leq k\leq k_0$,
			\begin{align}\label{Jan20-52}
				\left\|\frac{1}{n}\sum_{i=1}^{n-k}\left(\E(\mf x_{i+k,n}\mf x_{i,n}^\top)-\bs \Sigma_x(\frac{i}{n},k)\right)\right\|_F=O(\frac{p}{n}),\\
				\left\|\frac{1}{n}\sum_{i=1}^{n-k}\bs \Sigma_x(\frac{i}{n},k)-\int_0^1\bs \Sigma_x(t,k)dt\right\|_F=O(\frac{p}{n}),\label{Jan20-54}
			\end{align}
			Then by \eqref{newF3} to \eqref{Jan20-54} we have that
			\begin{align}
				\left\|\frac{1}{n}\sum_{i=1}^{n-k}\mf x_{i+k,n}\mf x_{i,n}^\top-\int_0^1\bs \Sigma_x(t,k)dt\right\|_F=O(\frac{p}{\sqrt n}),
			\end{align}
			which together with \eqref{11-29-eq7} in the main article and the definition of $\bs \Gamma_k$ proves \eqref{gamma_k}. Therefore the corollary holds.\hfill $\Box$

					\begin{corollary}\label{Jan20-Corol2}
						Assume conditions (A1), (A2), (M1), (M2), (M3) and conditions (S0), (S2)-(S4), then  under null hypothesis, there exist orthogonal matrices  $\hat {\mf O}_3\in \mathbb R^{d\times d}$ and   $\hat {\mf O}_4\in \mathbb R^{(p-d)\times (p-d)}$ , such that under the null hypothesis	\begin{align*}
							\|	\| \hat {\mf W}\hat {\mathbf O}_3-\tilde{\mf W}\|_F\|_{\mathcal L^1}=O(p^\delta/\sqrt n+p^{\delta-1}), \\
							\|	\| \hat {\mf F} \hat {\mathbf O}_4-\mf G\|_F\|_{\mathcal L^1}=O(p^\delta/\sqrt n+p^{\delta-1}),
						\end{align*}
						under (S1'),  provided that there exists $k', 1\leq k'\leq k_0$ such that $\sigma_d(\int \bs \Sigma_x(t,k')\geq \eta>0$.  If (S1) holds then trivially the term $p^{\delta-1}$ vanishes.
					\end{corollary}
					{\it Proof.} It suffices to prove the results under (S1'). Under (S1) $\tilde {\bs \Sigma}_x(t,k)=\bs\Sigma_x(t,k)$, the proof will be similar and simpler.
					By the proof of Theorem \ref{Thm-approx} and the definition of $\tilde {\bs \Sigma}_x(t,k)$, we have that 
					\begin{align*}
						\|{\bs \Sigma}_x(t,k)\|_2=O(p^{1-\delta}),	~~~~\|\tilde {\bs \Sigma}_x(t,k)-{\bs \Sigma}_x(t,k)\|_2=O(1)
					\end{align*}
					and consequently by triangle inequality, 
					\begin{align}
						\|\tilde{\bs \Gamma}-\bs \Gamma\|_2=O(p^{1-\delta}).
					\end{align}
					Together with Corollary \ref{Jan-Corol1} we have 
					\begin{align}
						\|\|\tilde{\bs \Gamma}-\hat{\bs \Gamma}\|_2\|_{\mathcal L^1}=O(p^{1-\delta}+p^{2-\delta}/\sqrt n).
					\end{align}
					
					Notice that under null hypothesis
					\begin{align}\label{new.c9}
						\tilde{\bs \Gamma}=
						\mf A\sum_{k=1}^{k_0}(\int (\bs \Sigma_z(t,k)\mf A^\top+\mf \Sigma_{ze}(t,k) dt)	(\int (\bs \Sigma_z(t,k)\mf A^\top+\mf \Sigma_{ze}(t,k) dt)^\top \mf A^\top.	
					\end{align}
					Since $\mf A$ is a $p\times  d$ matrix, we have $\lambda_{j}(\tilde{\bs \Gamma})=0$ for $j\geq d+1$. It remains to show that $\lambda_d(\tilde{\mf \Gamma})\gtrapprox(p^{2-2\delta})$ then the Corollary will follow from Theorem 2 of \cite{yu2015useful}.
					By condition (S4), it remains to show that
					\begin{align}
						\lambda_d\left(\mf A \sum_{k=1}^{k_0}\left(\int \bs \Sigma_z(t,k) dt\mf A^\top\mf A \int \bs \Sigma^\top_z(t,k) dt \right)\mf A^\top \right )\gtrapprox p^{2-2\delta}
					\end{align}
					By the QR decomposition argument of  in the proof of Theorem \ref{Space-Distance}, it suffices to prove that
					\begin{align}
						\lambda_{\min} \left( \sum_{k=1}^{k_0}\left(\int \bs \Sigma_z(t,k) dt\mf A^\top\mf A \int \bs \Sigma^\top _z(t,k)dt\right)\right)\gtrapprox p^{1-\delta}.
					\end{align}
					By Weyl's inequality, the LHS of the above is greater than
					\begin{align}
						\lambda_{\min} \left( \int \bs  \Sigma_z(t,k) dt\mf A^\top\mf A \int \bs \Sigma^\top _z(t,k)dt\right) =\|\int \bs \Sigma_z(t,k) dt\mf A^\top \|^2_m\geq \|\mf A\|_m^2\gtrapprox p^{1-\delta},
					\end{align}
					which finishes the proof. \hfill $\Box$

					\begin{corollary}\label{Corol4}
						Assume conditions of Corollary  \ref{Jan20-Corol2} hold, then under null hypothesis there exists an orthonormal basis $\{\mf f_i, 1\leq i\leq p-d\}$ of null space of $\mf A$, such that
						\begin{align}
							\|	\|\hat {\mf F}-\mf F\|_F\|_{\mathcal L^1}=O(\theta_0(n,p)),
						\end{align}
						where $\mf F=(\mf f_1,...\mf f_{p-d})$.
					\end{corollary}
					{\it Proof.} Notice that the null space of $\mf A$ is the same as the null space of  $\tilde{\mf \Gamma} $. Recall Corollary \ref{Jan20-Corol2} shows that
					\begin{align*}
						\|	\|\hat {\mf F} \hat{\mathbf O}_4-\mf G\|_F\|_{\mathcal L^1}=O(\theta_0(n,p)).
					\end{align*}
					Take $\mf F=\mf G\hat{\mathbf O}^\top_4$, notice that
					\begin{align*}
						\|\hat{\mf F}-\mf F\|_F=	\|\hat {\mf F}-\mf G\hat {\mf O}_4^\top\|_F
						&=\|(\hat {\mf F}\hat {\mf O}_4-\mf G)\hat {\mf O}_4^\top\|_F=tr^{1/2}((\hat {\mf F}\hat {\mf O}_4-\mf G)\hat {\mf O}_4^\top\hat {\mf O}_4(\hat {\mf F}\hat {\mf O}_4-\mf G)^\top)\\
						&=tr^{1/2}((\hat {\mf F}\hat {\mf O}_4-\mf G)(\hat {\mf F}\hat {\mf O}_4-\mf G)^\top)=\|\hat {\mf F}\hat {\mf O}_4-\mf G\|_F=O(\theta_0(n,p))
					\end{align*}
					and the corollary is proved. \hfill $\Box$\\
					The next two propositions are needed for the proof of Theorem \ref{Jan23-Thm4} in the main article. Recall the definition of $\tilde T_n$
					defined above the proof of Theorem \ref{Jan23-Thm4}.
					Recall that in \eqref{eq15hatTn} of the main article we have defined
					\begin{align}
						\hat T_n=\sqrt{m_n}\max_{1\leq h\leq N_n}\max_{1\leq i\leq p-\tilde d_n}|\hat {\mf f}_i^\top \mf S_h^X|\notag 
					\end{align}  
					Further, define that  
					\begin{align}	\label{tildeT} \tilde T_n=\sqrt{m_n}\max_{1\leq h\leq N_n}\max_{1\leq i\leq p-d}|\hat{\mf f}_i^\top\mf S_h^X|,
						\\T_n=\sqrt{m_n}\max_{1\leq h\leq N_n}\max_{1\leq i\leq p-d}|\mf f_i^\top\mf S_h^X|.\label{TnC5}
					\end{align}
					\begin{proposition}\label{Jan23-Lemma6}
						Suppose conditions of Proposition \ref{prop1} hold. In addition, 
						assume condition ($M2'$). Then there exists a set of orthnormal basis $\mf F$ of $\mf A$ such that for any sequence $g_n\rightarrow \infty$
						\begin{align}\label{ratehatT}
							\p(|\hat T_n- T_n|\geq g_nN_n^{1/l}\theta_0(n,p)p^{\frac{1}{2}})  =O(g_n^{-\frac{l}{l+1}}+\log^{-1/2} n+\theta(n,p)).
						\end{align}
						where $T_n$ is calculated using $\mf F$.
					\end{proposition}
					{\it Proof.}  Since on the event $\{\hat d_n=\tilde d_n\}$,  $\tilde T_n=\hat T_n$. By Proposition  \ref{prop1} it suffices to consider the event  $\{\hat d_n=\tilde d_n\}$. By the definition of $\tilde T_n$ and $T_n$,
					\begin{align}\label{tildeT-T}
						|\tilde T_n-T_n|&=\sqrt{m_n}\max_{1\leq h\leq N_n}\max_{1\leq i\leq p-d}|(\hat{\mf f}_i-\mf f_i)^\top \mf S_h^X|\notag
						\\&\leq \sqrt {m_n}\max_{1\leq h\leq N_n}\max_{1\leq i\leq p-d}|\hat {\mf f}_i- {\mf f}_i||\mf S_h^X|
						\notag\\&\leq \sqrt{m_n}\|\hat{\mf F}-\mf F\|_F\max_{1\leq h\leq N_n}|\mf S_h^X|
					\end{align}
					To shorten the notation,   write $\mf A_i$ for $\mf A(i/n)$. Then 
					\begin{align}\label{C19}
						\sqrt{m_n}\max_{1\leq h\leq N_n}|\mf S_h^X|=m_n^{-1/2}\max_{1\leq h\leq N_n}\Big(|\sum_{i\in b_h}\mf A_i\mf z_i|+|\sum_{i\in b_h}\mf e_i|\Big)
					\end{align}
					To deal with the above bound, first note that 
					\begin{align}\label{C21}
						|\sum_{i\in b_h}\mf e_i|=\Big|(\sum_{i\in b_h}e_{i1},...,\sum_{i\in b_h}e_{ip})^\top\Big|=\Big(\sum_{j=1}^p\big(\sum_{i\in b_h }e_{ij}\big)^2\Big)^{1/2}
					\end{align}
					Therefore \begin{align}\label{C22}
						\||\sum_{i\in b_h}\mf e_i|\|_{\mathcal L^l}=\Big(\E[\sum_{j=1}^p(\sum_{i\in b_h}e_{ij})^2]^{\frac{l}{2}}\Big)^{\frac{1}{l}}=\|\sum_{j=1}^p(\sum_{i\in b_h}e_{ij})^2\|_{\mathcal L^{l/2}}^{1/2}
						\leq \Big[\sum_{j=1}^p\|(\sum_{i\in b_h} e_{ij})^2\|_{\mathcal L^{l/2}}\Big]^{1/2}.
					\end{align}
					Meanwhile, since $l\geq 2$, by  Theorem 2 of \cite{wu2005nonlinear} and conditions (M2'), we have that for all $1\leq h\leq N_n$ and $1\leq j\leq p$, $\|\sum_{i\in b_h}e_{ij}\|_{\mathcal L^l}=O(\sqrt{m_n})$ and therefore
					\begin{align}
						\|(\sum_{i\in b_h}e_{ij})^2\|_{\mathcal L^{l/2}}=\|\sum_{i\in b_h}e_{ij}\|_{\mathcal L^l}^2=O(m_n).
					\end{align}
					Combining \eqref{C21} and \eqref{C22} we have that 
					\begin{align}
						\||\sum_{i\in b_h} \mf e_i|\|_{\mathcal L^l}=O(\sqrt{pm_n}).
					\end{align}
					On the other hand,  use the inequality that $\max_{1\leq i\leq n}|X_i|^l\leq \sum_{i=1}^n |X_i|^l$, which leads to the 
					\begin{align}\label{claim1}
						\|\max_{1\leq i\leq n}|X_i|\|_{\mathcal L^{l}}=O(n^{\frac{1}{l}})~~~
						\text{if}~~~\max_{1\leq i\leq n}\|X_i\|_{\mathcal L^{l}}=O(1).
					\end{align} 
					Now by \eqref{claim1}, we have
					\begin{align}\label{maxei}
						m_n^{-1/2}\|\max_{i\leq  h\leq N_n}|\sum_{i\in b_h} \mf e_i|\|_{\mathcal L^l}=O(N_n^{1/l}\sqrt{p}).
					\end{align}
					Similarly, since $\mf A_i=(a_{uv}(i/n))_{1\leq u\leq p, 1\leq v\leq d}$ and $\mf z_i=(z_{i,1},...,z_{i,d})^\top$, we have
					\begin{align}
						|\sum_{i\in b_h}\mf A_i\mf z_i|=\Big|(\sum_{i\in b_h}\sum_{j=1}^d a_{1j}(i/n)z_{ij},...,\sum_{i\in b_h}\sum_{j=1}^d a_{pj}(i/n)z_{ij})^\top\Big|=\sqrt{\sum_{s=1}^p(\sum_{i\in b_h}\sum_{j=1}^d a_{sj}(i/n)z_{ij})^2}.
					\end{align}
					As a consequence, we have by triangle inequality,
					\begin{align}\label{C26}
						\||\sum_{i\in b_h}\mf A_i\mf z_i|\|_{\mathcal L^l}\leq \Big[\sum_{s=1}^p\|(\sum_{i\in b_h}\sum_{j=1}^d a_{sj}(i/n)z_{ij})^2\|_{\mathcal L^{l/2}}\Big]^{1/2}\leq \Big[\sum_{s=1}^p\|\sum_{i\in b_h}\sum_{j=1}^d a_{sj}(i/n)z_{ij}\|_{\mathcal L^{l}}^2\Big]^{1/2}
						\notag	\\\leq\Big[\sum_{s=1}^p\big(\sum_{j=1}^d\|\sum_{i\in b_h} a_{sj}(i/n)z_{ij}\|_{\mathcal L^{l}}\Big)^2\Big]^{1/2}.
					\end{align}
					Notice that
					\begin{align*}
						\sum_{i\in b_h}a_{sj}(i/n)z_{ij}=\sum_{k=0}^\infty\sum_{i\in b_h}\mathcal P_{i-k}a_{sj}(i/n)z_{ij}.
					\end{align*}
					By Burkholder inequality (see \cite{wu2005nonlinear}) it follows that
					\begin{align*}
						\|\sum_{i\in b_h}\mathcal P_{i-k}a_{sj}(i/n)z_{ij}\|^2_{\mathcal L^l}\leq
						C_l\sum_{i\in b_h}\|\mathcal P_{i-k}a_{sj}(i/n)z_{ij}\|_{\mathcal L^l}^2
						=O\left(\sum_{i\in b_h}a_{sj}^2(i/n)(\delta^z_{l}(k))^2\right)
					\end{align*}
					where $C_l$ is a constant depending only on $l$. Combining the above two equations and via the triangle inequality,
					\begin{align}\label{C27}
						\|\sum_{j\in b_h}a_{sj}(i/n)z_{ij}\|_{\mathcal L^l}=O((\sum_{i\in b_h}a_{sj}^2(i/n))^{1/2}).
					\end{align}
					Combining \eqref{C26} and \eqref{C27}, we have
					\begin{align}
						\notag\|\sum_{i\in b_h}\mf A_i\mf z_i\|_{\mathcal L^l}&=O\Big( \Big(\sum_{s=1}^p\Big[\sum_{j=1}^d\big(\sum_{i\in b_h}a_{sj}^2(i/n)\big)^{\frac{1}{2}}\Big]^2\Big)^{\frac{1}{2}}\Big)
						\notag\\\notag&=O\Big(\big(\sum_{s=1}^pd\sum_{j=1}^d\sum_{i\in b_h}a_{sj}^2(i/n)\big)^{\frac{1}{2}}\Big)\\&=O\Big(\big(d\sum_{i\in b_h}\|\mf A_i\|_F^2\big)^{\frac{1}{2}}\Big)=O(d^{\frac{1}{2}}m_n^{\frac{1}{2}}p^{\frac{1-\delta}{2}}).
					\end{align}
					By \eqref{claim1}, we have
					\begin{align}
						\Big\|\frac{1}{\sqrt{m_n}}\max_{1\leq h\leq N_n}|\sum_{i\in b_h}\mf A_i\mf z_i|\Big\|_{\mathcal  L^l}=O(p^{\frac{1-\delta}{2}}N_n^{1/l}).
					\end{align}
					Combining with  \eqref{C19} and \eqref{maxei},   we have that 
					\begin{align}
						\|\sqrt{m_n}\max_{1\leq h\leq N_n}|\mf S_h^X|\|_{\mathcal L^l}=O(p^{\frac{1-\delta}{2}}N_n^{1/l}+N_n^{1/l}\sqrt{p})=O(N_n^{1/l}\sqrt{p}).
					\end{align}
					Use the fact that for any random variables $X$, $Y$ and positive constants $c_1$ and $c_2$,
					\begin{align}
						\p(|XY|\geq c_1c_2)\leq \p(|X|\geq c_1)+\p(|Y|\geq c_2)
					\end{align}
					and Corollary \ref{Corol4}, we have for any $g_n\rightarrow \infty$, via Markov inequality,
					\begin{align}
						&\p(|\tilde T_n-T_n|\geq g_nN_n^{1/l}\theta(n,p)p^{\frac{1}{2}})\notag
						\\&\leq \p(\sqrt{m_n}\max_{1\leq h\leq N_n}|\mf S_h^X|\geq g_n^{\frac{1}{l+1}}N^{\frac{1}{l}}\sqrt p)+\p(\|\hat{\mf  F}-{\mf F}\|_F\geq g_n^{\frac{l}{l+1}}\theta_0(n,p)) 
						\notag\\&=O(g_n^{-\frac{l}{l+1}})=o(1).
					\end{align}
					Therefore the proposition follows.
					\hfill $\Box$

					\begin{proposition}\label{Newprop}
						Under conditions (M2'), (M5), (M6) and (M7), we have that there exist constants $c'$ and $C'$ such that for all $1\leq j\leq n$ and $s'$ such that $s'g(n,p)/n\rightarrow 0$ and $s'\tilde g(n,p)/n\rightarrow 0$,  we have
						\begin{align}
							c'\leq \lambda_{min}(var(\frac{1}{\sqrt{s'}}\sum_{i=j}^{j+s'-1}\mf e_i))\leq 
							\lambda_{max}(var(\frac{1}{\sqrt{s'}}\sum_{i=j}^{j+s'-1}\mf e_i))\leq C'.
						\end{align}
					\end{proposition}
					{\it Proof.} Notice that
					\begin{align}\label{C41}
						var(\frac{1}{\sqrt{s'}}\sum_{i=j}^{j+s'-1}\mf e_i)=\frac{1}{s'}\sum_{i=j}^{j+s'-1}\sum_{l=j}^{j+s'-1}Cov(\mf H(i/n,\FF_i), \mf H(l/n,\FF_l))\notag\\
						=\frac{1}{s'}\sum_{i=j}^{j+s'-1}\sum_{l=j}^{j+s'-1}Cov(\mf H(i/n,\FF_i), \mf H(i/n,\FF_l))-\mf R
					\end{align}
					where 
					\begin{align}
						\mf R=\frac{1}{s'}\sum_{i=j}^{j+s'-1}\sum_{l=j}^{j+s'-1}\E\Big(\mf H(i/n,\FF_i)(\mf H(i/n,\FF_l)-\mf H(l/n,\FF_l))^\top\Big)
					\end{align}
					On the other hand noticing that
					\begin{align}
						\E\Big(\mf H(i/n,\FF_i)(\mf H(i/n,\FF_l)-\mf H(l/n,\FF_l))^\top\Big)=\E\Big(\mf H(i/n,\FF_i)(\int_{l/n}^{i/n} \frac{\partial}{\partial u}\mf H(u,\FF_l)du)^\top\Big)\notag\\
						=\int_{l/n}^{i/n} \E\Big(\mf H(i/n,\FF_i)(\frac{\partial}{\partial u}\mf H(u,\FF_l)du)^\top\Big).
					\end{align}
					Therefore together (M6) and triangle inequality we have
					\begin{align}\label{C44}
						\|\mf R\|_2=O(\frac{s'}{n}\frac{s'}{s'})=o(1).
					\end{align}
					Meanwhile,  
					\begin{align}\label{C45}
						\frac{1}{s'}\sum_{i=j}^{j+s'-1}\sum_{l=j}^{j+s'-1}Cov(\mf H(i/n,\FF_i),\mf H(i/n,\FF_l))=\E(\mf H(i/n,\FF_0)\mf H^\top(i/n,\FF_0))\notag\\+\sum_{k=1}^{s'}[\E(\mf H(i/n,\FF_0)\mf H^\top(i/n,\FF_k)+\mf H(i/n,\FF_0)\mf H^\top(i/n,\FF_{-k}))]\frac{s'-k}{s'}
						\notag\\=\sum_{k\in\mathbb Z,|k|\leq s'}\bs\Sigma_e(i/n,k)-\sum_{k=1}^{s'}\frac{k}{s'}\Big[\bs\Sigma_e(i/n,k)+\bs\Sigma_e(i/n,-k)\Big].
					\end{align}
					Notice that $\bs \Sigma_e(i/n,k)=\bs \Sigma_e^\top(i/n,-k)$ we have that
					\begin{align}
						\|\bs \Sigma_e(i/n,k)+\bs \Sigma_e(i/n,-k)\|_2\leq 2\|\bs \Sigma_e(i/n,k)\|_2.
					\end{align}
					Together with condition (M7), we have
					\begin{align}
						\|\sum_{k=1}^{s'}\frac{k}{s'}\Big[\bs\Sigma_e(i/n,k)+\bs\Sigma_e(i/n,-k)\|_2=o(1), \quad \|\sum_{|k|>s'}\bs\Sigma_e(i/n,k)\|_2=o(1).
					\end{align}
					Together with \eqref{C41}, \eqref{C44} and \eqref{C45} we have
					\begin{align}
						\|var(\frac{1}{\sqrt{s'}}\sum_{i=j}^{j+s'-1}\mf e_i)-\sum_{k\in\mathbb Z}\bs\Sigma_e(i/n,k)\|_2=o(1).
					\end{align}
					Using condition (M5), the Proposition follows. \hfill $\Box$
					
					Recall in Theorem \ref{Jan23-Thm4} we define the $p-d$ dimensional vector $\bs l_{j,s}$ for $1\leq s\leq N_n$ and the $(p-d)N_n$ dimensional vector $\bs l_i$ for $1\leq i\leq m_n$ as
					\begin{align*}
						\bs l_{j,s}=(\mf f_1^\top \mf e_{j+(s-1)m_n},..., \mf f_{p-d}^\top \mf e_{j+(s-1)m_n}), \quad \bs l_i=(\bs l_{i,1}^\top,...,\bs l_{i,N_n}^\top)^\top
					\end{align*}
					Recall that $\mf y_i$, $1\leq i\leq  m_n$ is a $(p-d)N_n$ centered Gaussian vector preserving the autocovariance structure of $\mf e_i$. Let $\mf y_i:=(\mf y^\top _{i,1},...,\mf y^\top_{i,(p-d)N_n})$.


					\begin{proposition}\label{definitioniota}
						Let the assumptions of Theorem \ref{Jan23-Thm4} be held. Define $\tilde T_n^{(M')}=\frac{1}{\sqrt{m_n-2M'}}\sum_{i=M'}^{m_n-M'}\mf l_i$ for some sequence $M'=o(m_n)$.  Then under null hypothesis, we have
						\begin{align}\label{Jan23-83}
							\sup_{t\in \mathbb R}|\p(\tilde T^{(M')}_n\leq t)-\p(|\mf y|_{\infty}\leq t)|\lessapprox \upsilon(m_n, N_n ,p,d,l),
						\end{align}
						where $\upsilon(m_n, N_n, p,d,l)=\iota(m_n-2M',N_n(p-d),l, (N_np)^{1/l})$ and $\iota(\cdot)$ is defined as
						\begin{align}\label{S8}\iota(n,p,q, D_n)=\min(n^{-1/8}M^{1/2}l_n^{7/8}+\gamma+(n^{1/8}M
							^{-1/2}l_n^{-3/8})^{q/(1+q)}(p \sum_{j=M}^\infty \Delta_q^q(j))^{1/(1+q)}\notag\\+\Xi_M^{1/3}(1\vee \log (p/\Xi_M))^{2/3}),\end{align}  where 
						$\Xi_{M}=\max_{1\leq j\leq p}$  $\sum_{j=M}^\infty j\Delta_{2}(j)$, and the minimum is taken over all possible values of $\gamma$ and $M$ subject to $$n^{3/8}M^{-1/2}l_n^{-5/8}\geq \max\{D_n(n/\gamma)^{1/l},l_n^{1/2}\}, \quad M\leq 2M'$$ with $l_n=\log (pn/\gamma)\vee 1$. Here $\Delta_q(j)$ is defined in condition (M8).
					\end{proposition}
					{\it Proof.} Observe that $\tilde T_n^{(M')}$ can be well approximated by sum of high dimensional $m$-dependent vectors in the sense of step 1 of Proof of Theorem 2 in \cite{zhang2018gaussian} for $m\leq M'$, and the summands are independent if the difference between corresponding indices are larger than $m$. Therefore we could invoke Theorem 2.1 of \cite{zhang2018gaussian}. For this purpose,
					We  have the following assertions. 
					\begin{description}
						\item (a)For $1\leq i\leq p-d$ and $1\leq j\leq p$, $\|\mf f_i^\top \mf e_j\|_{\mathcal L^l}\leq M_l \|\mf f_i^\top \mf e_j\|_{\mathcal L^2}=M_l \mf f_i^\top \E(\mf e_j\mf e_j^\top)\mf f_i\lessapprox 1$.
						\item (b) There exist constants $M_1$ and $M_2$ such that 
						for $1\leq q\leq p-d$ and $1\leq s\leq N_n$, we have  
						$M_1\leq \frac{1}{m_n}\sum_{i,l=1}^{m_n}Cov(\mf f_q^\top \mf e_{i+(s-1)m_n}, \mf f_q^\top \mf e_{l+(s-1)m_n})\leq M_2$.
					\end{description}
					Assertion (a) follows from (M4). Assertion (a) immediately implies that 
					\begin{align}
						\max_{1\le j\leq m_n}\E(\max_{\stackrel{1\leq i\leq p-d}{1\leq s\leq N_n} }|\mf f_i^\top \mf e_{j+(s-1)m_n}|^l)\lessapprox N_n(p-d)
					\end{align}
					such that
					\begin{align}
						\max_{1\le j\leq m_n}\Big\|\max_{\stackrel{1\leq i\leq p-d}{1\leq s\leq N_n} }\frac{|\mf f_i^\top \mf e_{j+(s-1)m_n}|}{(N_n(p-d))^{1/l}}\Big\|_{\mathcal L^l}\lessapprox 1.
					\end{align}
					Since $l\geq 4$, we have verified  condition (7) of Assumption 2.1 in \cite{zhang2018gaussian} by setting $h(x)=|x|^l$ and $\mathcal D_n=(N_n(p-d))^{1/l}$ there. Assertion (b) is in fact the Condition (9) of \cite{zhang2018gaussian}, and is indicated by Proposition \ref{Newprop} and the fact that $M'=o(m_n)$. By Condition (M8), Condition (10) of \cite{zhang2018gaussian} holds. Therefore the Proposition follows from Theorem 2.1 of \cite{zhang2018gaussian}. \hfill $\Box$\

					\subsection{ Proof of Theorem \ref{Jan23-Thm4}.}
					\label{should-recalculate} 
					To show Theorem \ref{Jan23-Thm4}, recall the quantity $T_n=\sqrt{m_n}\max_{1\leq h\leq N_n}\max_{1\leq i\leq p-d}|{\mf f}_i^\top\mf S_h^X|$ defined in \eqref{TnC5} using the true quantities $\mf F$ and $d$ to approximate $\hat T_n$.
					Recall the definition of $\hat{\mf F}$ and $\mf F$ in Section \ref{sec:test_loading}.
					Recall $\tilde T_n^{(M')}$ in Proposition \ref{definitioniota}. By condition (M8) and the fact that  $\max_{1\leq i\leq n}|X_i|^l\leq \sum_{i=1}^n |X_i|^l$, we have 
					\begin{align}
						\|\tilde T_n^{(M')}- T_n\|_{\mathcal L^l}=O(\sqrt{\frac{M'}{m_n}}(N_np)^{1/l})
					\end{align}
					Writing 
					$\delta_{0,1}=N_n^{1/l}\theta_0(n,p)p^{\frac{1}{2}}$, $\delta_{0,2}=\sqrt{\frac{M'}{m_n}}(N_np)^{1/l}$, $\delta_0=g_{n}(\delta_{0,1}+\delta_{0,2})$,
					where $g_{n}$ is a diverging  sequence such that $\delta_0=o(1)$. Hence by Proposition \ref{Jan23-Lemma6}, triangle inequality, and Markov inequality, we have that 
					\begin{align}\label{New.44}
						\p(|\tilde T^{(M')}_n-\hat T_n|\geq \delta_0)\leq \p(|\tilde T_n^{(M'}-T_n|\geq g_n\delta_{0,2})+\p(|T_n-\hat T_n|\geq g_n\delta_{0,1})\notag\\=O(g_{n}^{-\frac{l}{l+1}}+\log^{-1/2} n+\theta(n,p)+g^{-l}_{n})
						=O(g_{n}^{-\frac{l}{l+1}}+\log^{-1/2} n+\theta(n,p)).
					\end{align}
					Notice that
					\begin{align}\label{S95_Nov29}
						\p(\hat T_n\geq t)&\leq \p(\tilde T^{(M')}_n\geq t-\delta_0)+\p(|\tilde T^{(M')}_n-\hat T_n|\geq \delta_0),\\
						\p(\hat T_n\geq t)&\geq \p(\tilde T^{(M')}_n\geq t+\delta_0, |\tilde T^{(M')}_n-\hat T_n|\leq \delta_0)\notag\\&=
						\p(\tilde T^{(M')}_n\geq t+\delta_0)-\p(\tilde T^{(M')}_n\geq t+\delta_0, |\tilde T^{(M')}_n-\hat T_n|\geq \delta_0)\notag
						\\&\geq \p(\tilde T^{(M')}_n\geq t+\delta_0)-\p( |\tilde T^{(M')}_n-\hat T_n|\geq \delta_0).
					\end{align}
					Therefore following \eqref{New.44} and  Proposition \ref{definitioniota},  we have 
					\begin{align}\label{S97-Nov-21}
						&\sup_{t\in \mathbb R}|\p(\hat T_n\geq t)-\p(|\mf y|_\infty\geq t)|\leq
						\sup_{t}|\p(|\mf y|_\infty\geq t-\delta_0)-\p(|\mf y|_\infty\geq t+\delta_0)|\notag\\
						&+\sup_{t\in \mathbb R}|\p(|\mf y|_{\infty}\geq t)-\p(\tilde T^{(M')}_n\geq t)|+\p( |\tilde T^{(M')}_n-\hat T_n|\geq \delta_0)
						\notag\\&\leq  \sup_{t}|\p(|\mf y|_\infty\geq t-\delta_0)-\p(|\mf y|_\infty\geq t+\delta_0)|\notag\\&+O(g_n^{-\frac{l}{l+1}}+\log^{-1/2} n+\theta(n,p)+\upsilon(m_n-2M', N_n,p,d,l)).
					\end{align}
					Since $|\mf y|_\infty=\max(\mf y, -\mf y)$, by Corollary 1 of \cite{chernozhukov2015comparison},  we have that
					\begin{align}\label{S98-Nov-21}
						\sup_{t}|\p(|\mf y|_\infty\geq t-\delta_0)-\p(|\mf y|_\infty\geq t+\delta_0)|=O(\delta_0\sqrt{\log (n/\delta_0)}).
					\end{align}
					Combining \eqref{S97-Nov-21},\eqref{S98-Nov-21}, and letting 
					$g_n=\Omega_n^{-\frac{l+1}{2l+1}}$
					the second claim of Theorem \ref{Jan23-Thm4} follows. \hfill $\Box$

						\medskip
						\subsection{Proof of Theorem \ref{Boots-thm5}}
						Under the null hypothesis, $T_n=\sqrt{m_n}\max_{1\leq h\leq N_n}\max_{1\leq i\leq p-d}|\mf f_i^\top \mf S_h^e|$.
						Recall the $N_n(p-d)$ dimensional vector $\bs l_i$ defined above \ref{definitioniota}. In the following proofs, define\begin{align}
							{\mf s}_{j,w_n}=\sum_{r=j}^{j+w_n-1} {\bs l}_i ~~~\text{for}~~ 1\leq j\leq m_n-w_n+1, ~~\text{and}~{\mf s}_{m_n}=\sum_{i=1}^{m_n} {\bs l}_i.
						\end{align}
						\medskip
						Recall in the main article we have defined that $\bar \theta(n,p, l, N_n,w_n)=\sqrt{w_n}N_n^{1/l}\theta(n,p)p^{1/2}$. In the proof, to allow non-zero $\delta$, $\theta(n,p)$ is also a function of $\delta$. To stress the dependence on $\delta$ in the following we write  
						$\bar \theta(n,p, \delta, l, N_n,w_n)=\sqrt{w_n}N_n^{1/l}\theta(n,p)p^{1/2}$.
						
						\noindent {\bf Proof of Theorem \ref{Boots-thm5}} 
							
							Define \begin{align}
								\bs \upsilon_n=\frac{1}{\sqrt{w_n(m_n-w_n+1)}}\sum_{j=1}^{m_n-w_n+1}( {\mf s}_{j,w_n}-\frac{w_n}{m_n} {\mf s}_{m_n})R_j.
							\end{align}
							We shall show the following two assertions:
							\begin{align}
								\sup_{t\in \mathbb R}|\p(|\mf y|_\infty\leq t)-\p(|\bs \upsilon_n|_{\infty}\leq t|\FF_n)|=O_p(\Theta_n^{1/3}\log ^{2/3}(\frac{W_{n,p}}{\Theta_n})),\label{Junly14-S86}\\
								\sup_{t\in \mathbb R }|\p(|\bs \upsilon_n|_\infty\leq t|\FF_n)-\p(|\bs \kappa_n|_\infty\leq t|\FF_n)|=O_p(	\bar{\theta}(n,p,\delta,l,N_n,w_n)^{l/(l+1)}\log^{\frac{l}{2l+2}} n).
								\label{Junly15-S85}
							\end{align}
							The theorem then follows from \eqref{Junly14-S86}, \eqref{Junly15-S85} and Theorem \ref{Jan23-Thm4}, the fact that  $\{\mf x_{i,n}, 1\leq i\leq n\}$ is $\FF_n$ measurable and DCT.
							
							Step (i): Proof of \eqref{Junly14-S86}. To show \eqref{Junly14-S86}, we shall show that 
							\begin{align}\label{S.94}
								\|\max_{1\leq u,v\leq N_n(p-d)}|\sigma^\upsilon_{u,v}-\sigma^Y_{u,v}|\|_{\mathcal L^{q^*/2}}=O\left(w_n^{-1}+\sqrt{w_n/m_n}W_{n,p}^{2/{q^*}}\right),
							\end{align}
							where $\sigma^\upsilon_{u,v}$ and $\sigma^Y_{u,v}$ are the $(u,v)_{th}$ entry of the covariance matrix of $\bs\upsilon_n$ given $ \FF_n$ and covariance matrix of $\mf y$. Notice that  \eqref{S.94} together with claim (b) of  Proposition \ref{definitioniota} implies that there exists a constant $\eta_0>0$ such that 
							\begin{align}\label{S103-Nov-29}
								\p(\max_{1\leq u,v\leq N_n(p-d)}\sigma^\upsilon_{u,v}\geq \eta_0)\geq 1-O\left(\left(w_n^{-1}+\sqrt{w_n/m_n}W_{n,p}^{2/{q^*}}\right)^{q^*/2}\right).
							\end{align}
							Since by assumption $w_n^{-1}+\sqrt{w_n/m_n}W_{n,p}^{2/{q^*}}=o(1)$, it suffices to consider the conditional Gaussian approximation on the $\{\mf x_{i,n}\}$ measurable event $\{\max_{1\leq u,v\leq N_n(p-d)}\sigma^\upsilon_{u,v}\geq \eta_0\}$. Then by the construction of $\mf y$  and Theorem 2 of  \cite{chernozhukov2015comparison} (we consider the case $a_p=\sqrt{2\log p}$ in there), \eqref{Junly14-S86} will follow.
							
							Now we prove \eqref{S.94}.
							Let $ S_{j,w_n,s}$ and $ S_{m_n,s}$ be the $s_{th}$ element of the vectors $\mf s_{j,w_n}$ and $\mf s_{m_n}$, respectively. By our construction, we have
							\begin{align}
								\sigma_{u,v}^\upsilon=\frac{1}{w_n(m_n-w_n+1)}\left(\sum_{j=1}^{m_n-w_n+1}( S_{j,w_n,u}-\frac{w_n}{m_n} S_{m_n,u})( S_{j,w_n,v}-\frac{w_n}{m_n} S_{m_n,v})\right),
								\sigma_{u,v}^Y=\E{\frac{ S_{m_n,u} S_{m_n,v}}{m_n}}.
							\end{align}
							
							To simply the notation let $m_n=kw_n$ for some integer $k$. The case that $m_n=kw_n+q$ for some $1\leq q\leq w_n$ the proof will be the same but with more complicated notation.  Straightforward calculations show that
							\begin{align}
								\E(S_{m_n,u}S_{m_n,v})=\sum_{s=1}^{m_n/w_n}\E S_{(s-1)w_n+1,w_n,u}S_{(s-1)w_n+1,w_n,v}
								+\sum_{s_1=1}^{m_n/w_n}\sum_{s_2=1,s_2\neq s_1}^{m_n/w_n}\E(S_{(s_1-1)w_n+1,w_n,u}S_{(s_2-1)w_n+1,w_n,v})
							\end{align}
							Recall that $k=m_n-w_n$. Using the argument of the first inequality of Lemma 5 of \cite{zhou2010simultaneous} and condition (M1), with some calculations we can show that uniformly for all $u,v$,
							\begin{align}\label{f1}
								\sigma_{u,v}^Y=\E(S_{m_n,u}S_{m_n,v})/m_n=\sum_{s=1}^k\E S_{(s-1)w_n+1,w_n,u}S_{(s-1)w_n+1,w_n,v}/m_n+O(1/w_n)
							\end{align}
							We now study $\sigma^\upsilon_{u,v}$. First, using condition (i),  and similar argument of assertion (a) of Proposition \ref{definitioniota} we have $1\leq i\leq p-d$ and $1\leq j\leq p$,
							\begin{align}\label{fnew} \|\mf f_i^\top \mf e_j\|_{\mathcal L^{q^*}} \lessapprox 1.\end{align}
							Using \eqref{fnew} and condition (ii), via the triangle Cauchy inequality and Lemma 6 of \cite{zhou2013heteroscedasticity} and \eqref{claim1} we shall see that
							\begin{align}\label{f2-old}
								\|\sup_{u,v}|\sigma_{u,v}^\upsilon-\frac{1}{w_n(m_n-w_n+1)}\sum_{j=1}^{m_n-w_n+1} S_{j,w_n,u} S_{j,w_n,v}|\|_{q^*/2}=O(\sqrt{w_n/m_n}(W_{n,p})^{2/q^*}).
							\end{align}
							Again by \eqref{fnew} and condition (ii), using a similar argument to the proof of Lemma 1 of Zhou (2013), Cauchy-Schwartz inequality, and \eqref{claim1} we obtain that
							\begin{align}
								\bigg\|\max_{u,v}\bigg|\frac{1}{w_n(m_n-w_n+1)}\sum_{j=1}^{m_n-w_n+1} (S_{j,w_n,u} S_{j,w_n,v}-\E( S_{j,w_n,u} S_{j,w_n,v}))\bigg|\bigg\|_{\mathcal L^{q^*/2}}\label{Feb28-100-old}
								=O(\sqrt{w_n/m_n}W_{n,p}^{2/{q^*}})
							\end{align}
							Therefore 
							\begin{align}\label{f2}
								\|\sup_{u,v}|\sigma_{u,v}^\upsilon-\frac{1}{w_n(m_n-w_n+1)}\sum_{j=1}^{m_n-w_n+1} \E(S_{j,w_n,u} S_{j,w_n,v})|\|_{q^*/2}=O(\sqrt{w_n/m_n}(W_{n,p})^{2/q^*}).
							\end{align}
							We now study $\E(S_{j,w_n,u} S_{j,w_n,v})$. For this  aim, by condition (iii), it follows that for all $\mf c$ such that $|\mf c|=1$, $
							\mf c^\top \mf H(t,\FF_i)$ is locally stationary such that uniformly for $1\leq q\leq k$,
							\begin{align}
								\|\mf c^\top \mf H(t,\FF_i)-\mf c^\top \mf H(s,\FF_i)\|_{\mathcal L^2}
								=O(|t-s|).
							\end{align}
							Using this fact with Cauchy inequality,  triangle inequality and Lemma 6 of \cite{zhou2013heteroscedasticity} we shall see that
							\begin{align}\label{f3}
								V_{q,u,v}:=\sum_{j=(q-1)w_n+1}^{qw_n}\E S_{j,w_n,u} S_{j,w_n,v}=w_n\E S_{(q-1)w_n+1,w_n,u} S_{(q-1)w_n+1,w_n,v}+O(\sqrt{w_n}w^3_n/m_n)
							\end{align}
							
							Observe that $\sum_{j=1}^{m_n-w_n+1}\E(S_{j,w_n,u} S_{j,w_n,v})=\sum_{q=1}^{k-1}V_{q,u,v}+
							\E(S_{m_n-w_n+1,w_n,u} S_{m_n-w_n+1,w_n,v})$. Combining with \eqref{f2} and \eqref{f3} and the fact that $\E(S_{m_n-w_n+1,w_n,u} S_{m_n-w_n+1,w_n,v})=o(w_n)$, $w_n^2/m_n=o(1)$ we have that  
							\begin{align}\label{f2-1}
								\|\sup_{u,v}|\sigma_{u,v}^\upsilon-\frac{1}{(m_n-w_n+1)}\sum_{q=1}^{k-1} 
								\E S_{(q-1)w_n+1,w_n,u} S_{(q-1)w_n+1,w_n,v}
								|\|_{q^*/2}=O(\sqrt{w_n/m_n}(W_{n,p})^{2/q^*}+\frac{1}{m_n}).
							\end{align}
							
							Combining with \eqref{f1}, \eqref{S.94} follows.
							Step(ii).
							We now show \eqref{Junly15-S85}. It suffices to consider on the event $\{\tilde d_n=d\}$. We first show that for $\epsilon\in (0,\infty)$ \begin{align}\p(|\bs \upsilon_n-\bs \kappa_n|_{\infty}\geq \epsilon|\mf \FF_{n})=O_p((\epsilon^{-1}\bar{\theta}(n,p,\delta,l,N_n,w_n))^l)^l)\label{S111-Nov-29}.\end{align}
							After that we then show for $\epsilon\in (0,\infty)$ 
							\begin{align}
								\sup_{t\in \mathbb R}\p(|\bs \upsilon_n-t|
								\leq \epsilon|\FF_n)=O_p(\epsilon\sqrt{\log( n/\epsilon)} ).\label{S112-Nov-29}
							\end{align}
							Combining  \eqref{S111-Nov-29} and \eqref{S112-Nov-29}, and following the argument of \eqref{S95_Nov29} to \eqref{S97-Nov-21} in the main article, we have 
							\begin{align}\label{NewF39}
								\sup_{t\in \mathbb R }|\p(|\bs \upsilon_n|_\infty\leq t|\FF_n)-\p(|\bs \kappa_n|_\infty\leq t|\FF_n)|=O_p((\epsilon^{-1}	\bar{\theta}(n,p,\delta,l,N_n,w_n))^l+\epsilon\sqrt{\log( n/\epsilon)}).
							\end{align}
							Take $\epsilon=(	\bar{\theta}(n,p,\delta,l,N_n,w_n))^{l/(l+1)}\log^{-1/(2l+2)} n$, \eqref{Junly15-S85} follows.
							
							To show \eqref{S111-Nov-29} it suffices to prove that 
							\begin{align}
								\E\big(\big|\frac{1}{\sqrt{w_n(m_n-w_n+1)}}\sum_{j=1}^{m_n-w_n+1}( \hat{\mf s}_{j,w_n}- {\mf s}_{j,w_n})R_j\big|^l_\infty\big|\FF_n\big)^{1/l}=O_p(	\bar{\theta}(n,p,\delta,l,N_n,w_n)),\label{July15-S87}\\
								\E\big(\big|\frac{1}{\sqrt{w_n(m_n-w_n+1)}}\sum_{j=1}^{m_n-w_n+1}\frac{w_n}{m_n}( \hat{\mf s}_{m_n}- {\mf s}_{m_n})R_j\big|^l_\infty|\FF_n\big)^{1/l}=O_p(	\bar{\theta}(n,p,\delta,l,N_n,w_n)).\label{July15-S88}
							\end{align}
							We now show \eqref{July15-S87}, and \eqref{July15-S88} follows mutatis mutandis. Define $\hat S_{j,w_n,r}$ and $S_{j,w_n,r}$ as the $r_{th}$ component of the $N_n(p-d)$ dimensional vectors $\hat {\mf s}_{j,w_n}$ and $\mf s_{j,w_n}$. Using the notation of proof of Theorem \ref{Jan23-Thm4}, it follows that by triangle inequality,
							\begin{align}\label{July15-S89}
								&\big|\sum_{j=1}^{m_n-w_n+1}( \hat{\mf s}_{j,w_n}- {\mf s}_{j,w_n})R_j\big|_\infty\notag \\&=\max_{\substack{1\leq s_1\leq N_n,\\ 1\leq s_2\leq p-d}}\big|\sum_{j=1}^{m_n-w_n+1}R_j \sum_{i=j}^{j+w_n-1}(\hat{\mf f}_{s_2}^\top\hat{\mf e}_{i+(s_1-1)m_n}-\mf f_{s_2}^\top\mf e_{i+(s_1-1)m_n})\big|
								\notag\\&=\max_{\substack{1\leq s_1\leq N_n,\\ 1\leq s_2\leq p-d}}\Big|\sum_{j=1}^{m_n-w_n+1}R_j \sum_{i=j}^{j+w_n-1}\Big(\big(\hat{\mf f}_{s_2}-\mf f_{s_2}\big)^\top\hat{\mf e}_{i+(s_1-1)m_n}+\mf f_{s_2}^\top\big(\hat {\mf e}_{i+(s_1-1)m_n}-\mf e_{i+(s_1-1)m_n}\big)\Big)\Big|\notag\\
								&\leq \max_{\substack{1\leq s_1\leq N_n,\\ 1\leq s_2\leq p-d}} |I_{s_1,s_2}|+\max_{\substack{1\leq s_1\leq N_n,\\ 1\leq s_2\leq p-d}} |II_{s_1,s_2}|
							\end{align}
							where
							\begin{align}
								\label{Is1s2}
								I_{s_1,s_2}=\sum_{j=1}^{m_n-w_n+1}R_j \sum_{i=j}^{j+w_n-1}\big(\hat{\mf f}_{s_2}-\mf f_{s_2}\big)^\top\hat{\mf e}_{i+(s_1-1)m_n},\\
								II_{s_1,s_2}=\sum_{j=1}^{m_n-w_n+1}R_j \sum_{i=j}^{j+w_n-1}\mf f_{s_2}^\top\big(\hat {\mf e}_{i+(s_1-1)m_n}-\mf e_{i+(s_1-1)m_n}\big).
							\end{align}
							Furthermore, $|I_{s_1,s_2}|\leq |I_{s_1,s_2,1}|+|I_{s_1,s_2,2}|$ where
							\begin{align}
								I_{s_1,s_2,1}=\sum_{j=1}^{m_n-w_n+1}R_j \sum_{i=j}^{j+w_n-1}\big(\hat{\mf f}_{s_2}-\mf f_{s_2}\big)^\top(\hat{\mf e}_{i+(s_1-1)m_n}-{\mf e}_{i+(s_1-1)m_n}),\\
								I_{s_1,s_2,2}=\sum_{j=1}^{m_n-w_n+1}R_j \sum_{i=j}^{j+w_n-1}\big(\hat{\mf f}_{s_2}-\mf f_{s_2}\big)^\top{\mf e}_{i+(s_1-1)m_n}.
							\end{align}
							Let $M$ be a generic sufficiently large constant which varies from line to line. Then by Jansen's inequality and the triangle inequality, it is not hard to verify that
							the LHS of \eqref{July15-S87} is bounded by
							\begin{align*}
								\frac{M}{\sqrt{w_n(m_n-w_n+1)}}\Big(\E\max_{\substack{1\leq s_1\leq N_n,\\ 1\leq s_2\leq p-d}}(|I_{s_1,s_2,1}|^l|\FF_n)^{1/l}+	\E(\max_{\substack{1\leq s_1\leq N_n,\\ 1\leq s_2\leq p-d}}|I_{s_1,s_2,2}|^l|\FF_n)^{1/l}+	\E(\max_{\substack{1\leq s_1\leq N_n,\\ 1\leq s_2\leq p-d}}|II_{s_1,s_2}|^l|\FF_n)^{1/l}\Big)
							\end{align*}
							for some sufficiently large constant $M$.
							Notice that given $(\mf \FF_{n})$, $I_{s_1,s_2,1}$, $I_{s_1,s_2,2}$ and $II_{s_1,s_2}$ are Gaussian random variables, therefore  using the property of normal random variables, with probability tending to $1$, the above equation will be bounded by
							\begin{align}
								M\Big(\max_{\substack{1\leq s_1\leq N_n,\\ 1\leq s_2\leq p-d}}var^{1/2}(I_{s_1,s_2,1}|\FF_n)+	\max_{\substack{1\leq s_1\leq N_n,\\ 1\leq s_2\leq p-d}}var^{1/2}(I_{s_1,s_2,2}|\FF_n)+	\max_{\substack{1\leq s_1\leq N_n,\\ 1\leq s_2\leq p-d}}var^{1/2}(II_{s_1,s_2}|\FF_n)\Big)\notag\\\times (w_n(m_n-w_n+1))^{-1/2}(N_n(p-d))^{1/l}
							\end{align}
							where we have used \eqref{claim1}.
							First, by Jansen's inequality 
							\begin{align}\label{C.77}
								var(I_{s_1,s_2,1}|\FF_n)=\sum_{j=1}^{m_n-w_n+1}\Big( \sum_{i=j}^{j+w_n-1}\big(\hat{\mf f}_{s_2}-\mf f_{s_2}\big)^\top(\hat{\mf e}_{i+(s_1-1)m_n}-\mf e_{i+(s-1)m_n})\Big)^2\notag\\
								\leq \sum_{j=1}^{m_n-w_n+1}w_n\sum_{i=j}^{j+w_n-1}\big(\big(\hat{\mf f}_{s_2}-\mf f_{s_2}\big)^\top(\hat{\mf e}_{i+(s_1-1)m_n}-\mf e_{i+(s_1-1)m_n})\big)^2.
							\end{align}

							Further by Cauchy-Schwartz inequality,  notice that 
							\begin{align}
								\max_{\substack{1\leq s_1\leq N_n,\\ 1\leq s_2\leq p-d}}\sum_{j=1}^{m_n-w_n+1}\sum_{i=j}^{j+w_n-1}\big(\big(\hat{\mf f}_{s_2}-\mf f_{s_2}\big)^\top(\hat{\mf e}_{i+(s_1-1)m_n}-\mf e_{i+(s_1-1)m_n})\big)^2
								\notag\\\leq \|\hat{\mf F}-\mf F\|^2_F\max_{\substack{1\leq s_1\leq N_n}}\sum_{j=1}^{m_n-w_n+1}\sum_{i=j}^{j+w_n-1}\big\|\hat{\mf e}_{i+(s_1-1)m_n}-\mf e_{i+(s_1-1)m_n}\big\|_F^2.
							\end{align}
							
							Notice that $\hat {\mf e}_i-{\mf e}_i=\mf A(i/n)\mf z_i-\tilde{\mf V}(i/n)\tilde {\mf V}^\top(i/n)\mf x_i$. Write $i'$ for $(s_1-1)m_n+i$ for short. Thus by the proof of (ii) of Theorem \ref{Space-Distance}, 
							\begin{align}
								\sum_{j=1}^{m_n-w_n+1}\sum_{i=j}^{j+w_n-1}\big\|\hat{\mf e}_{i+(s_1-1)m_n}-\mf e_{i+(s_1-1)m_n}\big\|_F^2\leq \notag\\ 3\sum_{j=1}^{m_n-w_n+1}\sum_{i=j}^{j+w_n-1}(\|I_1(i'/n)\|_2^2+\|I_2(i'/n)\|_2^2+\|\tilde {\mf V}(i'/n)\tilde{\mf  V}^\top(i'/n)\mf e_{i',n}\|_2^2),
							\end{align}
							where we have used the fact that $\|\mf v\|_F=\|\mf v\|_2$ for any vector $\mf v$, and $I_1(\cdot)$ and $I_2(\cdot)$ are defined in the proof of (ii) of Theorem \ref{Space-Distance}.
							Write $\tilde {\mf \Delta}_i=\tilde {\mf V}(i/n)\hat{\mf O}_1(i/n)-\mf V^\top (i/n)$ where the rotation matrix $\hat{\mf O}_1(t)$ is also defined in Theorem \ref{Space-Distance}. Thus by definition, we have for all possible $s_1$ (or $i'$)
							\begin{align}\label{evaluateI}
								\sum_{j=1}^{m_n-w_n+1}\sum_{i=j}^{j+w_n-1}\|I_1(i'/n)\|^2_2\leq 
								\sum_{j=1}^{m_n-w_n+1}\sum_{i=j}^{j+w_n-1}\|\tilde{\mf \Delta}_{i'} \|_2^2\|\mf A(i'/n)\|_2^2|\mf z_{i'n}|^2\\\notag\leq O_p(p^{1-\delta}\theta^2(n,p))\sum_{j=1}^{m_n-w_n+1}\sum_{i=j}^{j+w_n-1}|\mf z_{i',n}|^2
							\end{align}
							where  for the second inequality we have used Theorem \ref{Space-Distance} with $\eta\equiv 1$ and $\delta\geq 0$.
							By  Jansen's inequality we have for $1\leq s_1\leq N_n$
							\begin{align}\label{GettingZi}
								\Big|\sum_{j=1}^{m_n-w_n+1}\sum_{i=j}^{j+w_n-1}|\mf z_{i'n}|^2\Big|^\frac{l}{2}
								\leq 
								(m_n-w_n+1)^{l/2-1}w_n^{l/2-1}\sum_{j=1}^{m_n-w_n+1}\sum_{i=j}^{j+w_n-1}|\mf z_{i,n}|^{l}
							\end{align}
							by (M2'), the above inequality implies that $$\|\sum_{j=1}^{m_n-w_n+1}\sum_{i=j}^{j+w_n-1}|\mf z_{i'n}|^2\|_{\mathcal L^\frac{l}{2}}\lessapprox (m_n-w_n+1)w_n.$$ Using \eqref{claim1} and \eqref{evaluateI}, we have that
							\begin{align}
								\max_{1\le s_1\le N_n}\sum_{j=1}^{m_n-w_n+1}\sum_{i=j}^{j+w_n-1}\|I_1(\frac{(s_1-1)m_n+i}{n})\|^2_2=O_p( N_n^{2/l} (m_n-w_n+1)w_n p^{1-\delta}\theta^2(n,p)).
							\end{align} 
							Similarly 
							\begin{align}
								\max_{1\le s_1\le N_n}\sum_{j=1}^{m_n-w_n+1}\sum_{i=j}^{j+w_n-1}\|I_2(\frac{(s_1-1)m_n+i}{n})\|^2_2=O_p( N_n^{2/l} (m_n-w_n+1)w_n p^{1-\delta}\theta^2(n,p)).
							\end{align} 
							For $\|\tilde {\mf V}(i/n)\tilde{\mf  V}^\top(i/n)\mf e_{i,n}\|_2$, first notice that 
							while for $1\leq s_1\leq N_n$
							\begin{align}
								\Big	\|\sum_{j=1}^{m_n-w_n+1}\sum_{i=j}^{j+w_n-1}\|{\mf e}_{i+(s_1-1)m_n}\|_F^2\Big \|_{\mathcal L^{l/2}}
								\leq \Big([(m_n-w_n+1)w_n]^{l/2-1} \sum_{j=1}^{m_n-w_n+1}\sum_{i=j}^{j+w_n-1}\E (\|\mf e_{i+(s_1-1)m_n}\|_F ^{l})\Big)^{2/l}\label{e_norm_0}
							\end{align}
							Conditions (M2') and (M4) yield that
							\begin{align}\label{e_norm}
								\|\|{\mf e}_{i+(s_1-1)m_n}\|_F\|_{\mathcal L^l}=O(p^{1/2}).
							\end{align}
							Then following \eqref{New.S16}, we have that
							\begin{align}
								\max_{1\le s_1\le N_n}\sum_{j=1}^{m_n-w_n+1}\sum_{i=j}^{j+w_n-1}\|\tilde {\mf V}(i'/n)\tilde{\mf  V}^\top(i'/n)\mf e_{i',n}\|^2_2\notag\\\leq 
								2	\max_{1\le s_1\le N_n}\sum_{j=1}^{m_n-w_n+1}\sum_{i=j}^{j+w_n-1}(\|\mf \Delta_{i'}\|^2_2|\mf e_{i',n}|^2_2+ (\sum_{s=1}^{d(i'/n)} ({\mf v}_s^\top(i'/n)\mf e_{i',n})^2))\notag\\
								\leq 2 \theta^2(n,p)	\max_{1\le s_1\le N_n}\sum_{j=1}^{m_n-w_n+1}\sum_{i=j}^{j+w_n-1}|\mf e_{i',n}|^2+
								\max_{1\le s_1\le N_n}\sum_{j=1}^{m_n-w_n+1}\sum_{i=j}^{j+w_n-1}(\sum_{s=1}^{d(i'/n)} ({\mf v}_s^\top(i'/n)\mf e_{i',n})^2))\notag\\
								=O_p(w_n(m_n-w_n+1)(p^{}\theta^2(n,p)+1)N_n^{2/l})
							\end{align} 
							where we have used 
							the fact that \begin{align}
								\|(\sum_{s=1}^{d(i/n)} ({\mf v}_s^\top(i/n)\mf e_{i,n})^2)^{1/2}\|_{\mathcal L^l}^2\leq \sum_{s=1}^d\|{\mf v}_s^\top(i/n)\mf e_{i,n}\|_{\mathcal L^l}^2=O(1) 
							\end{align}
								Therefore
								\begin{align}\label{summmarize1}
									\max_{\substack{1\leq s_1\leq N_n,\\ 1\leq s_2\leq p-d}}	var^{1/2}(I_{s_1,s_2,1}|\FF_n)=O_p(w_n(m_n-w_n+1)^{1/2}(p^{1/2}\theta(n,p)+1)\theta(n,p)N^{1/l}_n).
								\end{align}
								Similarly to \eqref{C.77}, 
								\begin{align}
									&	\max_{\substack{1\leq s_1\leq N_n,\\ 1\leq s_2\leq p-d}}var(I_{s_1,s_2,2}|\FF_n)\notag\\&\leq 
									w_n\max_{\substack{1\leq s_1\leq N_n,\\ 1\leq s_2\leq p-d}}\sum_{j=1}^{m_n-w_n+1}\sum_{i=j}^{j+w_n-1}\Big(\big(\hat{\mf f}_{s_2}-\mf f_{s_2}\big)^\top{\mf e}_{i+(s_1-1)m_n}\Big)^2.
								\end{align}
								And \begin{align}
									&\max_{\substack{1\leq s_1\leq N_n,\\ 1\leq s_2\leq p-d}}\sum_{j=1}^{m_n-w_n+1}\sum_{i=j}^{j+w_n-1}\Big(\big(\hat{\mf f}_{s_2}-\mf f_{s_2}\big)^\top{\mf e}_{i+(s_1-1)m_n}\Big)^2\notag\\&\leq 
									\max_{1\leq s_1\leq N_n}\|\hat {\mf F}-\mf F\|_F^2\sum_{j=1}^{m_n-w_n+1}\sum_{i=j}^{j+w_n-1}\|{\mf e}_{i+(s_1-1)m_n}\|_F^2
								\end{align}
								
								Combining with \eqref{e_norm_0},\eqref{e_norm} and \eqref{claim1}, 
								we  have  
								\begin{align}\label{summarize2}
									\max_{\substack{1\leq s_1\leq N_n,\\ 1\leq s_2\leq p-d}}	var^{1/2}(I_{s_1,s_2,2}|\FF_n)=O_p(w_n(m_n-w_n+1)^{1/2}p^{1/2}\theta(n,p)N^{1/l}_n).
								\end{align}
								Finally, 
								\begin{align}\label{C88}
									\max_{\substack{1\leq s_1\leq N_n,\\ 1\leq s_2\leq p-d}}	var(II_{s_1,s_2}|\FF_n)=\max_{\substack{1\leq s_1\leq N_n,\\ 1\leq s_2\leq p-d}}\sum_{j=1}^{m_n-w_n+1}\Big(\sum_{i=j}^{j+w_n-1}\mf f_{s_2}^\top\big(\hat {\mf e}_{i+(s_1-1)m_n}-\mf e_{i+(s_1-1)m_n}\big)\Big)^2
									\notag \\\leq  \max_{\substack{1\leq s_1\leq N_n,\\ 1\leq s_2\leq p-d}}
									w_n\sum_{j=1}^{m_n-w_n+1}\sum_{i=j}^{j+w_n-1}\|\mf f_{s_2}^\top\big(\hat {\mf e}_{i+(s_1-1)m_n}-\mf e_{i+(s_1-1)m_n}\big)\|_F^2
								\end{align}
								Notice  the summands $\|\mf f_{s_2}^\top\big(\hat {\mf e}_{i+(s_1-1)m_n}-\mf e_{i+(s_1-1)m_n}\big)\|_F^2$ can be written as
								\begin{align}
									\|\mf f_{s_2}^\top(\tilde{\mf V}(i'/n)\tilde{\mf V}^\top(i'/n)\mf x_{i',n}-\mf A\mf z_{i',n})\|_F^2=\|\mf f_{s_2}^\top\tilde{\mf V}(i'/n)\tilde{\mf V}^\top(i'/n)\mf x_{i',n}\|_F^2
								\end{align}
								for all $1\leq s_1\leq p-d$, where we have used the fact that $\mf A(i/n)\equiv \mf A$ under null hypothesis, and we write $i'=i+(s_1-1)m_n$ for short as we did in the evaluation of $I_{s_1,s_2,1}$. By Theorem \ref{Space-Distance} in the main article, for each $i$, there exists orthonormal matrices $\hat{\mathbf O}_1(i/n)$ such that
								\begin{align}\label{thetanu}
									\|\max_i\|\tilde {\mf V}(i/n) \hat{\mathbf O}_1(i/n)-\mf V\|_F\|_{\mathcal L^1}=O(\theta(n,p)).
								\end{align}
								where as before we also use the fact that under null hypothesis $\mf V(i/n)\equiv \mf V$. By definition,  there exists another $d\times d$ orthonormal matrices $\tilde{\mf O}$ such that 
								$\mf f_{s_1}^\top \mf V \tilde {\mf O}=0$. As a result,
								\begin{align}
									&	\mf f_{s_2}^\top\tilde{\mf V}(i'/n)\tilde{\mf V}^\top(i'/n)\mf x_{i',n}-0\notag\\&=\mf f_{s_2}^\top\tilde{\mf V}(i'/n)\hat{\mf O}_1(i'/n)\hat{\mf O}_1(i'/n)^\top \tilde{\mf V}^\top(i'/n)\mf x_{i',n}-
									\mf f_{s_2}^\top{\mf V}\tilde{\mf O}\tilde{\mf O}^\top {\mf V}^\top\mf x_{i',n}\notag\\
									&=\mf f_{s_2}^\top\tilde{\mf V}(i'/n)\hat{\mf O}_1(i'/n)\hat{\mf O}_1(i'/n)^\top \tilde{\mf V}^\top(i'/n)\mf x_{i',n}-
									\mf f_{s_2}^\top{\mf V} {\mf V}^\top\mf x_{i',n}
									\notag\\&=\mf f_{s_2}^\top \mf V\mf \Delta_{i'}^\top\mf x_{i'n}+\mf f_{s_2}^\top \mf \Delta_{i'}(\mf V+\mf \Delta_{i'})^\top \mf x_{i'n}
								\end{align} 
								where $
								\mf \Delta_{i'}=\tilde {\mf V}(i'/n)\hat{\mf O}_1(i'/n)-\mf V
								$. Notice that by \eqref{thetanu}, $\max_{1\leq i\leq n}\|\mf \Delta_i\|_F=O_p(\theta(n,p))$
								Together with \eqref{C88}, it follows that $\max_{\substack{1\leq s_1\leq N_n,\\ 1\leq s_2\leq p-d}}	var(II_{s_1,s_2}|\FF_n)$ is bounded by
								\begin{align}\label{C92}
									\max_{\substack{1\leq s_1\leq N_n,\\ 1\leq s_2\leq p-d}}
									w_n\sum_{j=1}^{m_n-w_n+1}\sum_{i=j}^{j+w_n-1}\|\mf f_{s_2}^\top \mf V\mf \Delta_{i'}^\top\mf x_{i'n}+\mf f_{s_2}^\top \mf \Delta_{i'}(\mf V+\mf \Delta_{i'})^\top \mf x_{i'n}\|_F^2
									\notag\\
									\leq  \max_{\substack{1\leq s_1\leq N_n,\\ 1\leq s_2\leq p-d}}
									2w_n\sum_{j=1}^{m_n-w_n+1}\sum_{i=j}^{j+w_n-1}\Big(\|\mf f_{s_2}^\top \mf V\mf \Delta_{i'}^\top\mf x_{i'n}\|_F^2+\|\mf f_{s_2}^\top \mf \Delta_{i'}(\mf V+\mf \Delta_{i'})^\top \mf x_{i'n}\|_F^2\Big)
									\notag\\\leq C_1\max_{1\leq s_1\leq N_n}
									2w_n\sum_{j=1}^{m_n-w_n+1}\sum_{i=j}^{j+w_n-1}\Big(\|\mf \Delta_{i'}^\top\mf x_{i'n}\|_F^2+\| \mf \Delta_{i'}(\mf V+\mf \Delta_{i'})^\top \mf x_{i'n}\|_F^2\Big)
									\notag\\\leq C_2(\max_{1\leq s_1\leq N_n}\|\mf \Delta_{i'}\|_F^2)\max_{1\leq s_1\leq N_n}
									2w_n\sum_{j=1}^{m_n-w_n+1}\sum_{i=j}^{j+w_n-1}\|\mf x_{i+(s_1-1)m_n}\|_F^2\notag\\=
									O_p(\theta^2(n,p))\Big(\max_{1\leq s_1\leq N_n}
									w_n\sum_{j=1}^{m_n-w_n+1}\sum_{i=j}^{j+w_n-1}\|\mf x_{i+(s_1-1)m_n}\|_F^2\Big).
								\end{align}
								for some large constants $C_1$ and $C_2$,
								where we have used $\|\mf V\|_F= d$.
								Notice that
								\begin{align*}
									\|\mf x_{i,n}\|_F\leq \|\mf Az_{i,n}\|_F+\|e_{i,n}\|_F\leq \|\mf A\|_F|\mf z_{i,n}|+|\mf e_{i,n}|.
								\end{align*}
								Using (A2), (M2') and (M4) we have that 	$\|\|\mf x_{i,n}\|_F\|_{\mathcal L^{l}}=O(p^{1/2})$. Using similar argument to \eqref{GettingZi} we can verify that 
								\begin{align}\label{New.C84}
									\max_{1\leq s_1\leq N_n}
									\sum_{j=1}^{m_n-w_n+1}\sum_{i=j}^{j+w_n-1}\|\mf x_{i+(s_1-1)m_n}\|_F^2=O_p(N_n^{2/l}w_n(m_n-w_n+1)p)
								\end{align}
								Together with \eqref{C92} we have
								\begin{align}\label{summarize3}
									\max_{\substack{1\leq s_1\leq N_n,\\ 1\leq s_2\leq p-d}}var(II_{s_1,s_2}|\FF_n)^{1/2}=O_p(w_n(m_n-w_n+1)^{1/2}N_n^{1/l}p^{1/2}\theta(n,p)).
								\end{align}
								The summarizing \eqref{summmarize1}, \eqref{summarize2} and \eqref{summarize3},  shall see that \eqref{July15-S87} holds.

								\hfill $\Box$

								\section{Proof of Theorem \ref{Power}}\label{Proof-Power}\setcounter{equation}{0}
								
								In th following, we write $\mf I$ for $\mf I_p$ for simplicity. Under the local alternative, since $\mf A(t)=\mf A+\rho_n\mf D(t)$, $\mf \Gamma$, $\tilde {\mf \Gamma}$ (which is defined in the beginning of Section \ref{Sec3proof}), $\mf\Lambda_1(t)$ and $\mf \Lambda(t)$ are functions of $\rho_n$  and therefore are denoted by $\mf \Gamma(\rho_n)$, $\tilde {\mf \Gamma}(\rho_n)$, $\mf\Lambda_1(t,\rho_n)$ and $\mf \Lambda(t,\rho_n)$, respectively. For simplicity we write $\mf \Gamma$, $\tilde {\mf \Gamma}$, $\mf\Lambda_1(t)$ and $\mf \Lambda(t)$ for $\mf \Gamma(0)$, $\tilde {\mf \Gamma}(0)$, $\mf\Lambda_1(t,0)$ and $\mf \Lambda(t,0)$. In the proof we prove  local alternatives with factor strength $\delta$: \begin{align}H_A: \mf A(t)=\mf A_n(t):=\mf A+\rho_n\mf D(t),\end{align}
								where $\rho_n=O(1)$, $\mf D(t)=(d_{ij}(t))$ is a $p\times d$ matrix satisfying (A1) and (A2') with $\eta_n\equiv 1$.
								
								
								
								\medskip
								\begin{proposition} \label{PropG1} Under conditions of Theorem \ref{Power}, if $\rho_n=O(1)$ there  exists an orthogonal basis of  null space of $\mf \Gamma(\rho_n)$, which is $\tilde {\mf F}_n=(\tilde{\mf f}_{1,n},...,\tilde{\mf f}_{p-d,n})$, such that 
									\begin{align}\label{S102}
										\|\|\hat{\mf F}-\tilde{\mf F}_n\|_F\|_{\mathcal L^1}=O(p^{\delta}/\sqrt{n}),
									\end{align}
									Furthermore, the exists a set of basis of null space of $\mf A$ which is  $\mf F=(\mf f_{1},...,\mf f_{p-d})$, such that
									\begin{align}\label{S103}
										\|\tilde {\mf F}_n-\mf F\|_F=O(\rho_n+p^{\delta-1}).
									\end{align}
									under (S1'), and $O(\rho_n)$ under (S1). 
								\end{proposition}
								
								We stress that that $\tilde {\mf f}_{i,n}s$ defined here are not the basis of $\tilde{\mf \Gamma}(\rho_n)$.

								{\it Proof.}
								By Corollary \ref{Jan-Corol1}, we have 
								\begin{align}\label{New.D3}
									\|\|\hat{\mf \Gamma}(\rho_n)-\mf \Gamma(\rho_n)\|_F\|_{\mathcal L^1}=O(\frac{p^{2-\delta}}{\sqrt n}).
								\end{align}
								Under (S1), $\bs \Gamma(\rho_n)=\tilde{\bs \Gamma}(\rho_n)$.
								Under (S1'), 
								\begin{align}\label{new.D4}
									\|\mf \Gamma(\rho_n)-\tilde{\mf \Gamma}(\rho_n)\|_2=O(p^{1-\delta}).
								\end{align}
								Meanwhile, elementary calculations show that
								\begin{align}
									\|\tilde{\mf \Gamma}(\rho_n)-\tilde{\mf \Gamma}(0) \|_2=O(\rho_np^{2(1-\delta)}).
								\end{align}
								Hence by the triangle inequality under (S1')
								\begin{align}\label{new.D7}
									\|{\mf \Gamma}(\rho_n)-\tilde{\mf \Gamma}(0) \|_2=O(\rho_np^{2(1-\delta)}+p^{1-\delta})
								\end{align}
								while the bound in the RHS of the above equation is reduced to $\rho_np^{2(1-\delta)}$ under (S1).
								Observe that $\tilde{\mf \Gamma}(0)$ is $\tilde{\mf \Gamma}$ in \eqref{new.c9}.
								By the proof of Corollary \ref{Jan20-Corol2},   \begin{align}\label{new.D5}\lambda_d(\tilde{\mf \Gamma}(0))\gtrapprox(p^{2-2\delta})\end{align}
								Then by Theorem 2 of \cite{yu2015useful},  the fact that the  null space of $\mf A$ is the null space of $\tilde {\mf \Gamma}(0)$,  \eqref{new.D7} and the similar argument to the proof of Corollary \ref{Corol4}, \eqref{S103} holds.

								Furthermore, by \eqref{new.D7} and \eqref{new.D5}, it follows that
								$\lambda_d({\mf \Gamma}(\rho_n))\gtrapprox(p^{2-2\delta})$. Then similarly by Theorem 2 of \cite{yu2015useful},  \eqref{New.D3} and the similar argument to the proof of Corollary \ref{Corol4}, \eqref{S102} holds.\hfill  $\Box$
								\begin{corollary}\label{CoroD1}
									Under the conditions  of Proposition \ref{PropG1}, 
									$$\p(\tilde d_n \neq d)=O\Big(\theta(n,p)\Big)+O(\log^{-1/2} n)=o(1).$$
								\end{corollary}
								{\it Proof.} 
								By \eqref{Final1} in the proof of  Theorem \ref{Thm-approx} we shall see that
								\begin{align}\label{D3}
									\|\sup_{t\in[0,1]}\| (\hat{\mf \Lambda}(t,\rho_n)-\mf \Lambda(t,\rho_n))\|_2\|_{\mathcal L^1}=O(p^{2-\delta}\nu_n).
								\end{align}
								Notice that under the local alternative, $\mf A(t)=\mf A+\rho_n\mf D(t)$. Define
								\begin{align}
									\bs \Sigma^\circ(t,k)=\bs \Sigma_z(t,k)\mf A^\top(t)+\bs \Sigma_{ze}(t,k),\quad
									\bs\Sigma^\diamond(t,k_0)=\sum_{k=1}^{k_0} \bs \Sigma^\circ(t,k)
									(\bs \Sigma^\circ(t,k))^\top. 
								\end{align} As a  consequence,
								\begin{align}
									\mf \Lambda(t,\rho_n)=(\mf A+\rho_n\mf D(t))\bs\Sigma^\diamond(t,k_0)(\mf A+\rho_n\mf D(t))^\top+\sum_{k=1}^{k_0}\mf A(t)\bs \Sigma^\circ(t,k)\bs \Sigma_e(t,k)^\top\notag\\+\sum_{k=1}^{k_0}\bs \Sigma_e(t,k)(\bs \Sigma^\circ(t,k))^\top\mf A^\top(t)+\sum_{k=1}^{k_0}\bs \Sigma_e(t,k)\bs \Sigma_e(t,k)^\top\notag\\
									:=\mf \Lambda_1(t,\rho_n)+\mf A(t)\sum_{k=1}^{k_0}\bs \Sigma^\circ(t,k)\bs \Sigma_e(t,k)^\top+\sum_{k=1}^{k_0}\bs \Sigma_e(t,k)(\bs \Sigma^\circ(t,k))^\top\mf A^\top(t)+\sum_{k=1}^{k_0} \bs \Sigma_e(t,k)\bs \Sigma_e(t,k)^\top.
								\end{align} 
								
								Therefore by \eqref{D3}  under (S1'), \begin{align}
									\|\sup_{t\in[0,1]}\| (\hat{\mf \Lambda}(t,\rho_n)-\mf \Lambda_1(t,\rho_n))\|_2\|_{\mathcal L^1}=O(p^{2-\delta}\nu_n+p^{1-\delta}).
								\end{align}
								If under (S1) the in the above estimate the term $p^{1-\delta}$ will varnish.
									Notice that by \eqref{lambdad} in the proof  of Theorem \ref{Space-Distance} \begin{align} \inf_{t\in [0,1]}\lambda_{d}(\mf \Lambda_1(t,\rho_n)) \gtrapprox  p^{2-2\delta}.\end{align} Also 
									$\lambda_{d+1}(\mf \Lambda_1(t,\rho_n))=0$.
									The Corollary follows exactly the proof of Proposition \ref{tildedrate}. \hfill $\Box$
									
									\medskip
									In the following, write 
									$\tilde{\mf f}_{i,n}$ as  $\tilde{\mf f}_{i}$ for short, 
									where $\tilde{\mf f}_{i,n}$ is defined in Proposition \ref{PropG1}.
									Define for $1\leq s\leq N_n$ and $1\leq j\leq m_n$,
									\begin{align}
										\tilde{\bs l}_{j,s}=\Big(\tilde {\mf f}_1^\top\hat{\mf e}_{j+(s-1)m_n},...,\tilde {\mf f}_{p- d}^\top 
										\hat {\mf e}_{j+(s-1)m_n}\Big)^\top,\\ \tilde{\bs l}^A_{j,s}=\Big(\tilde {\mf f}_1^\top(\mf I-\mf V_{j+(s-1)m_n}\mf V_{j+(s-1)m_n}^\top){\mf e}_{j+(s-1)m_n},...\notag\\,...,\tilde {\mf f}_{p- d}^\top 
										(\mf I-\mf V_{j+(s-1)m_n}\mf V_{j+(s-1)m_n}^\top) {\mf e}_{j+(s-1)m_n}\Big)^\top. 
									\end{align}
									where $\mf V_i$ is any $p\times d$ matrix with each column eigenvectors of kernel space of $\mf A(i/n)$, such that $\mf V_i^\top \mf V_i=\mf I_d$. Notice that $\mf V_i\mf V_i^\top$ is uniquely defined.  
									Further define 
									\begin{align}\label{tildell}
										\tilde {\bs l}_i=(	\tilde{\bs l}^\top_{i,1},...,	\tilde{\bs l}^\top_{i,N_n})^\top,
										\tilde {\bs l}^A_i=(	\tilde{\bs l}^{A,\top}_{i,1},...,	\tilde{\bs l}^{A,\top}_{i,N_n})^\top
									\end{align}
									for $1\leq i\leq m_n$, and that $
									{\tilde s}_{j,w_n}=\sum_{r=j}^{j+w_n-1}\tilde {\bs l}_r$, $
									\tilde {\mf s}_{m_n}=\sum_{r=1}^{m_n}\tilde {\bs l}_r$,
									$
									{\tilde {\mf s}}^A_{j,w_n}=\sum_{r=j}^{j+w_n-1}\tilde {\bs l}^A_r$ and $
									\tilde {\mf s}^A_{m_n}=\sum_{r=1}^{m_n}\tilde {\bs l}^A_r$
									for $1\leq j\leq m_n$ where $w_n=o(m_n)$ and $w_n\rightarrow \infty$ is the window size. Define 
									\begin{align}
										\tilde	 {\bs \kappa}_n=\frac{1}{\sqrt{w_n(m_n-w_n+1)}}\sum_{j=1}^{m_n-w_n+1}(\tilde {\mf s}_{j,w_n}-\frac{w_n}{m_n}\tilde {\mf s}_{m_n})R_j,\\
										\tilde	 {\bs \kappa}^A_n=\frac{1}{\sqrt{w_n(m_n-w_n+1)}}\sum_{j=1}^{m_n-w_n+1}(\tilde {\mf s}^A_{j,w_n}-\frac{w_n}{m_n}\tilde {\mf s}^A_{m_n})R_j
									\end{align}
									where $\{R_i\}_{i\in\mathbb Z}$ are $i.i.d.$ $N(0,1)$ independent of $\{ {\mf x}_{i,n},1\leq i\leq n\}$. 
									
									
									\begin{proposition}\label{PropD2}
										Under the conditions of  Proposition \ref{PropG1}, we have
										\begin{description}
											\item (a) (i) $\E (|\tilde \kappa_n-\tilde \kappa_n^A|_\infty|\FF_n)=O_p(\bar{\theta}(n,p,\delta,l,N_n,w_n))$, 
											and (ii) $\E (|\tilde \kappa_n-\kappa_n|_\infty|\FF_n)=O_p(\bar{\theta}(n,p,\delta,l,N_n,w_n))$. 
										\end{description}If further assume that $\max_{i,q}|\tilde{\mf f}^\top_q\mf V_i|\leq \eta<1$ we have the following (b) and (c).
										\begin{description}
											\item (b) Let $\mf y^A_i$ be Gaussian vector with the same auto-covariance of $\tilde{\mf l}^A_i$, and 
											$\mf y=m_n^{-1/2}\sum_{i=1}^{m_n}\mf y^A_i$. Let $\mf T^A=m_n^{-1/2}\sum_{i=1}^{m_n}\mf l_i^A$, then
											\begin{align}
												\sup_{x\in \mathbb R}|\p(|\mf y^A|_\infty<x)-\p(|\mf  T^A|_\infty<x)|\lessapprox \upsilon(m_n, N_n ,p,d,l)),
											\end{align}
											where $\upsilon(m_n, N_n ,p,d,l))$ is defined in Proposition \ref{definitioniota}.
											\item (c) 
											\begin{align}
												\sup_{x\in \mathbb R}|\p(|\mf y^A|_\infty<x)-\p(|\tilde {\bs \kappa}_n  |_\infty<x|\FF_n)|=O_p(\Theta_n^{1/3}\log ^{2/3}(\frac{W_{n,p}}{\Theta_n})).
											\end{align}
										\end{description}
									\end{proposition}
									{\it Proof.}
									We first show  (a). We start by evaluating $\E (|\tilde {\bs \kappa}_n-\tilde{\bs\kappa}^A_n|_\infty|\FF_n)$. Notice that 
									\begin{align*}
										\tilde {\bs\kappa}_n-\tilde{\bs \kappa}_n^A=|I-II|,
									\end{align*}
									where
									\begin{align}
										I=	\frac{1}{\sqrt{w_n(m_n-w_n+1)}}\sum_{j=1}^{m_n-w_n+1}(\tilde {\mf s}_{j,w_n}-\tilde{\mf s}_{j,w_n}^A)R_j,\\
										II=\frac{1}{\sqrt{w_n(m_n-w_n+1)}}\sum_{j=1}^{m_n-w_n+1}(\frac{w_n}{m_n}\tilde{\mf s}_{m_n}-\frac{w_n}{m_n}\tilde {\mf s}_{m_n}^A)R_j.
									\end{align}
									We now show that $(\E(|I|^l_\infty)|\FF_n)^{1/l}=O_p(\bar{\theta}(n,p,\delta,l,N_n,w_n))$, and the result that $\E(|II|^l_\infty|
									\mf x_{i,n})^{1/l}=O_p(\bar{\theta}(n,p,\delta,l,N_n,w_n))$ will follow similarly. Thus (a) will follow from the triangle inequality.
									Observe that given data, $I$ is a $N_n(p-d)$ dimensional Gaussian vector. Therefore, 
									\begin{align}\label{D.20}
										\E(|I|^l_\infty|\FF_n)^{1/l}=O_p(\frac{1}{\sqrt{w_n(m_n-w_n+1)}}|\sum_{j=1}^{m_n-w_n+1}(\tilde{\mf s}^A_{j,w_n}-\tilde{\mf s}_{j,w_n})^{\circ 2}|^{1/2}_\infty (N_np)^{1/l})
									\end{align}
									where $\circ$ represents the Hadamard product, and $\mf A^{\circ 2}=\mf A\circ \mf A$.
									On the other hand for $1\leq q\leq p-d$, $1\leq j\leq N_n$, we have that (we write $j'$ for $j+(s-1)m_n$ to shorten the notation)
									\begin{align}
										\tilde{\mf f}_q^\top\hat{\mf e}_{j'}&=
										\tilde{\mf f}_q^\top \mf e_{j'}+\tilde{\mf f}_q^\top(\hat{ \mf e}_{j'}-\mf e_{j'})\notag\\&=
										\tilde{\mf f}_q^\top \mf e_{j'}+\tilde{\mf f}_q^\top(\mf A(j'/n){ \mf z}_{j'}-\mf V_{j'}\mf V_j^\top \mf x_{j',n})+\tilde{\mf f}_q^\top (\mf V_{j'}\mf V_{j'}^\top \mf x_{j',n}-\tilde {\mf V}_{j'}\tilde{\mf V}_{j'}^\top \mf x_{j'n})\notag\\
										&=\tilde{\mf f}_q^\top (\mf I-\mf V_{j'}\mf V_{j'}^\top)\mf e_{j'}+
										\tilde{\mf f}_q^\top(\mf V_{j'}\mf V_{j'}^\top-\tilde {\mf V}_{j'}\tilde {\mf V}_{j'}^\top )\mf x_{j',n},
									\end{align}
									where for the last inequality we have used the $\mf V_{j'}\mf V_{j'}^\top \mf A(j'/n)=\mf A(j'/n)$ by the argument in proving Theorem \ref{Space-Distance}. Therefore by definition, each element of $\sum_{j=1}^{m_n-w_n+1}(\tilde{\mf s}^A_{j,w_n}-\tilde{\mf s}_{j,w_n})^{\circ 2}$ has the form of  (we write $r'$ for $r+(s-1)m_n$ for to shorten the notation)
									\begin{align}
										\sum_{j=1}^{m_n-w_n+1}\Big(\sum_{r=j}^{j+w_n-1}\tilde{\mf f}_q^\top(\hat{\mf e}_{r'}-\mf (\mf I-\mf V_{r'}\mf V_{r'}^\top)\mf e_{r'})\Big)^2
										=	\sum_{j=1}^{m_n-w_n+1}\Big(	\sum_{r=j}^{j+w_n-1}\tilde{\mf f}_q^\top(\mf V_{r'}\mf V_{r'}^\top-\tilde {\mf V}_{r'}\tilde {\mf V}_{r'}^\top )\mf x_{r',n}\Big)^2\notag
										\notag\\\leq w_n\sum_{j=1}^{m_n-w_n+1}\sum_{r=j}^{j+w_n-1}\|(\mf V_{r'}\mf V_{r'}^\top-\tilde {\mf V}_{r'}\tilde {\mf V}_{r'}^\top )\mf x_{r',n}\|_F^2\leq  w_n\sum_{j=1}^{m_n-w_n+1}\sum_{r=j}^{j+w_n-1}\|(\mf V_{r'}\mf V_{r'}^\top-\tilde {\mf V}_{r'}\tilde {\mf V}_{r'}^\top )\|_F^2\|\mf x_{r',n}\|_F^2\notag\\
										\leq w_n\max_r\|\mf V_{r}\mf V_{r}^\top-\tilde {\mf V}_{r}\tilde {\mf V}_{r}^\top\|_F^2
										\max_{s}\sum_{j=1}^{m_n-w_n+1}\sum_{r=j}^{j+w_n-1}\|\mf x_{r+(m_n-1)s,n}\|_F^2
									\end{align}
									
									Using \eqref{New.C84} and Theorem \ref{Space-Distance}, we have that 
									\begin{align}
										\max_{s,q}\Big|\sum_{j=1}^{m_n-w_n+1}\Big(\sum_{r=j}^{j+w_n-1}\tilde{\mf f}_q^\top(\hat{\mf e}_{r'}-\mf (\mf I-\mf V_{r'}\mf V_{r'}^\top)\mf e_{r'})\Big)^2\Big|=O_p(\theta(n,p)^2w_n(N_n^{2/l}w_n(m_n-w_n+1)p)).
									\end{align}
									Combing with \eqref{D.20} we have shown (a)(i). To show (a) (ii) note that
									\begin{align*}
										|	\tilde{\bs \kappa}_n-\bs \kappa_n|=|III-IV|,
									\end{align*}
									where
									\begin{align}
										III=	\frac{1}{\sqrt{w_n(m_n-w_n+1)}}\sum_{j=1}^{m_n-w_n+1}(\tilde {\mf s}_{j,w_n}-\hat{\mf s}_{j,w_n})R_j,\\
										IV=\frac{1}{\sqrt{w_n(m_n-w_n+1)}}\sum_{j=1}^{m_n-w_n+1}(\frac{w_n}{m_n}\tilde{\mf s}_{m_n}-\frac{w_n}{m_n}\hat {\mf s}_{m_n})R_j.
									\end{align}
									Similarly to the proof of (a)(i)
									We now show that $(\E(|III|^l_\infty)|\FF_n)^{1/l}=O_p(\bar{\theta}(n,p,\delta,l,N_n,w_n))$, and the result that $\E(|IV|^l_\infty|
									\mf x_{i,n})^{1/l}=O_p(\bar{\theta}(n,p,\delta,l,N_n,w_n))$ will follow similarly. 
									Again given data, $III$ is a $N_n(p-d)$ dimensional Gaussian vector. Therefore, 
									\begin{align}\label{D.201}
										\E(|III|^l_\infty|\FF_n)^{1/l}=O_p(\frac{1}{\sqrt{w_n(m_n-w_n+1)}}|\sum_{j=1}^{m_n-w_n+1}(\tilde{\mf s}_{j,w_n}-\hat{\mf s}_{j,w_n})^{\circ 2}|^{1/2}_\infty (N_np)^{1/l})
									\end{align}
									Moreover, 
									\begin{align}
										|\sum_{j=1}^{m_n-w_n+1}(\tilde{\mf s}_{j,w_n}-\hat{\mf s}_{j,w_n})^{\circ 2}|=\max_{q,s}\sum_{j=1}^{m_n-w_n+1}\Big(\sum_{r=j}^{j+w_n-1}(\hat{\mf f}_q-\tilde{\mf f}_q)^\top \hat{\mf e}_{r+(m-1)s}\Big)^2
									\end{align}
									Observe \eqref{Is1s2}. Use \eqref{S102} in Proposition \ref{PropG1}, with the same argument yielding \eqref{summmarize1} and \eqref{summarize2}, we show  $(\E(|III|^l_\infty)|\FF_n)^{1/l}=O_p(\bar{\theta}(n,p,\delta,l,N_n,w_n))$ and hence (a)(ii) follows.

									To show (b), we only need to verify assertion (a) and (b) in the proof of Proposition \ref{definitioniota}. Notice that
									$\mf I-\mf V_i\mf V_i^\top $ is a projection matrix, hence 
									\begin{align}\label{fivbound}
										\|\tilde{\mf f}_q^\top(\mf I-\mf V_i\mf V_i^\top)\|^2_2=
										\tilde{ \mf f}_q^\top (\mf I-\mf V_i\mf V_i^\top)\tilde{ \mf f}_q=1-\|\tilde{\mf f}^\top_q\mf V_i\|^2_2\in [1-\eta, 1]
									\end{align}
									for all $1\leq q\leq p$ and $1\leq i\leq n$, and therefore (a) in the proof of Proposition \ref{definitioniota} follows. Moreover, (b) in the proof of Proposition \ref{definitioniota} follows from \eqref{fivbound} and Proposition \ref{Newprop}. Thus we prove (b).
									
									Finally (c) follows exactly the proof of claim \eqref{Junly14-S86}. Details are ommitted for the sake of brevity. \hfill $\Box$
									\medskip
									\subsection{Proof of Theorem \ref{Power}}
									

									{\it Proof of (i)} Redefine $T_n$ in \eqref{TnC5} by replacing $\mf f_i's$ with $\tilde{\mf f_i}'s$.
									Using exactly the argument to the proof of proposition \ref{Jan23-Lemma6} (the only difference is $\tilde {\mf f}_q$ and  $\mf f_q$) and Corollary \ref{CoroD1} for estimating $d$ in the alternative guarantees  that
									\begin{align}\label{hatdn}
										\p(|\hat T_n- T_n|\geq g_nN_n^{1/l}p^{\delta+\frac{1}{2}}n^{-1/2})  =O(g_n^{-\frac{l}{l+1}}+\log^{-1/2} n).
									\end{align}
									Thus it suffices to consider $T_n$. Notice that $T_n\geq m_n^{-1/2} \tilde {\mf f}_v^\top\sum_{i=1}^{m_n}
									\mf x_{i+(s-1)m_n}=I+II$, where $s$ is the integer such that
									$(s-1)m_n+1\leq \lf nt\rf\leq sm_n$, and that 
									\begin{align}\label{newD58}
										I=m_n^{-1/2}\tilde {\mf f}_v^\top\sum_{i= 1}^{m_n}
										\mf A_{i+(s-1)m_n}\mf z_{i+(s-1)m_n},\\
										II=m_n^{-1/2} \tilde {\mf f}_v^\top\sum_{i= 1}^{m_n}
										\mf e_{i+(s-1)m_n}.\label{newD59}
									\end{align}
									By condition (M8) it follows that $II=O_p(1)$. For $I$, it can be written as $I_1+I_2$, where
									\begin{align}
										I_1=m_n^{-1/2}\tilde {\mf f}_v^\top	\mf A(t)\sum_{i= 1}^{m_n}
										\mf z_{i+(s-1)m_n},~~~
										I_2=m_n^{-1/2}\tilde {\mf f}_v^\top\sum_{i= 1}^{m_n}
										(\mf A_{i+(s-1)m_n}-\mf A(t))\mf z_{i+(s-1)m_n}.
									\end{align} 
									Further write $ z^\circ_{i+(s-1)m_n}(u,t)=\frac{\tilde{\mf f}_v^\top \mf A(t)}{|\tilde {\mf f}_v^\top \mf A(t)|}\mf Q(u,\FF_{i+(s-1)m_n})$. 
									Note that
									\begin{align}\label{I1}
										\frac{1}{|\tilde{\mf f}_v^\top \mf A(t)|}	I_1=m^{-1/2}_{n}\frac{\tilde {\mf f}_v^\top \mf A(t)}{|\tilde{\mf f}_v^\top \mf A(t)|}\sum_{i=1}^{m_n}\mf z_{i+(s-1)m_n}=
										m_n^{-1/2}\sum_{i=1}^{m_n} z^\circ_{i+(s-1)m_n}(\frac{i+(s-1)m_n}{n},t). 
									\end{align}
									On the other hand, since $|z^\circ_{i+(s-1)m_n}(u_1,t)-z^\circ_{i+(s-1)m_n}(u_2,t)|\leq |\mf Q(u_1,\FF_{i+(s-1)m_n})-\mf Q(u_2,\FF_{i+(s-1)m_n})|$,  by (M3) it follows that there exists a large constant $M$ such that
									\begin{align*}
										\|z_{i+(s-1)m_n}^\circ(u_1,t)-z_{i+(s-1)m_n}^\circ(u_2,t)\|_{\mathcal L^2}\leq M|u_1-u_2|
									\end{align*}
									therefore we have by the triangle inequality
									\begin{align}\label{D61}
										\|m_n^{-1/2}\sum_{i=1}^{m_n} z^\circ_{i+(s-1)m_n}(\frac{i+(s-1)m_n}{n},t)\|_{\mathcal L^2}=\|m_n^{-1/2}\sum_{i=1}^{m_n} z^\circ_{i+(s-1)m_n}(t,t)\|_{\mathcal L^2}+O(m_n^{3/2}/n)
									\end{align}
									Write  $z^\circ_{i+(s-1)m_n}(t):=z^\circ_{i+(s-1)m_n}(t,t):=Q^\circ(t,\FF_{i+(s-1)m_n})$ and also that $\Gamma^\circ(t,i-j)=\E(Q^\circ(t,\FF_{i+(s-1)m_n})Q^\circ(t,\FF_{j+(s-1)m_n}) )$. 
									Straightforward calculations show that $\sup_t\|Q^\circ(t,\FF_i)-Q^\circ(t,\FF^{(0)}_i)\|_{\mathcal L^l}=O(\delta^z_{l}(k))$
									Together with (M1) and Lemma 5 of \cite{zhou2010simultaneous},  it follows that for  $\Gamma^\circ(t, k)=O((|k|\log |k|)^{-2})$. As a consequence,
									\begin{align}\label{D62}
										\|\sum_{i=1}^{m_n} z^\circ_{i+(s-1)m_n}(t)\|^2_{\mathcal L^2}/m_n=\sum_{j=-{m_n}}^{m_n}(m_n-|j|)\Gamma^\circ(t,j)/m_n=\sum_{j={-m_n}}^{m_n}\Gamma^\circ(t,j)+O(1/m_n)\notag\\
										=\sum_{j=-m_n}^{m_n}\Gamma^\circ(t,j)+o(1)=\sum_{j=-\infty}^{\infty}\Gamma^\circ(t,j)+o(1)\geq 	\inf_{t\in [0,1]}\lambda_{\min}(\sum_{k=-\infty}^{\infty}\bs \Sigma_z(t,k))+o(1)>0
									\end{align}
									where for the first inequality we use the fact that the norm of $\frac{\tilde{\mf f}_v^\top \mf A(t)}{|\tilde {\mf f}_v^\top \mf A(t)|} $ is $1$, and for the second inequality we have used the condition \eqref{D.56}. 
									Together with \eqref{D61} we shall see that
									\begin{align}
										\|m_n^{-1/2}\sum_{i=1}^{m_n} z^\circ_{i+(s-1)m_n}(\frac{i+(s-1)m_n}{n},t)\|_{\mathcal L^2}>0
									\end{align}
									Combining with \eqref{I1} we have that $\|I_1\|_{\mathcal L^2}\gtrapprox \sqrt{\log (N_n p)}$.  On the other hand, using the condition that $|\frac{\partial}{\partial t}|\tilde{ \mf f}_q^\top \mf A(t)||\leq M |\tilde{ \mf f}_q^\top \mf A(t)|$ for all $t\in [0,1]$  and Burkholder inequality we have $\|I_2\|_{\mathcal L^2}=o(\|I_1\|_{\mathcal L^2})$. As a result, by the definition of $T_n$ and \eqref{newD58}, \eqref{newD59},  
									we have \begin{align}
										\p(T_n\geq \sqrt{\log (N_n p)}\iota_n^{-1})=1
									\end{align}
									where $\iota_n$ is a sequence diverging at an arbitrarily slowly rate. By \eqref{hatdn} we have 
									\begin{align}\label{NewD66}
										\p(|\hat T_n|\geq \sqrt{\log (N_n p)}\iota_n^{-1})\geq \p(T_n\geq \sqrt{\log (N_n p)}\iota_n^{-1}+g_nN_n^{1/l}\theta(n,p)p^{\frac{1}{2}})\notag\\-
										\p(|\hat T_n-T_n|\geq g_nN_n^{1/l}\theta(n,p)p^{\frac{1}{2}})\rightarrow 1
									\end{align}
									as $n\rightarrow \infty$.
									
									On the other hand, by Proposition \ref{PropD2} we have that
									\begin{align}
										\E (|\bs \kappa_n-\tilde {\bs\kappa}_n^A|_\infty|\FF_n)=O_p(\bar{\theta}(n,p,\delta,l,N_n,w_n)).\end{align}
									Moreover,  given $\FF_n$, $\tilde {\bs \kappa}_n^A$ is an $N_n(p-d)$ dimensional Gaussian process, with $v_{th}$ component has the variance of
									\begin{align}
										\tilde \sigma^{A,2}_v:= \frac{1}{w_n(m_n-w_n+1)}\sum_{j=1}^{m_n-w_n+1}(\tilde {\mf s}^{A}_{j,w_n,v}-\frac{w_n}{m_n}\tilde {\mf s}^{A}_{m_n,v})^{\circ 2}
									\end{align}
									where $\tilde {\mf s}^A_{j,w_n,v}$ and $\tilde {\mf s}^A_{m_n,v}$ is the $v_{th}$ entry of $\tilde {\mf s}^A_{j,w_n}$ and $\tilde {\mf s}^A_{m_n}$, respectively. By the proof of step (i) of Theorem \ref{Boots-thm5}, we have
									\begin{align}\label{E74}
										\Big\|\max_v|\tilde \sigma^{A,2}_v-\E \tilde \sigma^{A,2}_v| \Big\|_{\mathcal L^{q^*/2}}=O(\sqrt{w_n/m_n}(N_n(p-d))^{2/q^*})=o(1).
									\end{align}
									Observe that 
									\begin{align}
										\E(\tilde \sigma^{A,2}_v)
										=\frac{1}{w_n(m_n-w_n+1)}\sum_{j=1}^{m_n-w_n+1}\E(\tilde {\mf s}^A_{j,w_n,v}-\frac{w_n}{m_n}\tilde {\mf s}^A_{m_n,v})^2\notag\\
										\leq \frac{2}{w_n(m_n-w_n+1)}\sum_{j=1}^{m_n-w_n+1}\E(\tilde {\mf s}^A_{j,w_n})^ 2+\frac{2}{w_n(m_n-w_n+1)}\sum_{j=1}^{m_n-w_n+1}\E(\frac{w_n}{m_n}\tilde {\mf s}^A_{m_n})^2
										\notag\\:=2A_{1,v}+2A_{2,v}
									\end{align}
									where $A_{1,v}$ and $A_{2,v}$ are defined in an obvious way. 
									For $A_{1,v}$ it is bounded uniformly by 
									\begin{align}
										\max_{s,q}\frac{1}{w_n(m_n-w_n+1)}\sum_{j=1}^{m_n-w_n+1}
										\E\Big(\sum_{r=j}^{j+w_n-1}\tilde{\mf f}_q^\top (\mf I- {\mf V}_{r+(m_n-1)s} {\mf V}^\top_{r+(m_n-1)s})\mf e_{r+(m_n-1)s}\Big)^2\notag
										\notag\\\leq \max_{s,q}\frac{2}{w_n(m_n-w_n+1)}\sum_{j=1}^{m_n-w_n+1}
										\E\Big(	\sum_{r=j}^{j+w_n-1}\tilde{\mf f}_q^\top \mf e_{r+(m_n-1)s}\Big)^2\notag\\+
										\max_{s,q}\frac{2}{w_n(m_n-w_n+1)}\sum_{j=1}^{m_n-w_n+1}
										\E	\Big(\sum_{r=j}^{j+w_n-1}\tilde{\mf f}_q^\top {\mf V}_{r+(m_n-1)s}{\mf V}^\top_{r+(m_n-1)s}\mf e_{r+(m_n-1)s}\Big)^2.
									\end{align}
									Notice that by assumption
									\begin{align}
										|\tilde{\mf f}_q^\top  {\mf V}_{r+(m_n-1)s} {\mf V}^\top_{r+(m_n-1)s}|^2=|\tilde{\mf f}_q^\top  {\mf V}_{r+(m_n-1)s}|^2\in (0,1]
									\end{align}
									which leads to  that as $n\rightarrow \infty$, by the proof of Proposition \ref{Newprop}
									\begin{align}
										\max_v A_{1,v}\leq 4\sup_{t}\sum_k\|\bs \Sigma_e(t,k)\|_2+o(1)
									\end{align}
									By a similar argument $
									\max_v A_{2,v}\leq 4\sup_{t}\sum_k\|\bs \Sigma_e(t,k)\|_2+o(1)$. Hence for sufficiently large $n$,
									$\max_v\E(\tilde \sigma^{A,2}_v)\leq 17\sup_{t}\sum_k\|\bs \Sigma_e(t,k)\|_2$.
									Consider the event $$\mathcal E_n=\{\sup_v\tilde \sigma_v^{A,2}\leq 18\sup_{t}\sum_k\|\bs \Sigma_e(t,k)\|_2\}.$$ By \eqref{E74}, $\p(\mathcal E_n)\rightarrow 1$.
									Now apply Lemma 2.3.4 of \cite{gine_nickl_2015} we have that on $\mathcal E_n$, almost surely
									\begin{align}\label{E49}
										\E(|\tilde {\bs \kappa}^A_n|_\infty|\FF_n)\leq 18^{1/2}\sqrt{2\log(2(N_n(p-d'))}\sup_{t}(\sum_k\|\bs \Sigma_e(t,k)\|_2)^{1/2}.
									\end{align}
									Together with  \eqref{NewD66} (i) of the theorem holds.
									
									{\it Proof of (ii)}. Consider $\mathcal E_n$ defined in the proof of Step 2. 
									For any given $\alpha$, taking $u$ such that $2\exp(-\frac{1}{\pi^2}\frac{u^2}{2\bar \eta^2})=\alpha$ where $\bar \eta^2=18\sup_{t}\sum_k\|\bs \Sigma_e(t,k)\|_2$. 
									Notice that on $\mathcal E_n$ the conditional variance of each component of $\tilde{\bs\kappa}_n^A$ given $\FF_n$ will be smaller than $\bar \eta^2$ with probability going to $1$. Then by Example 2.1.19 of \cite{gine_nickl_2015}, and \eqref{E49}, we have that $a.s.,$
									\begin{align}\label{III.1}
										\p(|\tilde {\bs \kappa}_n^A|_\infty\geq (18\sup_{t}\sum_k\|\bs \Sigma_e(t,k)\|_2)^{1/2}\sqrt{2\log 2(N_n(p-d'))}+u,\mathcal E_n|\FF_n)\leq \alpha.
									\end{align} One the other hand is straightforward to see that \eqref{hatdn} holds. Therefore we shall prove that 
									\begin{align}\label{III.2}
										\p(T_n\geq  (18\sup_{t}\sum_k\|\bs \Sigma_e(t,k)\|_2)^{1/2}\sqrt{2\log 2(N_n(p-d'))}+u)\rightarrow 1.
									\end{align} Then (ii) will hold in view of taking expectation to \eqref{III.1}, and \eqref{III.2} and the fact that $\p(\mathcal E_n)\rightarrow 1$. 
									
									In the remaining proof, we shall focus on showing \eqref{III.2}.  
									Without loss of generality, consider $q=1$ in (ii), so $\tilde x_i=\tilde{\mf f}_1^\top\mf x_{i}.$ 
									Let index set $I_{m_n,n}$ denote $\{s\in \mathbb Z^+: \frac{(2s-1)m_n}{n}\in \mathcal I, 1\leq s\leq N_n\}$. Then
									\begin{align}\label{E82}
										T_n\geq m_n^{-1/2}\max_{1\leq s\leq N_n}
										|\sum_{i=1}^{m_n}\tilde x_{i+(s-1)m_n}|\geq  m_n^{-1/2}\max_{ \text{ $s\in I_{m_n,n}$}}
										|\sum_{i=1}^{m_n}\tilde x_{i+(2s-1)m_n}|
									\end{align}
									Define
									\begin{align}\tilde \iota(n,p,q, D_n)=\min(n^{-1/8}M^{1/2}l_n^{7/8}+\gamma+(n^{1/8}M
										^{-1/2}l_n^{-3/8})^{q/(1+q)}(p \sum_{j=M}^\infty (\delta^{\tilde G}_q(j))^q)^{1/(1+q)}\notag\\+\Xi_M^{1/3}(1\vee \log (p/\Xi_M))^{2/3}),\end{align}  where 
									$\Xi_{M}=\max_{1\leq j\leq p}$  $\sum_{j=M}^\infty j\delta^{\tilde G}_{2}(j)$, and the minimum is taken over all possible values of $\gamma$ and $M$ subject to $$n^{3/8}M^{-1/2}l_n^{-5/8}\geq \max\{D_n(n/\gamma)^{1/l},l_n^{1/2}\}$$ with $l_n=\log (pn/\gamma)\vee 1$. By Theorem 2.1 of \cite{zhang2018gaussian}, there exists a sequence of centered Gaussian random variable 
									$\tilde{\mf y}_i=((\tilde y_{i,s})_{1\leq s\leq |I_{m_n,n}|})^\top$
									where $\tilde{\mf y}_i$ preserve the autocovariance structure of 
									$\tilde{\mf x}_i:=((\tilde {x}_{i+(2s-1)m_n})_{s\in I_{m_n,n}})^\top$ 
									such that 
									\begin{align}\label{E83}
										\sup_{x\in \mathbb R}|\p(m_n^{-1/2}|\sum_{i=1}^{m_n} \tilde{\mf y}_i|_\infty \leq x)-\p(m_n^{-1/2}|\max_{s\in I_{m_n,n}}\sum_{i=1}^{m_n}\tilde x_{i+(2s-1)m_n}  |\leq x)|\notag\\=
										\sup_{x\in \mathbb R}|\p(m_n^{-1/2}|\sum_{i=1}^{m_n} \tilde{\mf y}_i|_\infty \leq x)-\p(m_n^{-1/2}|\sum_{i=1}^{m_n} \tilde{\mf x}_i|_\infty \leq x)|=\tilde \iota(m_n, N_n,l,(N_n^{1/l})).
									\end{align} 
									Here the term $N_n^{1/l}$ can be obtained
									by the same argument yielding $(N_np)^{1/l}$ in the proof of Proposition \ref{definitioniota},
									and we have used the fact that $|\mathcal I|\geq \eta_1>0$, and  the fact that $p$ is fixed.
									By condition (b) $\beta>2$, $l\geq 8$, $m_n\asymp n^\alpha$ for some $\alpha>\frac{16}{5l}$, it can be verified that  $\tilde \iota(m_n, N_n,l,N_n^{1/l})=o(1).$
									
									Consider centered Gaussian vector  $\mf y=( y_1,... y_{|I_{m_n,n}|})^\top$ such that $\E(y_iy_j)=0$ for $i\neq j$, and for $1\leq s'\leq |I_{m_n,n}|$ let $s$ be the $s'_{th}$  element of $I_{m_n,n}$, and then
									\begin{align}
										&\E y_{s'}^2=\frac{1}{m_n}\E((\sum_{i=1}^{m_n}\tilde y_{i,s'})^2)=\frac{1}{m_n}\sum_{i=1}^{m_n}\sum_{j=1}^{m_n}
										\tilde{\mf f}_1^\top \E(\mf x_{i+(2s-1)m_n}\mf x^\top_{j+(2s-1)m_n})\tilde{\mf f}_1\notag\\&=
										\sum_{k=-\infty}^\infty \tilde {\mf f}_1^\top \mf A(\frac{(2s-1)m_n}{n})\bs \Sigma_z(\frac{(2s-1)m_n}{n},k) \mf A(\frac{(2s-1)m_n}{n})^\top \tilde{\mf f}_1+\sum_{k=-\infty}^\infty \tilde {\mf f}_1^\top \bs \Sigma_e(\frac{(2s-1)m_n}{n},k)\tilde{\mf f}_1+o(1)
										\notag\\&\geq |\tilde{\mf f}_1^\top \mf A(\frac{(2s-1)m_n}{n})|^2\lambda_{\min}(\sum_{k=-\infty}^\infty \bs \Sigma_z(\frac{(2s-1)m_n}{n},k))\geq \min_{t\in \mathcal I}|\tilde {\mf f}_1^\top \mf A(t)|^2\underline \lambda_z(1-o(1)):=\eta^2.
									\end{align}
									where the $o(1)$ term in the last line is positive, and we obtain the second line by similar but easier argument to the proof of Proposition \ref{Newprop}, utilizing the fact that $p$ is fixed.
									Moreover, by Lemma 5 of \cite{zhou2010simultaneous}, $\E(\tilde G(t,\FF_i)\tilde G(t,\FF_j))=|i-j|^{-(1+\beta)}$.  Therefore, 
									\begin{align}
										\max_{u,v}|\frac{1}{m_n}\E((\sum_{i=1}^{m_n}\tilde y_{i,u})(\sum_{i=1}^{m_n}\tilde y_{i,v}))-\E(y_uy_v)|=O(m_n^{-\beta}).
									\end{align}
									As a consequence by Theorem 2 of \cite{chernozhukov2015comparison},
									\begin{align}\label{E85}
										\sup_{x\in \mathbb R}|\p(m_n^{-1/2}|\sum_{i=1}^{m_n} \tilde{\mf y}_i|_\infty \leq x)-\p(|{\mf y}_i|_\infty \leq x)|=O\big(m_n^{-\beta/3}\big(1\vee \log (N_nm_n^\beta)\big)^{2/3}\big)=o(1)
									\end{align}
									By Lemma \ref{New.LemmaE3},
									there exists $i.i.d$ $N(0,1)$ random variables $Z_1,...,Z_{|I_{m_n,n}|}$ such that
									\begin{align}\label{E58-new}
										\p(|{\mf y}_i|_\infty \leq x)|\leq \p(\eta \max_{1\leq i\leq |I_{m_n,n}|}|Z_i|\leq x)=\p(max_{1\leq i\leq |I_{m_n,n}|}|Z_i|\leq x/\eta).
									\end{align}
									Consider 
									\begin{align}
										a_n=\sqrt{2\log |I_{m_n,n}|},~~ b_n=a_n-\frac{\log \log |I_{m_n,n}|+\log \pi}{2a_n}
									\end{align}
									and notice that $|I_{m_n,n}|=N_n|\mathcal I|/2+O(1)$.
									Then by Theorem 2.7.1 of \cite{gine_nickl_2015},  for any $x$ such that $a_n(x/\eta-b_n)\rightarrow z$ as $n\rightarrow \infty$.
									\begin{align}\label{E60-new}
										\lim_{n\rightarrow \infty}  \p(max_{1\leq i\leq |I_{m_n,n}|}|Z_i|\leq x/\eta)
										=\p(a_n(max_{1\leq i\leq|I_{m_n,n}|}|Z_i|-b_n)\leq a_n(x/\eta-b_n))=\exp(-e^{-z}).
									\end{align}
									Taking $x=18\sup_{t}\sum_k\|\bs \Sigma_e(t,k)\|_2\sqrt{2\log 2(N_n(p-d'))}+u$. Since $\|\bs \Sigma_{e}(t,k)\|_F=0$ for $k\geq 1$ and that   $$\min_{t\in \mathcal I}(|\tilde{\mf f}^\top_q\mf A(t)|)> (18+\gamma_0)^{1/2}	\underline \lambda^{-1/2}_z\sup_t \|Var(\mf H(t,\FF_0))\|^{1/2}_2 ,$$ 
									we have as $n\rightarrow \infty$, $a_n(x/\eta-b_n)\rightarrow -\infty$, which combining with \eqref{E58-new} and \eqref{E60-new} leads to 
									\begin{align}
										\p(|{\mf y}_i|_\infty \leq (18\sup_{t}\sum_k\|\bs \Sigma_e(t,k)\|_2)^{1/2}\sqrt{2\log 2(N_n(p-d'))}+u)\rightarrow 0.
									\end{align}
									Combining with \eqref{E82},\eqref{E83} and \eqref{E85}, \eqref{III.2} follows and (ii) of the theorem holds.
									\hfill $\Box$
									\begin{lemma}\label{New.LemmaE3}
										Let $x_i$, $1\leq i\leq n$ be independent normal random variables with variance $\sigma_1^2$,...,$\sigma_n^2$. Let $y_i$, $1\leq i\leq n$ be $i.i.d.$ $N(0,1)'s$.  Let $\underline{\sigma}^2=\min_{1\leq i\leq n}\sigma_i^2$.  Then we have for any $x\in \mathbb R$
										\begin{align}
											\p(\max_{1\leq i\leq n}|x_i|\leq x)\leq \p(\underline \sigma \max_{1\leq i\leq n}|y_i|\leq x).
										\end{align}
										
									\end{lemma}
									{\it Proof.} For any $i$, we have
									\begin{align}
										\p(|x_i|\leq x)=\p(|y_i|\leq x/\sigma_i)\leq 
										\p(|y_i|\leq x/\underline \sigma)=\p(\underline \sigma |y_i|\leq x).
									\end{align}
									Then the lemma follows from independence. \hfill $\Box$
									
									\medskip
									\bibliographystyle{apalike}
									\bibliography{Ref}

\end{document}